# Degrees of Freedom Region of the MIMO $X$ channel with an Arbitrary Number of Antennas

Adrian Agustin, *Member, IEEE*, and Josep Vidal, *Member, IEEE*

***Abstract*— We characterize the total degrees of freedom (DoF) of the full-rank MIMO $X$ channel with arbitrary number of antennas at each node. We elucidate that the existing outer bound is tight for any antenna configuration and provide transmit and receive filter designs that attain this outer bound. The proposed achievable scheme exploits channel extensions in terms of both, symbol and asymmetric complex signaling when the communication is carried out over a constant channel case, and is also applicable to time varying channels. The proposed scheme represents a general framework for the derivation of the total DoF of any two-by-two multiuser channels. Furthermore, the rank-deficient MIMO channels case is naturally addressed, and it is shown that the total DoF of the interference (IC) and MIMO $X$ channels are in general superior to the full rank MIMO case.**

Manuscript received September 26, 2012. This work has been done in the framework of the project TROPIC FP7 ICT-2011-8-318784, funded by the European Community. Also by the Spanish Science and Technology Commissions and EC FEDER funds through projects: TEC2010-19171/TCM and CONSOLIDER INGENIO CSD2008-00010 COMONSENS, and by project 2009SGR1236 (AGAUR) of the Catalan Administration. The material in this paper was presented in part at the IEEE International Conference on Communications, Kyoto (ICC), Japan, June 2011.

A. Agustin and J. Vidal are with Signal Processing in Communications Group at the Department of Signal Theory and Communications, Universitat Politècnica de Catalunya (UPC), Barcelona, Spain (e-mail: {adrian.agustin, josep.vidal}@upc.edu).

## I. INTRODUCTION

Wireless communications have become an important research field in recent years, spurred basically by two factors: the scarcity of the wireless medium and a worldwide increase of cellular data traffic demand driven by high penetration of smartphones and internet-based social networks. Consequently, wireless cellular networks have to be redesigned in order to incorporate efficient multiuser techniques able to satisfy these traffic demands. The use of of multi-antenna terminals (multiple-input multiple-output, MIMO) in these systems offers the possibility of even further increasing the data rate communications and/or combating the wireless channel impairments.

The most common types of multiuser MIMO channels that come up in wireless communications (beyond the trivial point-to-point MIMO (PTP) or single-user) are the MIMO multiple access channel (MAC), the MIMO broadcast channel (BC) and the MIMO interference channel (IC). The properties of all of them can be analyzed in the high signal-to-noise ratio regime by means of *multiplexing gain* or *degrees of freedom* (DoF), i.e. how the system rate scales in the high power regime. In general, the maximum DoF are obtained by means of spatial zero forcing linear filters, [1], except for the IC with more than two users. In this latter case, a new transmission technique has been introduced: *interference alignment* (IA). This technique pursues to design the transmit filters of different terminals in such a way that each receiver observes the interfering signals overlapped on the same spatial subspace, while the intended signal comes up in a different subspace. The IA concept was first employed in example 7 of [2] and by *index coding* literature. Later on, [3][4] observed it for the $X$ channel and crystallized the concept. An overview of the interference alignment benefits can be found in [6] and latest advances in the 3-user IC in [5].

The two-user MIMO $X$ channel (XC) consists of two sources/transmitters ($T_1$, $T_2$) and two destinations/receivers ($R_1$, $R_2$), where each transmitter has independent messages to both receivers. This channel generalizes the two-by-two user communications and subsumes all the previous multiuser channels (PTP, MAC, BC and IC). However, its performance in terms of DoF goes beyond all of them. For example, when all terminals are equipped with the same number of antennas $M$ and channel matrices are full-rank, the two-user IC, BC, MAC get $M$ DoF [1][3][4], while the MIMO $X$ has $4M/3$, [7][8]. Fig. 1 illustrates a MIMO XC where each node is equipped with $M$=3 antennas, i.e. it can exploit up to three spatial dimensions. There are two messages intended to receiver $R_1$: $W_{11}$ from transmitter $T_1$ and $W_{12}$ from transmitter $T_2$, and two messages intended to receiver $R_2$: $W_{21}$ and and $W_{22}$ from transmitter $T_1$ and $T_2$, respectively. The messages are precoded in order to define a common overlapped subspace at the unintended receiver, while keeping the intended messages and interference in different subspaces. For example, messages $W_{11}$, $W_{12}$ at $R_2$ and messages $W_{21}$, $W_{22}$ at $R_1$ are enclosed in the same subspace at the unintended receiver, see dotted lines in Fig. 1, but occupy different subspaces at the desired receiver, as solid lines show in Fig. 1. In this regard, each receiver employs two spatial dimensions for its intended independent messages, while the third one is reserved for the aligned interference signal.

An important feature shown by the MIMO XC and IC is that the optimal DoF might be non-integer, which means that the communication has to be carried out over multiple channel extensions.

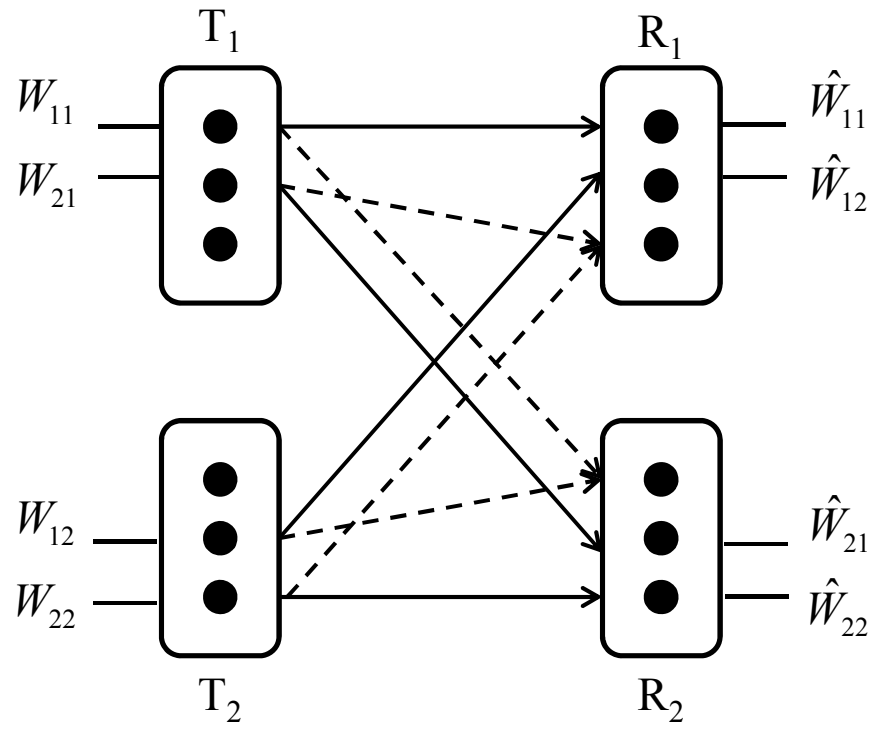

Fig. 1. MIMO $X$ channel with 2 sources-destinations, where sources $T_1$, $T_2$ and destinations $R_1$, $R_2$ are equipped with $M_1$=$M_2$=$N_1$=$N_2$ =3 antennas. $T_1$ sends messages $W_{11}$ and $W_{21}$ to $R_1$ and $R_2$, respectively, while $T_2$ sends messages $W_{12}$ and $W_{22}$. In this regard, $R_1$ has to decode messages $W_{11}$, $W_{12}$ and $R_2$ messages $W_{21}$, $W_{22}$. Intended signals are referenced by solid lines, while unintended signals are denoted by dotted lines.

The $X$ channel has been analyzed for different number of single antenna transmitters and receivers, see for example [9]. The transmission scheme is based on arbitrarily long channel extensions in most cases. Likewise, a new IA scheme named *layered interference alignment* is introduced in [10] which attains the total DoF for $K$ transmitters ($K$>2) and 2 receivers, all terminals equipped with $M$ antennas. Similarly, in the same type of $X$ network ($K$×2), a scheme is proposed in [11] to design rank-one precoders (i.e. beamformers) in order to improve the total sum-rate of the system. When there are $K$=2 transmitters/receivers, the outer bound region for the DoF is provided in [7], where it is shown that the outer bound becomes tight when terminals have $M$ antenna elements and non-degenerate channel coefficients vary over time/frequency. On the other hand, if channel coefficients are fixed during the communication, then the transmission scheme proposed in [7] is only optimal for $M$>1. The constant channel case for $M$=1 is addressed in [8] where the channel extensions needed to deal with the non-integer DoF are performed by means of symbol extension and *asymmetric complex signaling*. This latter concept assumes that input covariance matrices are chosen to be complex, not circularly symmetric. This new concept allows exploiting the real and imaginary dimensions of complex channels independently. For example, the transmit precoders could be designed in such a way that each receiver observes the desired signal in the real dimension, while interference is enclosed in the imaginary dimension.

In spite of all these progresses, there are still open issues regarding the DoF in the two-user MIMO $X$ channel: a) arbitrary number of antennas at terminals and b) rank-deficient channels. When all terminals have arbitrary number of antennas a precoder design is proposed in section III of [7] that achieves an integer inner bound on DoF, but no proof of optimality is provided when the DoF are non-integers. In [12] we present a precoding scheme that is able to get the outer bound DoF for certain cases involving non-integer DoF per message, but without any proof of optimality in general. The precoding scheme used in [12] is based on the generalized singular value decomposition (GSVD) defined in [15][16]. On the other hand, the DoF's literature usually assumes uncorrelated channel coefficients, i.e. full-rank matrices associated to a rich propagation scattering that might be present in wireless scenarios when transmit and receivers have a non-line of sight (NLOS). However, such assumption is not always valid in wireless networks where poor propagation scattering might produce rank-deficient channels, for example when there is a line-of-sight situation (LOS). In this regard, the DoF of the rank-deficient MIMO IC have been investigated in [13][14], elucidating that the attained DoF are larger than in full rank case. Nevertheless, up to author's knowledge, there are not rank-deficient studies for other type of multiuser channels.

### A. Contributions

The objective of the present paper is to provide a general framework for the derivation of the optimal DoF of the two-user MIMO XC and its subsumed multiuser channels by means of the GSVD-based precoding considered in [12]. The GSVD has been adopted for defining a coordinated beamforming strategy in [17] and for deriving MIMO linear precoders [18]. Additionally, GSVD-based precoding [12] is employed for maximizing the rate gain of MIMO XC at low-medium signal-to-noise ratio regime in [19]. The nice property of the GSVD is that given two channel matrices, this decomposition naturally provides the generating basis of the overlapped and null subspaces, which will be used to design the precoders aimed to either align the interference or avoid interference at the unintended receiver. The main contributions are:

- We prove that the outer bound on the total DoF for the MIMO $X$ channel derived in [7] is also tight for any antenna configuration when the channel coefficients are constant over the communication and channels are full-rank. Thus, we fully characterize two-user MIMO $X$ channel in terms of total DoF.

- The GSVD-based precoding proposed in section VI uses a finite number of channel extensions and combines symbol extensions with asymmetric complex signaling. The obtained precoders are based on the interference alignment concept and null-steering transmission. We conjecture, supported by computational experiments, that the achievable DoF region of the proposed scheme attains the outer bound region of DoF. Moreover, the scheme addresses the DoF of the MIMO BC, MAC and IC as a function of the number of messages to be transmitted.

- The proposed scheme permits to derive the achievable DoF of the rank-deficient MIMO $X$ channel and the subsumed multiuser channels. It turns out that the attained DoF in rank-deficient channels can be superior to those ones obtained with full-rank channels. Additionally, we show that transmit cooperation is beneficial in terms of DoF when the rank of the channel matrices is large. Otherwise, the same DoF can be attained with and without cooperation.

### B. Organization

The present paper is organized as follows. Section II introduces the system model that describes the received and transmitted signals in the MIMO XC. It also presents the feasibility conditions required to have linear receive and transmit filters. Finally, section II discusses the reciprocal MIMO $X$ network, a required approach to prove the optimality of the attained DoF with arbitrary number of antennas. Section III describes the outer bounds in terms of DoF for the MIMO XC. In section IV the main results obtained in this work are presented, elucidating the optimal total DoF of the full-rank MIMO XC and describing the achievable region of DoF. Section V describes the Generalized Singular Value Decomposition, the tool used to obtain the generating basis of the overlapped and null-steering spaces. Such basis will be used to derive the transmit precoders that exploit the interference alignment or null-steering concepts. Afterwards, section VI introduces the proposed achievable scheme intended for the full rank channel case. For ease of exposition we have considered the constant channel case, but the achievable scheme can be easily applied to time varying channels as it is described in section VI.E. Section VII is devoted to show closed-form DoF results for some particular antenna configurations. Moreover, section VIII reveals that the proposed scheme is a general framework for deriving the DoF of two-by-two multiuser channels. Finally, section IX generalizes the proposed achievable scheme in order to deal with rank-deficient channels.

### C. Notation

Matrix $\mathbf{I}_{m\times n}$ has *m* rows and *n* columns with elements of the main diagonal equal to one and all remaining elements equal to zero. Similarly, $\mathbf{0}_{m\times n}$ denotes a matrix with *m* rows and *n* columns where all elements are equal to zero. Assuming that matrices **A**, **B** have dimensions (*m*,*d*) and (*n*,*m*), operator $\mathbf{A}\perp\mathbf{B}$ denotes that $\mathbf{BA}=\mathbf{0}_{n\times d}$. *span*(**A**) is the subspace generated by the columns of **A**. *rank* returns the rank of a matrix, $\otimes$ stands for the Kronecker operator. Function max($x$,0) is denoted by $(x)^+$. Operator $[\mathbf{A}]^{1:i}$ selects the columns of **A** denoted by vector [1:1:*i*]. Operator $\lfloor a \rfloor$ rounds number *a* to the nearest integer towards minus infinity. $\mathbb{C},\mathbb{R}$ represent the complex and real numbers, while $\mathbb{Z}_+$ denotes the non-negative integers.

## II. System Model

The two-user MIMO *X* channel introduced in Fig. 1 consists of four terminals, two transmitters equipped with $M_1$, $M_2$ antennas, and two receivers with $N_1$, $N_2$ antennas, respectively. This channel is described by the following input-output equations,

$$\begin{aligned}\mathbf{y}_1&=\mathbf{H}_{11}\mathbf{x}_1+\mathbf{H}_{12}\mathbf{x}_2+\mathbf{n}_1\\ \mathbf{y}_2&=\mathbf{H}_{21}\mathbf{x}_1+\mathbf{H}_{22}\mathbf{x}_2+\mathbf{n}_2\end{aligned}\qquad(1)$$

where $\mathbf{y}_1\in\mathbb{C}^{N_1\times1},\mathbf{y}_2\in\mathbb{C}^{N_2\times1}$ are received signal vectors at each destination, $\mathbf{n}_1\in\mathbb{C}^{N_1\times1}$, $\mathbf{n}_2\in\mathbb{C}^{N_2\times1}$ stand for the received noise, which is assumed additive white Gaussian noise (AWGN), $\mathbf{x}_1\in\mathbb{C}^{M_1\times1},\mathbf{x}_2\in\mathbb{C}^{M_2\times1}$ are the input vectors at the transmitters and finally, the channel matrix between the *i*th receiver and *j*th transmitter ($R_i$-$T_j$ in Fig. 1) is defined by $\mathbf{H}_{ij}\in\mathbb{C}^{N_i\times M_j}$.

Every transmitter sends two independent messages, one intended to each destination. Consequently, we decompose the input vector at each transmitter as a linear combination of the symbols associated to different messages as,

$$\mathbf{x}_j=\mathbf{P}_{1j}\mathbf{s}_{1j}+\mathbf{P}_{2j}\mathbf{s}_{2j}\qquad(2)$$

where $\mathbf{P}_{ij}\in\mathbb{C}^{M_j\times d_{ij}}$ is the linear symbol precoder employed by the message sent by the *j*th transmitter and intended to the *i*th receiver, $\mathbf{s}_{ij}\in\mathbb{C}^{d_{ij}\times1}$ contains the $d_{ij}$ symbol streams extracted from a Gaussian codebook, used to convey message $W_{ij}$.

Each intended message is decoded independently with a linear receiver, assuming the effect of remaining messages as additive noise. Because nodes are equipped with multiple antennas, we can assume that some of the symbols at each message could be transmitted in such a way they do not generate any kind of interference to the non-intended receiver, i.e. *null-steering* transmission. The remaining ones do generate interference but minimizing the affected spatial dimensions at unintended receivers thanks to *interference alignment*-based techniques as suggested in [4]. Hence, the transmitted symbol streams per message become,

$$d_{ij}=d_{ij}^{(IA)}+d_{ij}^{(NS)}\qquad(3)$$

where $d_{ij}^{(IA)},d_{ij}^{(NS)}$ denote the number of symbol streams using interference alignment and null-steering concepts at each precoder. In this regard, the input vector at the *j*th transmitter defined in (2) can be decomposed as,

$$\mathbf{x}_j=\underbrace{\begin{bmatrix}\mathbf{V}_{1j} & \mathbf{Z}_{1j}\end{bmatrix}}_{\mathbf{P}_{1j}}\underbrace{\begin{bmatrix}\mathbf{s}_{1j}^{(IA)}\\ \mathbf{s}_{1j}^{(NS)}\end{bmatrix}}_{\mathbf{s}_{1j}}+\underbrace{\begin{bmatrix}\mathbf{V}_{2j} & \mathbf{Z}_{2j}\end{bmatrix}}_{\mathbf{P}_{2j}}\underbrace{\begin{bmatrix}\mathbf{s}_{2j}^{(IA)}\\ \mathbf{s}_{2j}^{(NS)}\end{bmatrix}}_{\mathbf{s}_{2j}},\quad j=1,2\qquad(4)$$

where $\mathbf{V}_{ij}\in\mathbb{C}^{M_j\times d_{ij}^{(IA)}},\mathbf{Z}_{ij}\in\mathbb{C}^{M_j\times d_{ij}^{(NS)}}$ are the linear *interference alignment* and the null-steering precoders used by message $W_{ij}$, respectively. Taking into account these definitions, the received signal is given by,

$$\mathbf{y}_i=\mathbf{H}_{i1}\mathbf{V}_{i1}\mathbf{s}_{i1}^{(IA)}+\mathbf{H}_{i2}\mathbf{V}_{i2}\mathbf{s}_{i2}^{(IA)}+\mathbf{H}_{i1}\mathbf{Z}_{i1}\mathbf{s}_{i1}^{(NS)}+\mathbf{H}_{i2}\mathbf{Z}_{i2}\mathbf{s}_{i2}^{(NS)}+\mathbf{i}_i+\mathbf{n}_i\quad i=1,2\qquad(5)$$

where the interfering signal at *i*th receiver due to the non-intended messages is defined by,

$$\mathbf{i}_i=\mathbf{H}_{i1}\mathbf{V}_{k1}\mathbf{s}_{k1}^{(IA)}+\mathbf{H}_{i2}\mathbf{V}_{k2}\mathbf{s}_{k2}^{(IA)}\qquad i,k=1,2\quad i\neq k\qquad(6)$$

*A. Constant Channel Extensions*

The outer bound region of the MIMO *X* channel is characterized in general by non-integer DoF [7], which imposes certain drawbacks for defining the number of symbol streams per message and checking when the outer bound becomes tight. Symbol extension [7] and asymmetric complex signaling [8] are techniques that allow addressing such issue. Since our objective is to provide an achievable scheme for arbitrary number of antennas, we assume that the transmission is carried out over $2T$ channel extensions using both techniques when the channel remains constant over the $T$ channel uses. Hence, the new channel matrices to be considered in our signal model are defined by,

$$\hat{\mathbf{H}}_{ij} = \mathbf{I}_T \otimes \tilde{\mathbf{H}}_{ij}, \quad \tilde{\mathbf{H}}_{ij} = \begin{bmatrix} \mathrm{Re}\{\mathbf{H}_{ij}\} & -\mathrm{Im}\{\mathbf{H}_{ij}\} \\ \mathrm{Im}\{\mathbf{H}_{ij}\} & \mathrm{Re}\{\mathbf{H}_{ij}\} \end{bmatrix}, \tag{7}$$

where $T$ denotes the symbol extension, $\mathbf{H}_{ij} \in \mathbb{C}^{N_i \times M_j}$ are the channel matrices considered in (1), $\tilde{\mathbf{H}}_{ij} \in \mathbb{R}^{2N_i \times 2M_j}$ are the channel matrices obtained after applying the asymmetric complex signaling and $\hat{\mathbf{H}}_{ij} \in \mathbb{R}^{2TN_i \times 2TM_j}$ are the channel matrices when both transformations, $T$-symbol extension and asymmetric complex signaling are applied. Thus, the received signal presented in (5) is modified as follows,

$$\hat{\mathbf{y}}_i = \begin{bmatrix} \hat{\mathbf{H}}_{i1}\hat{\mathbf{V}}_{i1} & \hat{\mathbf{H}}_{i1}\hat{\mathbf{Z}}_{i1} \end{bmatrix} \begin{bmatrix} \mathbf{s}_{i1}^{(IA)} \\ \mathbf{s}_{i1}^{(NS)} \end{bmatrix} + \begin{bmatrix} \hat{\mathbf{H}}_{i2}\hat{\mathbf{V}}_{i2} & \hat{\mathbf{H}}_{i2}\hat{\mathbf{Z}}_{i2} \end{bmatrix} \begin{bmatrix} \mathbf{s}_{i2}^{(IA)} \\ \mathbf{s}_{i2}^{(NS)} \end{bmatrix} + \hat{\mathbf{i}}_i + \hat{\mathbf{n}}_i, \qquad i = 1,2 \tag{8}$$

where $\hat{\mathbf{y}}_i \in \mathbb{R}^{2TN_i \times 1}$, $\hat{\mathbf{n}}_i \in \mathbb{R}^{2TN_i \times 1}$ stand for the received signal and noise at the $i$th destination, $\hat{\mathbf{V}}_{ij} \in \mathbb{R}^{2TM_j \times d_{ij}^{(IA)}}$, $\hat{\mathbf{Z}}_{ij} \in \mathbb{R}^{2TM_j \times d_{ij}^{(NS)}}$ are the transmit filters used to transmit the symbol streams from the $j$th transmitter to the $i$th receiver using the interference alignment and null-steering transmission, respectively. Finally, the received interference becomes,

$$\hat{\mathbf{i}}_i = \begin{bmatrix} \hat{\mathbf{H}}_{i1}\hat{\mathbf{V}}_{k1} & \hat{\mathbf{H}}_{i2}\hat{\mathbf{V}}_{k2} \end{bmatrix} \begin{bmatrix} \mathbf{s}_{k1}^{(IA)} \\ \mathbf{s}_{k2}^{(IA)} \end{bmatrix}, \quad i \neq k, \quad i,k = 1,2 \tag{9}$$

At each receiver the symbols are estimated independently using linear receiver filters that project the received signal into the orthogonal space defined by the total interference, so that the signal considered to estimate the different parts of the intended message becomes,

$$\begin{cases} \hat{\mathbf{y}}_{ij}^{(IA)} = \hat{\mathbf{L}}_{ij}\hat{\mathbf{y}}_i \\ \hat{\mathbf{y}}_{ij}^{(NS)} = \hat{\mathbf{F}}_{ij}\hat{\mathbf{y}}_i \end{cases}, \quad i,j = 1,2 \tag{10}$$

where $\hat{\mathbf{L}}_{ij} \in \mathbb{R}^{d_{ij}^{(IA)} \times 2TN_j}$, $\hat{\mathbf{F}}_{ij} \in \mathbb{R}^{d_{ij}^{(NS)} \times 2TN_j}$ are the receive filters employed to decode symbols $\mathbf{s}_{ij}^{(IA)}, \mathbf{s}_{ij}^{(NS)}$, respectively.

*B. Linear Feasibility*

If employing linear transmit and receive filter in our system model, the following conditions have to be satisfied:

$$\begin{cases} \left.\begin{aligned} &\hat{\mathbf{L}}_{ij}\hat{\mathbf{H}}_{ij}\hat{\mathbf{V}}_{kj} = \mathbf{0}_{d_{ij}^{(IA)} \times d_{kj}^{(IA)}}, \quad \hat{\mathbf{L}}_{ij}\hat{\mathbf{H}}_{ij}\hat{\mathbf{Z}}_{pj} = \mathbf{0}_{d_{ij}^{(IA)} \times d_{kj}^{(NS)}} \\ &\hat{\mathbf{F}}_{ij}\hat{\mathbf{H}}_{ij}\hat{\mathbf{V}}_{pj} = \mathbf{0}_{d_{ij}^{(NS)} \times d_{kj}^{(IA)}}, \quad \hat{\mathbf{F}}_{ij}\hat{\mathbf{H}}_{ij}\hat{\mathbf{Z}}_{kj} = \mathbf{0}_{d_{ij}^{(NS)} \times d_{kj}^{(NS)}} \end{aligned}\right\}, \quad k \neq i \\ \left.\begin{aligned} &\hat{\mathbf{L}}_{ij}\hat{\mathbf{H}}_{iq}\hat{\mathbf{V}}_{pq} = \mathbf{0}_{d_{ij}^{(IA)} \times d_{pq}^{(IA)}}, \quad \hat{\mathbf{L}}_{ij}\hat{\mathbf{H}}_{iq}\hat{\mathbf{Z}}_{pq} = \mathbf{0}_{d_{ij}^{(IA)} \times d_{pq}^{(NS)}} \\ &\hat{\mathbf{F}}_{ij}\hat{\mathbf{H}}_{iq}\hat{\mathbf{V}}_{pq} = \mathbf{0}_{d_{ij}^{(NS)} \times d_{pq}^{(IA)}}, \quad \hat{\mathbf{F}}_{ij}\hat{\mathbf{H}}_{iq}\hat{\mathbf{Z}}_{pq} = \mathbf{0}_{d_{ij}^{(NS)} \times d_{pq}^{(NS)}} \end{aligned}\right\}, \quad q \neq j \end{cases}, \quad i,j,p = 1,2 \tag{11}$$

where transmitter-receiver design must be able to remove all the generated interference by undesired signals and still decode the desired symbols,

$$\begin{cases} rank\left(\hat{\mathbf{L}}_{ij}\hat{\mathbf{H}}_{ij}\hat{\mathbf{V}}_{ij}\right) = d_{ij}^{(IA)} \\ rank\left(\hat{\mathbf{F}}_{ij}\hat{\mathbf{H}}_{ij}\hat{\mathbf{V}}_{ij}\right) = d_{ij}^{(NS)} \end{cases}, \quad \forall i,j \tag{12}$$

Notice that transmit and receive filters have to be jointly designed in order to meet the previous linear feasibility conditions, (11) and (12). Nevertheless, we follow an approach where only the transmit filters are designed for exploiting the *interference alignment* concept and null-steering transmission, while receive filters are obtained from those transmitters. We will show in section II.C that this one-side approach is optimal for the MIMO *X* channel in case the reciprocal MIMO *X* network is also considered. The conditions for designing the transmit filters assuming this one-side approach are defined by,

$$\begin{array}{lll} & \hat{\mathbf{Z}}_{i1} \perp \hat{\mathbf{H}}_{k1},\quad \mathbf{Z}_{i2} \perp \hat{\mathbf{H}}_{k2} & \text{free-interf. at } k\text{th rx} \\ d_{k1}^{(IA)} < d_{k2}^{(IA)} & \begin{cases} \operatorname{span}\left(\hat{\mathbf{H}}_{i1}\hat{\mathbf{V}}_{k1}\right) \subseteq \operatorname{span}\left(\hat{\mathbf{H}}_{i2}\hat{\mathbf{V}}_{k2}\right) \\ \mathbf{G}_i = \left[\hat{\mathbf{H}}_{i1}\hat{\mathbf{V}}_{i1}\ \hat{\mathbf{H}}_{i2}\hat{\mathbf{V}}_{i2}\ \hat{\mathbf{H}}_{i1}\hat{\mathbf{Z}}_{i1}\ \hat{\mathbf{H}}_{i2}\hat{\mathbf{Z}}_{i2}\ \hat{\mathbf{H}}_{i1}\hat{\mathbf{V}}_{k1}\right] \end{cases} & \begin{array}{l} \text{interf. alignment at } i\text{th rx} \\ \text{full column-rank} \end{array} \\ d_{k1}^{(IA)} \geq d_{k2}^{(IA)} & \begin{cases} \operatorname{span}\left(\hat{\mathbf{H}}_{i1}\hat{\mathbf{V}}_{k1}\right) \supseteq \operatorname{span}\left(\hat{\mathbf{H}}_{i2}\hat{\mathbf{V}}_{k2}\right) \\ \mathbf{G}_i = \left[\hat{\mathbf{H}}_{i1}\hat{\mathbf{V}}_{i1}\ \hat{\mathbf{H}}_{i2}\hat{\mathbf{V}}_{i2}\ \hat{\mathbf{H}}_{i1}\hat{\mathbf{Z}}_{i1}\ \hat{\mathbf{H}}_{i2}\hat{\mathbf{Z}}_{i2}\ \hat{\mathbf{H}}_{i2}\hat{\mathbf{V}}_{k2}\right] \end{cases} & \begin{array}{l} \text{interf. alignment at } i\text{th rx} \\ \text{full column-rank} \end{array} \end{array} ,\quad i \neq k,\quad i,k = 1,2 \tag{13}$$

where the free-interference constraint is needed to design null steering precoders $\hat{\mathbf{Z}}_{i1}, \hat{\mathbf{Z}}_{i2}$, and the interference alignment constraint imposes that precoders $\hat{\mathbf{V}}_{k1}, \hat{\mathbf{V}}_{k2}$ associated to the messages intended to *k*th destination have to be overlapped at the *i*th receiver. Notice that by exploiting the overlapping concept we reduce the dimension of the interference subspace. Finally, the signal space matrix $\mathbf{G}_i \in \mathbb{R}^{2TN_i \times \omega}$ with $\omega = d_{i1}^{(IA)} + d_{i1}^{(NS)} + d_{i2}^{(IA)} + d_{i2}^{(NS)} + \max\left(d_{k1}^{(IA)}, d_{k2}^{(IA)}\right)$ contains the desired and interference signals at receiver *i*, and it must be full column rank in order to ensure that linear receive filters cancel undesired symbols, so receivers have to be designed to match the equations,

$$\begin{array}{l} \begin{cases} \hat{\mathbf{L}}_{i1}\left[\hat{\mathbf{H}}_{i1}\hat{\mathbf{Z}}_{i1} \quad \hat{\mathbf{H}}_{i2}\hat{\mathbf{V}}_{i2} \quad \hat{\mathbf{H}}_{i2}\hat{\mathbf{Z}}_{i2} \quad \hat{\mathbf{H}}_{i1}\hat{\mathbf{V}}_{k1} \quad \hat{\mathbf{H}}_{i2}\hat{\mathbf{V}}_{k2}\right] = \mathbf{0} \\ \hat{\mathbf{F}}_{i1}\left[\hat{\mathbf{H}}_{i1}\hat{\mathbf{V}}_{i1} \quad \hat{\mathbf{H}}_{i2}\hat{\mathbf{V}}_{i2} \quad \hat{\mathbf{H}}_{i2}\hat{\mathbf{Z}}_{i2} \quad \hat{\mathbf{H}}_{i1}\hat{\mathbf{V}}_{k1} \quad \hat{\mathbf{H}}_{i2}\hat{\mathbf{V}}_{k2}\right] = \mathbf{0} \end{cases} \\ \begin{cases} \hat{\mathbf{L}}_{i2}\left[\hat{\mathbf{H}}_{i2}\hat{\mathbf{Z}}_{i2} \quad \hat{\mathbf{H}}_{i1}\hat{\mathbf{V}}_{i1} \quad \hat{\mathbf{H}}_{i1}\hat{\mathbf{Z}}_{i1} \quad \hat{\mathbf{H}}_{i1}\hat{\mathbf{V}}_{k1} \quad \hat{\mathbf{H}}_{i2}\hat{\mathbf{V}}_{k2}\right] = \mathbf{0} \\ \hat{\mathbf{F}}_{i2}\left[\hat{\mathbf{H}}_{i2}\hat{\mathbf{V}}_{i2} \quad \hat{\mathbf{H}}_{i1}\hat{\mathbf{V}}_{i1} \quad \hat{\mathbf{H}}_{i1}\hat{\mathbf{Z}}_{i1} \quad \hat{\mathbf{H}}_{i1}\hat{\mathbf{V}}_{k1} \quad \hat{\mathbf{H}}_{i2}\hat{\mathbf{V}}_{k2}\right] = \mathbf{0} \end{cases} \end{array} ,\quad i \neq k,\quad i,k = 1,2 \tag{14}$$

and still be able to decode the intended symbols. Hence, conditions (11) and (12) are satisfied.

Assuming that all messages are independent with a probability of error $P_e$, achievable rates and capacities for each message are defined in the standard Shannon sense and conditions given in (13) and (14) are fulfilled, then the achievable DoF are,

$$\tilde{d}_{ij} = \frac{d_{ij}^{(IA)} + d_{ij}^{(NS)}}{2T} \tag{15}$$

## C. Reciprocal MIMO X Network

The reciprocal network is obtained by switching the direction of communication, see for example Fig. 2 where receivers become the transmitters of the different messages while transmitters must receive the intended messages. Network duality implies having the same set of signal-to-noise plus interference ratio in the original and reciprocal network given a total transmit power constraint, [21]. In [22] it was shown that the feasibility conditions needed in the *K*-user IC are identical in both networks.

Due to the convention employed for identifying the DoF per message (first sub-index indicates the receiver and second sub-index denotes the transmitter) the following identification must be done between the symbol streams in the original and reciprocal network,

$$\begin{cases} d_{11} = \underleftarrow{d}_{11}, & d_{22} = \underleftarrow{d}_{22} \\ d_{21} = \underleftarrow{d}_{12}, & d_{12} = \underleftarrow{d}_{21} \end{cases} \tag{16}$$

where $\underleftarrow{d}_{ij}$ is the number of symbol streams of message transmitted from the *j*th source to the *i*th destination in the reciprocal network (message $\underleftarrow{W}_{ij}$), which is message $W_{ji}$ in the original network with $d_{ji}$ symbol streams, as it is shown in Fig. 2.

The received signal in the reciprocal network is given by,

$$\underleftarrow{\hat{\mathbf{y}}}_i = \left[\underleftarrow{\hat{\mathbf{H}}}_{i1}\underleftarrow{\hat{\mathbf{V}}}_{i1} \quad \underleftarrow{\hat{\mathbf{H}}}_{i1}\underleftarrow{\hat{\mathbf{Z}}}_{i1}\right]\begin{bmatrix} \underleftarrow{\mathbf{s}}_{i1}^{(IA)} \\ \underleftarrow{\mathbf{s}}_{i1}^{(NS)} \end{bmatrix} + \left[\underleftarrow{\hat{\mathbf{H}}}_{i2}\underleftarrow{\hat{\mathbf{V}}}_{i2} \quad \underleftarrow{\hat{\mathbf{H}}}_{i2}\underleftarrow{\hat{\mathbf{Z}}}_{i2}\right]\begin{bmatrix} \underleftarrow{\mathbf{s}}_{i2}^{(IA)} \\ \underleftarrow{\mathbf{s}}_{i2}^{(NS)} \end{bmatrix} + \left[\underleftarrow{\hat{\mathbf{H}}}_{i1}\underleftarrow{\hat{\mathbf{V}}}_{k1} \quad \underleftarrow{\hat{\mathbf{H}}}_{i2}\underleftarrow{\hat{\mathbf{V}}}_{k2}\right]\begin{bmatrix} \underleftarrow{\mathbf{s}}_{k1}^{(IA)} \\ \underleftarrow{\mathbf{s}}_{k2}^{(IA)} \end{bmatrix} + \underleftarrow{\hat{\mathbf{n}}}_i, \quad i \neq k, i,k = 1,2 \tag{17}$$

where $\underleftarrow{\hat{\mathbf{y}}}_i \in \mathbb{R}^{2TM_i \times 1}$, $\underleftarrow{\hat{\mathbf{n}}}_i \in \mathbb{R}^{2TM_i \times 1}$, $\underleftarrow{\hat{\mathbf{V}}}_{ij} \in \mathbb{R}^{2TN_j \times \underleftarrow{d}_{ij}^{(IA)}}$, $\underleftarrow{\hat{\mathbf{Z}}}_{ij} \in \mathbb{R}^{2TN_j \times \underleftarrow{d}_{ij}^{(NS)}}$ and $\underleftarrow{\hat{\mathbf{H}}}_{ij} \in \mathbb{R}^{2TM_i \times 2TN_j}$. Following the same convention, the channel matrices in both networks are connected by $\underleftarrow{\hat{\mathbf{H}}}_{ji} = \hat{\mathbf{H}}_{ij}^T$. Similarly to (10) and (15) we can apply linear receive filters to decode the intended symbols and the attained DoF in the reciprocal network become $\underleftarrow{\tilde{d}}_{ij} = \left(\underleftarrow{d}_{ij}^{(IA)} + \underleftarrow{d}_{ij}^{(NS)}\right)/2T$.

*Lemma 1*: (*Reciprocal network*) The attained DoF ($\underleftarrow{\tilde{d}}_{ij}^{(IA)}, \underleftarrow{\tilde{d}}_{ij}^{(NS)}$) in a reciprocal MIMO *X* network with $\underleftarrow{M}_1 = N_1, \underleftarrow{M}_2 = N_2$ transmitting and $\underleftarrow{N}_1 = M_1, \underleftarrow{N}_2 = M_2$ receiving antennas can also be achieved in the original MIMO *X* network consisting in $M_1$, $M_2$ transmitting and $N_1$, $N_2$ receiving antennas if we adopt as the transmit filter (resp. receiver) the receive filter (resp. transmitter) derived in the reciprocal network,

$$\hat{\mathbf{V}}^o_{ij} = \underleftarrow{\mathbf{L}}^T_{ji} \quad \hat{\mathbf{Z}}^o_{ij} = \underleftarrow{\mathbf{F}}^T_{ji} \quad \mathbf{L}^o_{ij} = \underleftarrow{\mathbf{V}}^T_{ji} \quad \mathbf{F}^o_{ij} = \underleftarrow{\mathbf{Z}}^T_{ji}, \quad i,j = 1,2, \tag{18}$$

where $\underleftarrow{\hat{\mathbf{V}}}_{ij} \in \mathbb{R}^{2TN_j \times \underleftarrow{d}^{(IA)}_{ij}}$ and $\underleftarrow{\hat{\mathbf{Z}}}_{ij} \in \mathbb{R}^{2TN_j \times \underleftarrow{d}^{(ZI)}_{ij}}$ are the transmit filters, $\underleftarrow{\hat{\mathbf{L}}}_{ij} \in \mathbb{R}^{\underleftarrow{d}^{(IA)}_{ij} \times 2TM_i}$ and $\underleftarrow{\hat{\mathbf{F}}}_{ij} \in \mathbb{R}^{\underleftarrow{d}^{(ZI)}_{ij} \times 2TN_j}$ denote the receive filters obtained in the reciprocal network, while $\hat{\mathbf{V}}^o_{ij}, \hat{\mathbf{Z}}^o_{ij}$ and $\mathbf{L}^o_{ij}, \mathbf{F}^o_{ij}$ are the transmit and receive filters in the original network, respectively.

*Proof.* The achievable scheme derived in the reciprocal network is designed to satisfy conditions (13) and (14) in that network. In such a case, the condition associated to $rank\left(\underleftarrow{\hat{\mathbf{L}}}_{ii} \underleftarrow{\hat{\mathbf{H}}}_{ii} \underleftarrow{\hat{\mathbf{V}}}_{ii}\right) = \underleftarrow{d}^{(NS)}_{ii}$ is also satisfied. Since, the rank of a matrix does not change after transposition, hence $rank((\underleftarrow{\hat{\mathbf{L}}}_{ii} \underleftarrow{\hat{\mathbf{H}}}_{ii} \underleftarrow{\hat{\mathbf{V}}}_{ii})^T) = rank\left(\underleftarrow{\hat{\mathbf{V}}}^T_{ii} \underleftarrow{\hat{\mathbf{H}}}^T_{ii} \underleftarrow{\hat{\mathbf{L}}}^T_{ii}\right) = \underleftarrow{d}^{(IA)}_{ii}$. Furthermore, $\underleftarrow{\hat{\mathbf{H}}}_{ii} = \hat{\mathbf{H}}^T_{ii}$, hence, using (18), we identify the transmitters in the original network as the receivers obtained in the reciprocal one. The proof for the $\underleftarrow{d}^{(IA)}_{ij}$ follows similar guidelines, by taking into account that the cross messages are exchanged ( $d_{ij} = \underleftarrow{d}_{ji}$, ). □

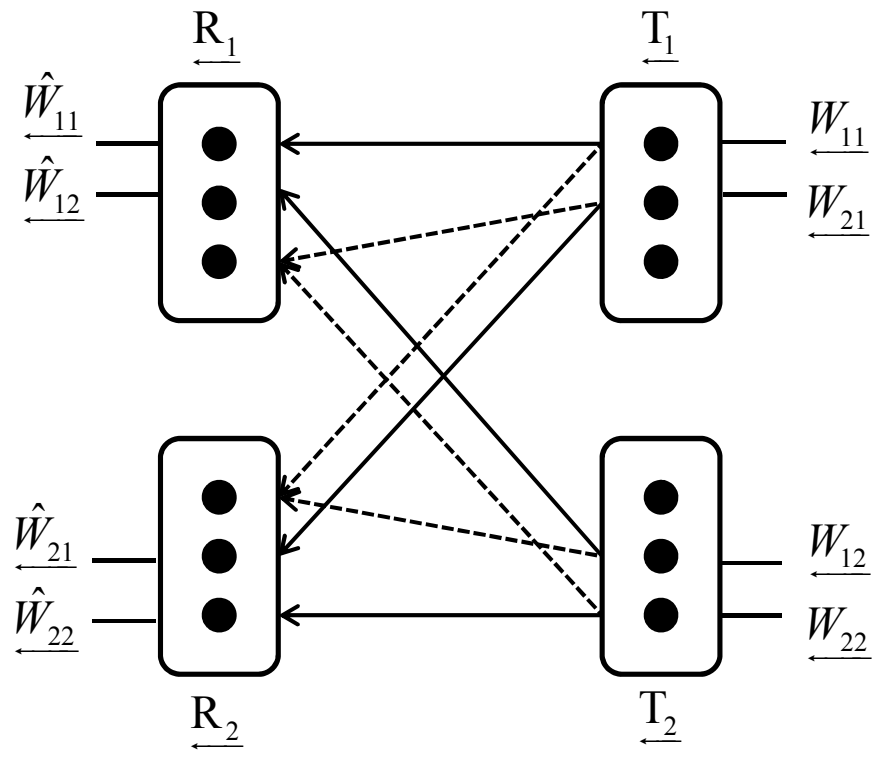

Fig. 2. Reciprocal MIMO *X* channel. Intended signals are referenced by solid lines, while unintended signals are denoted by dotted lines. Compared with the original MIMO *X* channel shown in Fig. 1 the following identification has to be done: $W_{ij} = \underleftarrow{W}_{ji}$, $i,j = 1,2$ .

*Remark*: Lemma 1 establishes that the attained DoF in the original MIMO *X* network (or reciprocal) are also achievable in the reciprocal (resp. original) network by switching the obtained transmit (resp. receive) filters into receive (resp. transmit) filters. Nevertheless, the one-side approach defined in (13) might produce that the attained DoF when transmit filters are designed in the original network could be different of the attained ones when transmit filters are designed in the reciprocal network, but not obtained by switching the receiver into a transmitter. In order to illustrate this issue, let us assume a MIMO *X* network where transmitters are equipped with $M_1$=$M_2$=2 antennas while receivers have $N_1$=$N_2$=5 antennas. If we try to design the transmit filters according to the conditions presented in (13), we observe that we cannot exploit neither the null-steering or interference alignment concepts. However, designing the transmit filters in the reciprocal network ( $\underleftarrow{M}_1 = \underleftarrow{M}_2 = 5$ and $\underleftarrow{N}_1 = \underleftarrow{N}_2 = 2$ ) the total achievable DoF become 4, each message carries one DoF and the transmit filter is obtained by exploiting the null-steering transmission. Notice that working in the reciprocal network is equivalent to design the receive filters and then obtain the transmit precoders accordingly.

*Theorem 1*: (*Maximum achievable DoF*) The maximum achievable DoF provided by a certain transmitter-receiver design is obtained by evaluating the transmit filter design in the original MIMO *X* network (with $M_1$, $M_2$ transmitting and $N_1$, $N_2$ receiving antennas) and reciprocal MIMO *X* network (with $N_1$, $N_2$ transmitting and $M_1$, $M_2$ receiving antennas), keeping the design with the highest number of DoF.

*Proof.* Let us assume that $M_1$=$M_2$=$M$ and $N_1$=$N_2$=$N$. If $M$>$N$, by designing first the transmit precoders we can exploit the ($M$-$N$) spatial dimensions at each transmitter that do not generate any kind of inference to unintended receivers. However, if $M$<$N$, this is not possible, unless we design first the receivers that block the transmitted signal of some sources, and afterwards get the transmit precoders. This latter approach is met when we work in the reciprocal network. If the achievable scheme in the reciprocal network provides a higher number of DoF, then we obtain the corresponding precoders and receivers in the original network by means of Lemma 1. □

## III. Outer Bounds on Degrees of Freedom (DoF)

In the definition of the outer bounds on DoF for the MIMO $X$ channel, the analysis of the MIMO $Z$ channel depicted in Fig. 3 is needed. A MIMO $Z$ channel presents the same input-output equations as the $X$ channel in (1), but with the constraint that one of the four messages is removed, and additionally, the channel associated of the link for transmitting that message has zero gain. In the MIMO XC there are up to four MIMO $Z$ channels, denoted by $Z(ij)$, obtained by imposing $W_{ij}=\varnothing$ and $\mathbf{H}_{ij}=\mathbf{0}$. It has to be emphasized that the MIMO $Z$ channel considered here differs from the one usually analyzed in the literature (see for example [20]), where every transmitter only has one message to one intended receiver. Here, one of the transmitters has messages to both receivers, see for example transmitter $T_2$ in the MIMO $Z(21)$ shown in Fig. 3.

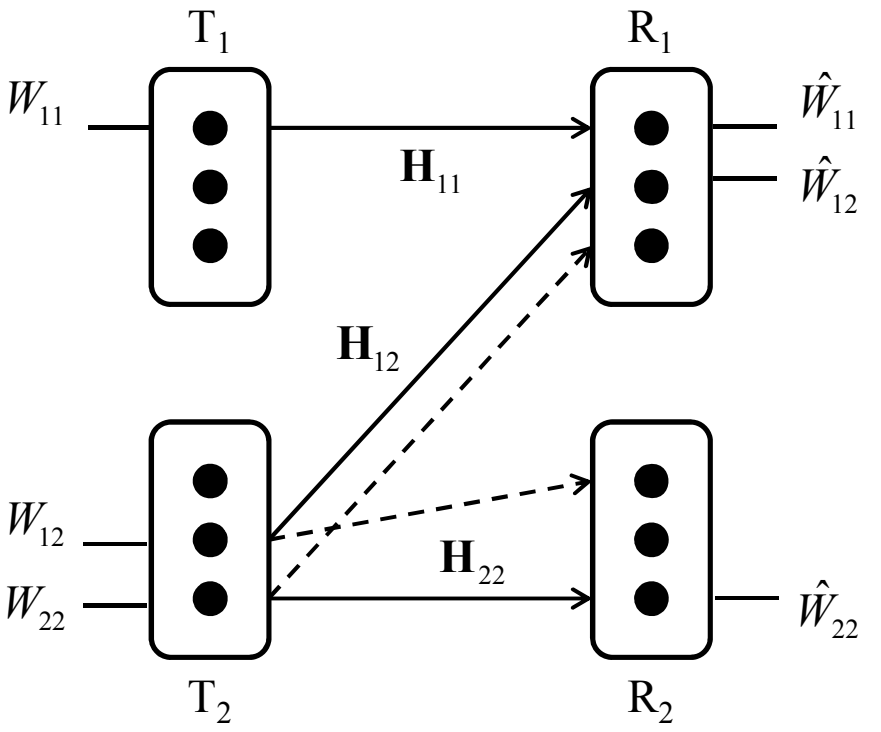

Fig. 3. MIMO $Z(21)$ channel with 2 sources-destinations, where sources $T_1$, $T_2$ and destinations $R_1$, $R_2$ are equipped with $M_1=M_2=N_1=N_2=3$ antennas. $T_1$ sends message $W_{11}$ to $R_1$, while $T_2$ sends messages $W_{12}$ and $W_{22}$ to $R_1$ and $R_2$. Transmitter $T_1$ does not generate any interference to $R_2$ because the channel matrix of the corresponding link, $\mathbf{H}_{21}$, is zero. Intended signals are referenced by solid lines, while unintended signals are denoted by dotted lines.

Section II in [7] derives an upper bound on the sum rate of the MIMO $Z$ channel. In the following we characterize the outer bound on DoF region for this channel.

*Theorem 2:* An outer bound DoF region for the full-rank MIMO $Z$ channel $Z(ij)$ is defined as follows,

$$\begin{aligned} D_{out}^{Z(ij)} = \Big\{ & \left(\hat{d}_{11},\hat{d}_{12},\hat{d}_{21},\hat{d}_{22}\right)\in\mathbb{R}_{+}^{4}, \quad \hat{d}_{ij}=0 \\ & \hat{d}_{11}+\hat{d}_{12}+\hat{d}_{21}+\hat{d}_{22}\le \max\left(N_k,M_q\right), \\ & \hat{d}_{11}+\hat{d}_{12}\le N_1 \\ & \hat{d}_{21}+\hat{d}_{22}\le N_2 \\ & \hat{d}_{11}+\hat{d}_{21}\le M_1 \\ & \hat{d}_{12}+\hat{d}_{22}\le M_2, \quad k\ne i, q\ne j \Big\} \end{aligned} \tag{19}$$

where $\hat{d}_{nm}$ with $n,m$=1,2 denote the DoF that carriers message $W_{nm}$ and message $W_{ij}=\varnothing$ .

*Proof.* The theorem is proved using Corollary 1 in [7] (first constraint) along with the outer bounds due to the multiple access and broadcast channels present in the $Z$ channel (last four constraints in (19)). □

Since the DoF derived for the MIMO $Z$ channel are larger or equal to the ones attained in a MIMO $X$ channel, see Lemma 1 in [7], it turns out that an outer bound DoF region for the MIMO $X$ channel can be obtained by the union of the outer bound DoF regions obtained by the different MIMO $Z$ channels obtained by removing one of the messages and setting the corresponding channel matrix coefficients to zero: $Z(11)$, $Z(12)$, $Z(21)$, $Z(22)$.

*Theorem 3* (Theorem 2 in [7]): An outer bound region of DoF for the full-rank MIMO $X$ channel is defined by

$$\begin{aligned} D_{out}^{X} = \Big\{ & \left(\hat{d}_{11},\hat{d}_{12},\hat{d}_{21},\hat{d}_{22}\right)\in\mathbb{R}_{+}^{4} \\ & \hat{d}_{11}+\hat{d}_{12}+\hat{d}_{21}\le \max\left(N_1,M_1\right), \ \hat{d}_{11}+\hat{d}_{12}\le N_1 \\ & \hat{d}_{11}+\hat{d}_{21}+\hat{d}_{22}\le \max\left(N_2,M_1\right), \ \hat{d}_{11}+\hat{d}_{21}\le M_1 \\ & \hat{d}_{11}+\hat{d}_{12}+\hat{d}_{22}\le \max\left(N_1,M_2\right), \ \hat{d}_{21}+\hat{d}_{22}\le N_2 \\ & \hat{d}_{12}+\hat{d}_{21}+\hat{d}_{22}\le \max\left(N_2,M_2\right), \ \hat{d}_{12}+\hat{d}_{22}\le M_2 \ \Big\} \end{aligned} \tag{20}$$

These outer bound DoF regions provide bounds for all possible DoF combinations for the MIMO $Z$ and $X$ channel. In the following we define the outer bound on the total DoF, as defined by the following theorems:

*Theorem 4*: The outer bound on the total DoF in a full-rank MIMO $Z(ij)$ channel is given by,

$$\eta_{Z(ij)}^{out} \triangleq \max_{D_{out}^{Z(ij)}} \left( \hat{d}_{11} + \hat{d}_{12} + \hat{d}_{21} + \hat{d}_{22} \right) = \\ = \begin{cases} \min\left(M_1 + M_2, N_k\right) & if \quad M_q < N_k \\ \min\left(N_1 + N_2, M_q\right) & if \quad M_q \geq N_k \end{cases}, \quad k \neq i, \quad q \neq j \tag{21}$$

where $D_{out}^{Z(ij)}$ is defined in Theorem 2.

*Proof.* The theorem is proved by solving the dual problem of the linear programming problem, [23], with constraints imposed by Theorem 2. □

*Theorem 5* (Theorem 4 in [7]): An outer bound on the total DoF in a full-rank MIMO $X$ channel is defined by

$$\eta_X^{out} \triangleq \max_{D_{out}^X} \left( \hat{d}_{11} + \hat{d}_{12} + \hat{d}_{21} + \hat{d}_{22} \right) = \\ = \min \left\{ \begin{array}{c} M_1 + M_2, N_1 + N_2, \\ \dfrac{\varepsilon_{11} + \varepsilon_{12} + M_2}{2}, \dfrac{\varepsilon_{21} + \varepsilon_{22} + M_1}{2} \\ \dfrac{\varepsilon_{11} + \varepsilon_{21} + N_2}{2}, \dfrac{\varepsilon_{12} + \varepsilon_{22} + N_1}{2} \\ \dfrac{\varepsilon_{11} + \varepsilon_{12} + \varepsilon_{21} + \varepsilon_{22}}{3} \end{array} \right\}, \quad \begin{array}{l} \varepsilon_{ij} = \max\left(M_i, N_j\right) \\ i, j = 1, 2 \end{array} \tag{22}$$

where $D_{out}^X$ is defined in Theorem 3.

## IV. MAIN RESULTS

The principal result presented in this work is Theorem 7 which shows that the known outer bound on the total DoF of the full rank MIMO $X$ channel [7] presented in Theorem 5 is, in fact, tight for any antenna configuration. Basically, we introduce a scheme in section VI that allows deriving the transmit filters and designing the number of symbol streams per message in order to maximize the achievable total DoF, valid for time-varying and constant channel cases.

Likewise, in case the total number of transmitted messages in a MIMO $X$ channel is 3 (one transmitter just sends to one receiver), then the proposed scheme attains the outer bound on the total DoF defined for the MIMO $Z$ channel and presented in Theorem 4. This result presented in Theorem 6 is obtained thanks to the interference alignment and null-steering concepts. In the first case, all interferers overlap on a single space, i.e. same as one of the interfering links had zero gain. Additionally, those streams associated to the null-steering precoders do not introduce any interference to unintended receivers.

The main results derived in this work are enunciated in the following.

*Theorem 6*: (*Total DoF of a MIMO X channel with 3 messages*) For the 2-user MIMO $X$ channel with $M_1$, $M_2$ transmitting and $N_1$, $N_2$ receiving antennas and full-rank channel matrices, the outer bound on the total DoF, defined as the MIMO $Z(ij)$ channel by Theorem 4, is attained when corresponding message $W_{ij}$ is not transmitted and precoding is based on null-steering and interference alignment.

*Proof.* See appendix B. □

*Theorem 7*: (*Total DoF of a MIMO X channel*) For the 2-user MIMO $X$ channel with $M_1$, $M_2$ transmitting and $N_1$, $N_2$ receiving antennas and full-rank channel matrices, the outer bound on the total DoF defined by Theorem 5 is attained tightly when precoding is based on null-steering and interference alignment.

*Proof.* The outer bound region of DoF on the MIMO $X$ channel defined by Theorem 2 is obtained by the union of the outer bound regions of the existing MIMO $Z$ channels ($Z(11)$, $Z(12)$, $Z(21)$, $Z(22)$) given by Theorem 4. According to Theorem 6, the total DoF of the MIMO Z channel are attained in the MIMO $X$ channel. Consequently, the outer bound on the total DoF of the MIMO $X$ channel is achievable. Additionally, section VII provides an analytical solution of how the outer bound is attained for some antenna configurations. □

The achievable region of DoF obtained by the proposed scheme is given by

*Theorem 8*: (*Inner bound region of DoF*) The achievable DoF region for the 2-user MIMO $X$ channel with $M_1$, $M_2$ transmitting and $N_1$, $N_2$ receiving antennas and full-rank channel matrices when precoding is based on null-steering and interference alignment over a constant channel with $T$ symbol extensions is

$$D_{in}^{X}(T)=\left(D_{out}^{X}\cap\Delta(T)\right)\cup\left(D_{out}^{X}\cap\underline{\Delta}(T)\right) \tag{23}$$

where $D_{out}^{X}$ is the outer bound DoF region of the MIMO $X$ channel defined in Theorem 3, $T$ denotes the symbol extension and $\Delta(T)$, $\underline{\Delta}(T)$ stand for the achievable DoF region defined as a function of the symbol streams allocated to the null steering ($d_{ij}^{NS}$) and interference alignment ($d_{ij}^{IA}$) when the scheme proposed in section VI is applied in the original or in the reciprocal network respectively,

$$\begin{aligned}\Delta(T)=\Big\{&\left(d_{11}^{(IA)},d_{12}^{(IA)},d_{21}^{(IA)},d_{22}^{(IA)}\right)\in\mathbb{Z}_{+}^{4},\quad\left(d_{11}^{(NS)},d_{12}^{(NS)},d_{21}^{(NS)},d_{22}^{(NS)}\right)\in\mathbb{Z}_{+}^{4},\\&\tilde{d}_{ij}=\frac{d_{ij}^{(IA)}+d_{ij}^{(NS)}}{2T},\\&d_{ij}^{(IA)}\le 2\left(\min\left(N_k,M_1\right)+\min\left(N_k,M_2\right)-\min\left(N_k,M_1+M_2\right)\right),\\&d_{ij}^{(NS)}\le 2T\left(\max\left(M_j,N_k\right)-N_k\right),\quad i,j,k=1,2,\quad i\neq k\Big\}\end{aligned} \tag{24}$$

$$\begin{aligned}\underline{\Delta}(T)=\Big\{&\left(d_{11}^{(IA)},d_{12}^{(IA)},d_{21}^{(IA)},d_{22}^{(IA)}\right)\in\mathbb{Z}_{+}^{4},\quad\left(d_{11}^{(NS)},d_{12}^{(NS)},d_{21}^{(NS)},d_{22}^{(NS)}\right)\in\mathbb{Z}_{+}^{4},\\&\tilde{d}_{ij}=\frac{d_{ij}^{(IA)}+d_{ij}^{(NS)}}{2T},\\&d_{ij}^{(IA)}\le 2\left(\min\left(M_k,N_1\right)+\min\left(M_k,N_2\right)-\min\left(M_k,N_1+N_2\right)\right),\\&d_{ij}^{(NS)}\le 2T\left(\max\left(M_k,N_i\right)-N_i\right),\quad i,j,k=1,2,\quad j\neq k\Big\}\end{aligned} \tag{25}$$

*Proof.* See Appendix C. □

The proof of the tightness of the outer bound of total DoF for the MIMO $X$ channel (see Theorem 7) cannot be used for the outer bound region of DoF presented in Theorem 3. However, since the region of DoF can be obtained by maximizing the weighted sum of DoF, both regions can be evaluated and compared. We have performed several computational simulations with different number of antennas and weights, and we have observed that same result when the maximization is carried out over $D_{out}^{X}$ or over $D_{in}^{X}$ , where symbol stream per message and symbol extension, $T$, are designed. Hence, we conjecture that the outer bound region is also tight.

Another important result of this paper is that the proposed achievable GSVD-based scheme introduced in section VI for the MIMO $X$ channel provides a general framework for addressing the optimal DoF of different two-user channels such as MIMO MAC, BC or IC, in addition to cope with rank-deficient MIMO channels. In this latter case, it may be shown that inner bound under rank-deficient MIMO channels can be superior to the outer bound defined for full-rank MIMO channels (shown in section IX). As a result, derivation of better outer bounds for reduced-rank channels is needed, but is beyond the scope of the paper.

## V. Signal Subspace Decomposition

This section is devoted to introduce the Generalized Singular Value Decomposition (GSVD) because it is a key tool to show achievability of the outer bounds of the MIMO $X$ channel and presents interesting properties that can be exploited by multiuser channels through the interference alignment concept. The GSVD decomposes two channel matrices providing the subspace where signals can be overlapped (aligned) and the subspace where null steering transmissions can be spanned, both defined with orthonormal basis. The GSVD (as introduced in section 8.7.3 of [15] and in [16]) is able to connect two matrices $\mathbf{A}\in\mathbb{C}^{p\times m}$ and $\mathbf{B}\in\mathbb{C}^{p\times m}$ through a non-singular matrix. Note that we keep the same number of rows for $\mathbf{A}$, $\mathbf{B}$ in contrast to [15], [16], and hence we enunciate the GSVD accordingly:

*Theorem 9* (Theorem 1 in [16]): Take any two matrices $\mathbf{A} \in \mathbb{C}^{p\times m}$ with $r_A = rank(\mathbf{A})$ and $\mathbf{B} \in \mathbb{C}^{p\times m}$ with $r_B = rank(\mathbf{B})$. Then the GSVD defines unitary matrices $\mathbf{U}_A \in \mathbb{C}^{m\times m}$ and $\mathbf{U}_B \in \mathbb{C}^{n\times n}$, a non-singular matrix $\mathbf{W} \in \mathbb{C}^{p\times q}$ and diagonal matrices $\mathbf{C}_A \in \mathbb{C}^{q\times m}, \mathbf{C}_B \in \mathbb{C}^{q\times n}$ such that,

$$\mathbf{A} = \mathbf{W}\mathbf{C}_A\mathbf{U}_A^H, \quad \mathbf{B} = \mathbf{W}\mathbf{C}_B\mathbf{U}_B^H, \quad \mathbf{C}_A\mathbf{C}_A^H + \mathbf{C}_B\mathbf{C}_B^H = \mathbf{I}_q \tag{26}$$

where $q = rank([\mathbf{A}\ \mathbf{B}])$ defines the number of rows in $\mathbf{C}_A$, $\mathbf{C}_B$ which contain the generalized singular values. Notice that variable $q$ describes the dimension of the space sum of the spaces generated by columns of matrices **A** and **B**. Furthermore, matrices $\mathbf{U}_A$, $\mathbf{U}_B$ stand for the generalized singular vectors whose columns define the generating basis of three different subspaces: *overlapped, null-steering* and *non-overlapping*. The dimensions of each subspace is defined by,

- Overlapping (OV) dimensions: $s = r_A + r_B - q$
- Null-steering (NS) dimensions from A: $\phi_A = m - r_A$,
- Null-steering (NS) dimensions from B: $\phi_B = n - r_B$
- Non-overlapping (NOV) dimensions from A: $\nu_A = q - r_B$
- Non-overlapping (NOV) dimensions from B: $\nu_B = q - r_A$

The matrices containing the generalized singular values and generalized singular vectors exhibit the following structure,

$$\mathbf{C}_A = \begin{bmatrix} \mathbf{0}_{\nu_B\times s} & \mathbf{0}_{\nu_B\times\nu_A} & \mathbf{0}_{\nu_B\times\phi_A} \\ \mathbf{\Gamma}_A & \mathbf{0}_{s\times\nu_A} & \mathbf{0}_{s\times\phi_A} \\ \mathbf{0}_{\nu_A\times s} & \mathbf{I}_{\nu_A\times\nu_A} & \mathbf{0}_{\nu_A\times\phi_A} \end{bmatrix}, \quad \mathbf{U}_A = \left[\mathbf{U}_A^{OV}\ \mathbf{U}_A^{NOV}\ \mathbf{U}_A^{NS}\right]$$
$$\mathbf{\Gamma}_A = diag([\gamma_1 \ldots \gamma_s]), \quad 0 < \gamma_1 \leq \cdots \leq \gamma_s < 1, \tag{27}$$

and,

$$\mathbf{C}_B = \begin{bmatrix} \mathbf{I}_{\nu_B\times\nu_B} & \mathbf{0}_{\nu_B\times s} & \mathbf{0}_{\nu_B\times\phi_B} \\ \mathbf{0}_{s\times\nu_B} & \mathbf{\Gamma}_B & \mathbf{0}_{s\times\phi_B} \\ \mathbf{0}_{\nu_A\times\nu_B} & \mathbf{0}_{\nu_A\times s} & \mathbf{0}_{\nu_A\times\phi_B} \end{bmatrix}, \quad \mathbf{U}_B = \left[\mathbf{U}_B^{NOV}\ \mathbf{U}_B^{OV}\ \mathbf{U}_B^{NS}\right]$$
$$\mathbf{\Gamma}_B = diag([\sigma_1 \ldots \sigma_s]), 1 > \sigma_1 \geq \cdots \geq \sigma_s > 0 \tag{28}$$

where matrices $\mathbf{U}_A^{OV} \in \mathbb{C}^{m\times s}$, $\mathbf{U}_B^{OV} \in \mathbb{C}^{n\times s}$ denote the column generalized eigenvectors of the overlapped subspace and they are associated to the generalized singular values in $\mathbf{\Gamma}_A$, $\mathbf{\Gamma}_B$; on the other hand, $\mathbf{U}_A^{MS} \in \mathbb{C}^{m\times\phi_A}$, $\mathbf{U}_B^{NS} \in \mathbb{C}^{n\times\phi_B}$ contain the eigenvectors of the null-space of **A**, **B**, so that they generate the null-steering subspaces and finally, matrices $\mathbf{U}_A^{NOV} \in \mathbb{C}^{m\times\nu_A}$, $\mathbf{U}_B^{NOV} \in \mathbb{C}^{n\times\nu_B}$ contain the eigenvectors of the non-overlapped subspaces.

In the following we elucidate how the GSVD help us to design the transmit filters devoted either to align the interference or null the interference at an unintended receiver. Let us assume two interfering transmitters equipped with $M_1$ and $M_2$ antennas and one unintended receiver with $N$ receive antennas. The channel matrices are denoted by $\mathbf{H}_1 \in \mathbb{C}^{N\times M_1}$ and $\mathbf{H}_2 \in \mathbb{C}^{N\times M_2}$. Taking into account the GSVD of $\mathbf{H}_1$ and $\mathbf{H}_2$, it turns out that both transmitters align their signals at the receiver side when precoders are selected to be paired $\left[\mathbf{U}_{\mathbf{H}_1}^{OV}\right]^{(k)}$ $\left[\mathbf{U}_{\mathbf{H}_2}^{OV}\right]^{(k)}$ for some $k$=1,...,$s$. Otherwise, their signal is not aligned,

$$\begin{cases} span\left(\mathbf{H}_1\left[\mathbf{U}_{\mathbf{H}_1}^{OV}\right]^{(k)}\right) = span\left(\mathbf{W}\left[\mathbf{0}_{1\times\nu_B+k-1}\quad \gamma_k \quad \mathbf{0}_{1\times\nu_A+s-k}\right]^T\right) = span\left(\mathbf{H}_2\left[\mathbf{U}_{\mathbf{H}_2}^{OV}\right]^{(k)}\right) = span\left(\mathbf{W}\left[\mathbf{0}_{1\times\nu_B+k-1}\quad \sigma_k \quad \mathbf{0}_{1\times\nu_A+s-k}\right]^T\right) \\ span\left(\mathbf{H}_1\left[\mathbf{U}_{\mathbf{H}_1}^{OV}\right]^{(k)}\right) = span\left(\mathbf{W}\left[\mathbf{0}_{1\times\nu_B+k-1}\quad \gamma_k \quad \mathbf{0}_{1\times\nu_A+s-k}\right]^T\right) \neq span\left(\mathbf{H}_2\left[\mathbf{U}_{\mathbf{H}_2}^{OV}\right]^{(j)}\right) = span\left(\mathbf{W}\left[\mathbf{0}_{1\times\nu_B+j-1}\quad \sigma_j \quad \mathbf{0}_{1\times\nu_A+s-j}\right]^T\right) \\ \quad \text{with} \quad j \neq k \end{cases} \tag{29}$$

On the other hand, both transmitters do not generate any interference to the given receiver if their precoders are selected to be $\left[\mathbf{U}_{\mathbf{H}_1}^{NS}\right]^{(m)}$ $\left[\mathbf{U}_{\mathbf{H}_2}^{NS}\right]^{(n)}$ for $m$=1.. $\phi_{\mathbf{H}_1}$, $n$=1.. $\phi_{\mathbf{H}_2}$.

$$\begin{cases} \left[\mathbf{U}_{\mathbf{H}_1}^{OV}\right]^{(m)} \perp \mathbf{H}_1 & \text{with} \quad m = 1\ldots\phi_{\mathbf{H}_1} \\ \left[\mathbf{U}_{\mathbf{H}_2}^{OV}\right]^{(n)} \perp \mathbf{H}_2 & \text{with} \quad n = 1\ldots\phi_{\mathbf{H}_2} \end{cases} \tag{30}$$

## VI. Achievable MIMO *X* Scheme

Adopting the overlapped and null-steering subspaces defined from the GSVD and assuming the signal model from section II, it is possible to derive a general achievable scheme for the MIMO *X* with an arbitrary number of antennas. Section VI.B proposes the precoder definition in terms of the number of transmitted symbol streams and its associated linear filters. Section VI.C shows that the proposed solution satisfies the necessary conditions stated in section II.B. The maximum achievable DoF are defined in section VI.D and finally, section VI.E is devoted to introduce the changes in the generating basis of the overlapped and null-steering subspaces in case the channel is time varying over the channel uses.

### *A. Overlapped and Null-steering subspaces*

With the help of the GSVD we introduce two lemmas that define the DoF-achieving precoders for the MIMO *X* channel.

*Lemma 2:* (*Basis of the overlapped subspace*) The paired-generating basis of the transmit filters associated to symbol streams intended to the *i*th receiver, $\mathbf{s}_{i1}^{(IA)},\mathbf{s}_{i2}^{(IA)}$, that define a common overlapped space at the *k*th receiver (*i*≠*k*) are given by,

$$\mathbf{\Omega}_{i1}=\mathbf{U}_{k1}^{OV}\mathbf{\Gamma}_{k1}^{-1},\quad \mathbf{\Omega}_{i2}=\mathbf{U}_{k2}^{OV}\mathbf{\Gamma}_{k2}^{-1},\quad i\neq k,\quad i,k=1,2 \tag{31}$$

where $\mathbf{\Omega}_{ij}\in\mathbb{C}^{M_j\times s_k}$ and $\mathbf{U}_{kj}^{OV}\in\mathbb{C}^{M_j\times s_k}, \mathbf{\Gamma}_{kj}\in\mathbb{C}^{s_k\times s_k}$ are obtained by the GSVD over $\mathbf{H}_{k,1},\mathbf{H}_{k,2}$. The dimension of this space is,

$$s_k=rank\left(\mathbf{H}_{k1}\right)+rank\left(\mathbf{H}_{k2}\right)-rank\left(\left[\mathbf{H}_{k1}\ \mathbf{H}_{k2}\right]\right) \tag{32}$$

In case the channel elements are generated from a continuous probability distribution then the overlapped space dimension becomes,

$$s_k=\min\left(N_k,M_1\right)+\min\left(N_k,M_2\right)-\min\left(N_k,M_1+M_2\right)$$

*Proof.* Take into account (27) and (28) when GSVD is carried out on channel matrices $\mathbf{H}_{k,1},\mathbf{H}_{k,2}$ .The concept of paired-basis comes up because for defining the final transmit precoders, both transmitters must apply the same linear combination to its respective basis with the objective of preserving the signal overlapping, as it is illustrated in equation (29). □

*Lemma 3:* (*Basis of the null-steering subspace*) The generating basis of the transmit filters associated to symbol streams intended to the *i*th receiver, $\mathbf{s}_{i1}^{(NS)},\mathbf{s}_{i2}^{(NS)}$, that do not add interference at the *k*th receiver (*i*≠*k*) are given by,

$$\mathbf{\Psi}_{i1}=\mathbf{U}_{k1}^{NS},\quad \mathbf{\Psi}_{i2}=\mathbf{U}_{k2}^{NS},\quad i\neq k,\quad i,k=1,2 \tag{33}$$

where $\mathbf{\Psi}_{ij}\in\mathbb{C}^{M_j\times\phi_{kj}}$ and $\mathbf{U}_{kj}^{NS}\in\mathbb{C}^{M_j\times\phi_{kj}}$ are obtained from the GSVD over $\mathbf{H}_{k,1},\mathbf{H}_{k,2}$. The dimension of the null-steering space is

$$\phi_{kj}=M_j-rank\left(\mathbf{H}_{kj}\right) \tag{34}$$

In case the channel coefficients are generated from a continuous probability distribution, the dimension of the basis becomes,

$$\phi_{kj}=M_j-\min\left(N_k,M_j\right)=\left(M_j-N_k\right)^+$$

*Proof.* Take into account the matrix structure presented in (27) and (28) when GSVD is carried out over $\mathbf{H}_{k,1},\mathbf{H}_{k,2}$. □

The signal model introduced in section III assumes that the transmission is done over multiple channel extension using asymmetric complex signaling and symbol extension where the channel remains fixed. The lemmas can be easily extended to

*Lemma 4:* (*Basis in the constant channel extension*) The channel extension based on asymmetric complex signaling modifies the generating basis of the overlapping and null-steering spaces according to,

$$\tilde{\mathbf{\Omega}}_{ij}=\begin{bmatrix}\mathrm{Re}\left(\mathbf{\Omega}_{ij}\right) & -\mathrm{Im}\left(\mathbf{\Omega}_{ij}\right)\\ \mathrm{Im}\left(\mathbf{\Omega}_{ij}\right) & \mathrm{Re}\left(\mathbf{\Omega}_{ij}\right)\end{bmatrix},\qquad \tilde{\mathbf{\Psi}}_{ij}=\begin{bmatrix}\mathrm{Re}\left(\mathbf{\Psi}_{ij}\right) & -\mathrm{Im}\left(\mathbf{\Psi}_{ij}\right)\\ \mathrm{Im}\left(\mathbf{\Psi}_{ij}\right) & \mathrm{Re}\left(\mathbf{\Psi}_{ij}\right)\end{bmatrix} \tag{35}$$

where $\tilde{\mathbf{\Omega}}_{ij}\in\mathbb{R}^{2M_j\times 2s_k}$, $\tilde{\mathbf{\Psi}}_{ij}\in\mathbb{R}^{2M_j\times 2\phi_{kj}}$ and $\mathbf{\Omega}_{ij},\mathbf{\Psi}_{ij}$ are defined by Lemma 2 and Lemma 3, respectively. Likewise, when a *T* symbols extension is considered, the generating basis becomes,

$$\tilde{\tilde{\mathbf{\Omega}}}_{ij}=\left(\mathbf{I}_T\otimes\tilde{\mathbf{\Omega}}_{ij}\right),\qquad \tilde{\tilde{\mathbf{\Psi}}}_{ij}=\left(\mathbf{I}_T\otimes\tilde{\mathbf{\Psi}}_{ij}\right) \tag{36}$$

where $\tilde{\tilde{\mathbf{\Omega}}}_{ij}\in\mathbb{R}^{2TM_j\times 2Ts_k}$, $\tilde{\tilde{\mathbf{\Psi}}}_{ij}\in\mathbb{R}^{2TM_j\times 2T\phi_{kj}}$.

*Proof.* Results are obtained by applying (7) over a complex matrix in order to get a real matrix. Afterwards, because we are dealing with the constant channel case, the total basis over the *T* channel uses is obtained by means of the Kronecker product, (36).□

Motivated by the fact that channel extensions are required to address the possible non-integer DoF of the MIMO *X* channel, we propose to keep the same dimension of the overlapped space while increasing the null-steering space when we increase the number of channel extensions (i.e. we deal with spatial and temporal dimensions):

*Definition 1*: The selected generating basis of the overlapping and null-steering spaces for the achievable scheme are defined to be

$$\hat{\mathbf{\Omega}}_{ij} = \tilde{\tilde{\mathbf{\Omega}}}_{ij} \begin{bmatrix} \mathbf{T}_1^i \\ \vdots \\ \mathbf{T}_T^i \end{bmatrix}, \qquad \hat{\mathbf{\Psi}}_{ij} = \tilde{\tilde{\mathbf{\Psi}}}_{ij} \tag{37}$$

where $\hat{\mathbf{\Omega}}_{ij} \in \mathbb{R}^{2TM_j \times 2s_k}$, $\hat{\mathbf{\Psi}}_{ij} \in \mathbb{R}^{2TM_j \times 2T\phi_{kj}}$ and $\tilde{\tilde{\mathbf{\Omega}}}_{ij}, \tilde{\tilde{\mathbf{\Psi}}}_{ij}$ are given by Lemma 4. Additionally, matrix $\mathbf{T}_n^i \in \mathbb{R}^{2s_k \times 2s_k}$ is employed by all transmit filters conveying messages to the *i*th receiver at the *n*th symbol extension with *n*=1,…,*T*. This is required in order to preserve the overlapping space at the unintended receiver. Notice that if precoders of messages $W_{i1}$, $W_{i2}$ employ $\tilde{\tilde{\mathbf{\Omega}}}_{i1}\mathbf{A}$ and $\tilde{\tilde{\mathbf{\Omega}}}_{i2}\mathbf{B}$ they generate a common overlapped space at the *k*th receiver when $\mathbf{A} = \mathbf{B}$. These matrices are independently drawn with random elements

$$\begin{cases} \mathbf{T}_n^i \neq \mathbf{T}_m^i \\ \mathbf{T}_n^1 \neq \mathbf{T}_n^2 \end{cases} \quad n, m = 1, \ldots, T \quad i = 1, 2 \tag{38}$$

The importance of matrices $\mathbf{T}_n^i$ will be elucidated in section V.C, where the proposed structure allows us to prove the achievability of the precoder and receiver design.

## B. *Inner bound*

The transmit precoders for messages originated at the *j*th transmitter and intended to the *i*th receiver are selected according to

$$\hat{\mathbf{V}}_{ij} = \left[\hat{\mathbf{\Omega}}_{ij}\right]^{1:d_{ij}^{(IA)}}, \qquad \hat{\mathbf{Z}}_{ij} = \left[\hat{\mathbf{\Psi}}_{ij}\right]^{1:d_{ij}^{(NS)}} \quad i, j = 1, 2 \tag{39}$$

where $\hat{\mathbf{V}}_{ij} \in \mathbb{R}^{2TM_j \times d_{ij}^{(IA)}}$, $\hat{\mathbf{Z}}_{ij} \in \mathbb{R}^{2TM_j \times d_{ij}^{(NS)}}$ are the precoders exploiting the interference alignment and null-steering concepts, respectively, and overlapping and null-steering basis $\hat{\mathbf{\Omega}}_{ij}, \hat{\mathbf{\Psi}}_{ij}$ are introduced in Definition 1, (37).

*Theorem 10* (*Symbol stream optimization*) : Given a 2*T* full rank channel matrix obtained from asymmetric complex signaling and *T*-symbol extension, the number of transmitted symbol streams per precoder, denoted by $d_{ij}^{(IA)}, d_{ij}^{(NS)} \in \mathbb{Z}_+$, are obtained as the solution of the following integer linear programming problem which maximizes the weighted sum achievable DoF of the MIMO *X* channel, see (15),

$$\begin{aligned} &(\mathrm{P}_0) \quad \underset{\{d_{ij}^{(IA)}\},\{d_{ij}^{(NS)}\}}{\text{maximize}} \quad \frac{1}{2T} \sum_{i=1}^{2} \sum_{j=1}^{2} \mu_{ij} \left( d_{ij}^{(IA)} + d_{ij}^{(NS)} \right) \\ &\textit{subject to} \\ &\begin{cases} d_{i1}^{(NS)} + d_{i2}^{(NS)} + d_{i1}^{(IA)} + d_{i2}^{(IA)} + d_{k1}^{(IA)} \leq 2TN_i \\ d_{i1}^{(NS)} + d_{i2}^{(NS)} + d_{i1}^{(IA)} + d_{i2}^{(IA)} + d_{k2}^{(IA)} \leq 2TN_i \\ d_{1j}^{(NS)} + d_{2j}^{(NS)} + d_{1j}^{(IA)} + d_{2j}^{(IA)} \leq 2TM_j \\ d_{ij}^{(IA)} \leq 2s_k \\ d_{ij}^{(NS)} \leq 2T\phi_{kj} \\ d_{ij}^{(IA)}, d_{ij}^{(NS)} \in \mathbb{Z}_+ \end{cases} \quad , \quad i \neq k, \quad i, k, j = 1, 2 \end{aligned} \tag{40}$$

where $\mu_{ij}$ are the weight factors for the different messages, $s_k$ is the dimension of the overlapped space according to Lemma 2 and $\phi_{kj}$ is the dimension of the null-steering space used by the *j*th transmitter (Lemma 3).

*Proof*: The problem formulated in (40) maximizes the number of total symbol streams taking into account the available spatial dimensions. The first two constraints describe the maximum number of spatial dimensions occupied at the *i*th destination, that depend on the dimension of interference space which is $\max\left(d_{k1}^{(IA)}, d_{k2}^{(IA)}\right)$ with ($k \neq i$). On the other hand, the null-steering symbol streams intended to the *k*th destination do not generate any interference at the *i*th destination by definition. The third constraint in (40) states that the maximum number of transmitted symbol streams is limited by the number of antennas at the *j*th transmitter. Finally, the fourth and fifth constraints are associated to the dimension of the overlapped and null-steering spaces. □

*Remark*: The optimization problem presented in Theorem 9 has to be solved for different values of $T$ in order to get the maximum achievable DoF, according to (15). Nevertheless, the dimension of the basis of the overlapped space introduced in Definition 1 is independent of the symbol extension $T$ in contrast to the other constraints of the optimization problem. Consequently, the number of times that problem $P_0$ have to be evaluated for different values of $T$ is finite.

Finally, the receivers are obtained by means of (14). The linear receive filters exist because the transmit precoders given in (39) satisfy the necessary properties presented in (13). The ensuing section illustrates the achievability proof.

### C. *Achievability Proof*

Here we show that the linear feasibility conditions introduced in (13) in section II.B are fulfilled. The first two conditions of (13) are satisfied by construction as shown in section VI.A. In order to complete the achievability proof, we have to prove the full-column rank condition of signal space matrix $\mathbf{G}_i$ defined in (13). Let us assume that the number of transmitted symbol streams are defined by $d_{ij}^{(IA)} \neq 0$, $d_{ij}^{(NS)} \neq 0$ with $i,j=1,2$ and without loss of generality $d_{22}^{(IA)} \leq d_{21}^{(IA)}$, so the dimension of the interfering space should be $d_{21}^{(IA)}$, according to (13). In the following, we show the full-column rank of matrix $\mathbf{G}_1$ at the first receiver (the proof for $\mathbf{G}_2$ can be obtained following the same steps),

$$\mathbf{G}_1 = \Big[ \underbrace{\underbrace{\hat{\mathbf{H}}_{11}\hat{\mathbf{V}}_{11}}_{d_{11}^{(IA)}} \; \underbrace{\hat{\mathbf{H}}_{12}\hat{\mathbf{V}}_{12}}_{d_{12}^{(IA)}} \; \underbrace{\hat{\mathbf{H}}_{11}\hat{\mathbf{Z}}_{11}}_{d_{11}^{(NS)}} \; \underbrace{\hat{\mathbf{H}}_{12}\hat{\mathbf{Z}}_{12}}_{d_{12}^{(NS)}}}_{\text{desired signal}} \quad \underbrace{\underbrace{\hat{\mathbf{H}}_{11}\hat{\mathbf{V}}_{12}}_{d_{21}^{(IA)}}}_{\text{interference}} \Big]$$

where $\mathbf{G}_1 \in \mathbb{R}^{2TN_1 \times \omega}$ with $\omega = d_{11}^{(IA)} + d_{11}^{(NS)} + d_{12}^{(IA)} + d_{12}^{(NS)} + d_{21}^{(IA)}$. We have identified the following three basic cases of $\mathbf{G}_1$:

- There is not symbol extension, $T$=1. (Lemma 5).
- There is symbol extension $T$, but there are not symbols transmitted in the null-steering dimensions. (Lemma 6).
- There is symbol extension $T$, and overlapping and null-steering dimensions are considered. (Lemma 7).

*Lemma 5*: Assume $T$=1 and the symbol streams that maximize problem $P_0$ in (40) are $d_{ij}^{(IA)} \neq 0$, $d_{ij}^{(NS)} \neq 0$ with $i,j=\{1,2\}$. The signal space matrix containing the desired signal and the interference is full-column rank with probability 1.

*Proof.* Assume $d_{22}^{(IA)} \leq d_{21}^{(IA)}$. The signal space matrix $\mathbf{G}_1$ is given by,

$$\begin{gathered}\mathbf{G}_1 = \left[ \tilde{\mathbf{H}}_{11}\tilde{\mathbf{\Omega}}_{11}\hat{\mathbf{T}}_1^1 \quad \tilde{\mathbf{H}}_{12}\tilde{\mathbf{\Omega}}_{12}\hat{\hat{\mathbf{T}}}_1^1 \quad \tilde{\mathbf{H}}_{11}\hat{\mathbf{\Psi}}_{11} \quad \tilde{\mathbf{H}}_{12}\hat{\hat{\mathbf{\Psi}}}_{12} \quad \tilde{\mathbf{H}}_{11}\tilde{\mathbf{\Omega}}_{21}\hat{\mathbf{T}}_1^2 \right] \\ \hat{\mathbf{T}}_1^1 = \left[\mathbf{T}_1^1\right]^{1:d_{11}^{(IA)}}, \quad \hat{\hat{\mathbf{T}}}_1^1 = \left[\mathbf{T}_1^1\right]^{1:d_{12}^{(IA)}}, \quad \hat{\mathbf{\Psi}}_{11} = \left[\tilde{\mathbf{\Psi}}_{11}\right]^{1:d_{11}^{(NS)}}, \quad \hat{\hat{\mathbf{\Psi}}}_{12} = \left[\tilde{\mathbf{\Psi}}_{12}\right]^{1:d_{12}^{(NS)}}, \quad \hat{\mathbf{T}}_1^2 = \left[\mathbf{T}_1^2\right]^{1:d_{21}^{(IA)}}\end{gathered} \tag{41}$$

where $\tilde{\mathbf{H}}_{ij}$ is the channel matrix when asymmetric complex signaling is envisioned using (7), matrix $\tilde{\mathbf{\Omega}}_{ij}$ is defined in (35) and $\hat{\mathbf{T}}_t^1 \in \mathbb{R}^{2s_2 \times d_{11}^{(IA)}}$, $\hat{\hat{\mathbf{T}}}_t^1 \in \mathbb{R}^{2s_2 \times d_{12}^{(IA)}}$, $\hat{\mathbf{T}}_t^2 \in \mathbb{R}^{2s_1 \times d_{21}^{(IA)}}$, $\hat{\mathbf{\Psi}}_{11} \in \mathbb{R}^{2\phi_{11} \times d_{11}^{(NS)}}$, $\hat{\mathbf{\Psi}}_{21} \in \mathbb{R}^{2\phi_{21} \times d_{21}^{(NS)}}$ stand for the column-wise selection of matrices $\mathbf{T}_t^i, \tilde{\mathbf{\Psi}}_{ij}$ defined in (37).

Matrix $\mathbf{G}_1$ is either a square or a tall matrix because of the first two constraints in problem $P_0$: the number of columns ($d_{11}^{(IA)} + d_{11}^{(NS)} + d_{12}^{(IA)} + d_{12}^{(NS)} + d_{21}^{(IA)}$) is equal or smaller than the number of rows ($2N_1$). Moreover, matrices $\tilde{\mathbf{\Omega}}_{ij}$ and $\tilde{\mathbf{\Psi}}_{ij}$ satisfy by construction $\tilde{\mathbf{\Omega}}_{ij} \perp \tilde{\mathbf{\Psi}}_{ij}$. Finally, if matrices $\tilde{\mathbf{H}}_{ij}$ are independent, so are matrices $\tilde{\mathbf{H}}_{ij}$, $\tilde{\mathbf{\Omega}}_{ij}$, $\tilde{\mathbf{\Psi}}_{ij}$. Therefore we can conclude that matrix $\mathbf{G}_1$ is full column rank with probability 1.

The proof for $d_{22}^{(IA)} > d_{21}^{(IA)}$ follows the same steps. □

*Lemma 6*: Assume $T$>1 and the symbol streams that maximize problem $P_0$ in (40) are $d_{ij}^{(IA)} \neq 0$, $d_{ij}^{(NS)} = 0$ with $i,j=\{1,2\}$. The signal space matrix $\mathbf{G}_1$ is full-column rank with probability 1.

*Proof.* Assume $d_{22}^{(IA)} \leq d_{21}^{(IA)}$. The signal space matrix $\mathbf{G}_1$ is given by,

$$\mathbf{G}_1=\begin{bmatrix}\tilde{\mathbf{H}}_{11}\tilde{\mathbf{\Omega}}_{11}\hat{\mathbf{T}}_1^1 & \tilde{\mathbf{H}}_{12}\tilde{\mathbf{\Omega}}_{12}\hat{\hat{\mathbf{T}}}_1^1 & \tilde{\mathbf{H}}_{11}\tilde{\mathbf{\Omega}}_{21}\hat{\mathbf{T}}_1^2\\ \vdots & \vdots & \vdots\\ \tilde{\mathbf{H}}_{11}\tilde{\mathbf{\Omega}}_{11}\hat{\mathbf{T}}_t^1 & \tilde{\mathbf{H}}_{12}\tilde{\mathbf{\Omega}}_{12}\hat{\hat{\mathbf{T}}}_t^1 & \tilde{\mathbf{H}}_{11}\tilde{\mathbf{\Omega}}_{21}\hat{\mathbf{T}}_t^2\\ \vdots & \vdots & \vdots\\ \tilde{\mathbf{H}}_{11}\tilde{\mathbf{\Omega}}_{11}\hat{\mathbf{T}}_T^1 & \tilde{\mathbf{H}}_{12}\tilde{\mathbf{\Omega}}_{12}\hat{\hat{\mathbf{T}}}_T^1 & \tilde{\mathbf{H}}_{11}\tilde{\mathbf{\Omega}}_{21}\hat{\mathbf{T}}_T^2\end{bmatrix}\qquad \hat{\mathbf{T}}_t^1=\left[\mathbf{T}_t^1\right]^{1:d_{11}^{(IA)}},\quad \hat{\hat{\mathbf{T}}}_t^1=\left[\mathbf{T}_t^1\right]^{1:d_{12}^{(IA)}},\quad \hat{\mathbf{T}}_t^2=\left[\mathbf{T}_t^2\right]^{1:d_{21}^{(IA)}} \tag{42}$$

where $\tilde{\mathbf{H}}_{ij}$ is the channel matrix after the asymmetric complex signaling is applied, matrix $\tilde{\mathbf{\Omega}}_{ij}\in\mathbb{R}^{2M_j\times 2s_k}$ ($k\neq i$) is defined in (35) and $\hat{\mathbf{T}}_t^1\in\mathbb{R}^{2s_2\times d_{11}^{(IA)}}$, $\hat{\hat{\mathbf{T}}}_t^1\in\mathbb{R}^{2s_2\times d_{12}^{(IA)}}$, $\hat{\mathbf{T}}_t^2\in\mathbb{R}^{2s_1\times d_{21}^{(IA)}}$ are the column-wise selection of matrices $\mathbf{T}_t^1,\mathbf{T}_t^2$ introduced in Definition 1. Matrix $\mathbf{G}_1$ is full-column rank with probability 1 whenever matrices $\mathbf{T}_t^1,\mathbf{T}_t^2$ are randomly and independently generated if $s_k>1$, as imposed by Definition 1. When $s_k=1$, it remains to show that those columns of $\mathbf{G}_1$ that use $\mathbf{T}_t^1$ by means of $\hat{\mathbf{T}}_1^1$ and $\hat{\hat{\mathbf{T}}}_1^1$ (first and second block column of $\mathbf{G}_1$) are linearly independent, because the third block column of $\mathbf{G}_1$ is linearly independent of the first two block columns thanks to use $\mathbf{T}_t^2$. The first two block columns of $\mathbf{G}_1$ are linearly independent by using Lemma 16 defined in Appendix A.

The proof for $d_{22}^{(IA)}>d_{21}^{(IA)}$ follows the same steps. □

*Lemma 7*: Assume $T>1$ and the symbol streams that maximize problem $P_0$ in (40) are $d_{ij}^{(IA)}\neq 0$, $d_{ij}^{(NS)}\neq 0$ with $i,j=\{1,2\}$. The signal space matrix $\mathbf{G}_1$ is full-column rank with probability 1.

*Proof.* Signal space matrix $\mathbf{G}_1$ when $d_{22}^{(IA)}\leq d_{21}^{(IA)}$, $d_{12}^{(NS)}\leq d_{11}^{(NS)}$ is given by,

$$\mathbf{G}_1=\Bigg[\underbrace{\begin{matrix}\tilde{\mathbf{H}}_{11}\tilde{\mathbf{\Omega}}_{11}\hat{\mathbf{T}}_1^1\\ \vdots\\ \\ \tilde{\mathbf{H}}_{11}\tilde{\mathbf{\Omega}}_{11}\hat{\mathbf{T}}_t^1\\ \vdots\\ \tilde{\mathbf{H}}_{11}\tilde{\mathbf{\Omega}}_{11}\hat{\mathbf{T}}_T^1\end{matrix}}_{d_{11}^{(IA)}}\;\underbrace{\begin{matrix}\tilde{\mathbf{H}}_{12}\tilde{\mathbf{\Omega}}_{12}\hat{\hat{\mathbf{T}}}_1^1\\ \vdots\\ \\ \tilde{\mathbf{H}}_{12}\tilde{\mathbf{\Omega}}_{12}\hat{\hat{\mathbf{T}}}_t^1\\ \vdots\\ \tilde{\mathbf{H}}_{12}\tilde{\mathbf{\Omega}}_{12}\hat{\hat{\mathbf{T}}}_T^1\end{matrix}}_{d_{12}^{(IA)}}\;\underbrace{\begin{matrix}\tilde{\mathbf{H}}_{11}\tilde{\mathbf{\Psi}}_{11} & \mathbf{0} & \mathbf{0} & \mathbf{0} & \mathbf{0}\\ \mathbf{0} & \ddots & & \vdots & \vdots\\ \vdots & & \tilde{\mathbf{H}}_{11}\tilde{\mathbf{\Psi}}_{11} & \mathbf{0} & \\ \mathbf{0} & \mathbf{0} & \mathbf{0} & \ddots & \mathbf{0}\\ \vdots & \vdots & \vdots & \vdots & \tilde{\mathbf{H}}_{11}\hat{\mathbf{\Psi}}_{11}\\ \mathbf{0} & \mathbf{0} & \mathbf{0} & \mathbf{0} & \mathbf{0}\end{matrix}}_{d_{11}^{(NS)}}\;\underbrace{\begin{matrix}\tilde{\mathbf{H}}_{12}\tilde{\mathbf{\Psi}}_{12} & \mathbf{0} & \mathbf{0}\\ \vdots & \ddots & \vdots\\ \mathbf{0} & \mathbf{0} & \tilde{\mathbf{H}}_{12}\hat{\hat{\mathbf{\Psi}}}_{12}\\ \mathbf{0} & \mathbf{0} & \mathbf{0}\\ \vdots & \vdots & \vdots\\ \mathbf{0} & \mathbf{0} & \mathbf{0}\end{matrix}}_{d_{12}^{(NS)}}\;\underbrace{\begin{matrix}\tilde{\mathbf{H}}_{11}\tilde{\mathbf{\Omega}}_{21}\hat{\mathbf{T}}_1^2\\ \vdots\\ \\ \tilde{\mathbf{H}}_{11}\tilde{\mathbf{\Omega}}_{21}\hat{\mathbf{T}}_t^2\\ \vdots\\ \tilde{\mathbf{H}}_{11}\tilde{\mathbf{\Omega}}_{21}\hat{\mathbf{T}}_T^2\end{matrix}}_{d_{21}^{(IA)}}\Bigg]$$

$$\hat{\mathbf{T}}_t^1=\left[\mathbf{T}_t^1\right]^{1:d_{11}^{(IA)}},\quad \hat{\hat{\mathbf{T}}}_t^1=\left[\mathbf{T}_t^1\right]^{1:d_{12}^{(IA)}},\quad \hat{\mathbf{T}}_t^2=\left[\mathbf{T}_t^2\right]^{1:d_{21}^{(IA)}},\quad \hat{\mathbf{\Psi}}_{11}=\left[\tilde{\mathbf{\Psi}}_{11}\right]^{1:m_{11}},\quad \hat{\hat{\mathbf{\Psi}}}_{12}=\left[\tilde{\mathbf{\Psi}}_{12}\right]^{1:m_{12}} \tag{43}$$

$$m_{ij}=d_{ij}^{(NS)}-\left\lfloor\frac{d_{ij}^{(NS)}}{2\phi_{kj}}\right\rfloor 2\phi_{kj},\quad d_{ij}^{(NS)}\leq 2T\phi_{kj},\quad k\neq i$$

where $\tilde{\mathbf{H}}_{ij}$ is the channel matrix after the asymmetric complex signaling is applied, matrix $\tilde{\mathbf{\Omega}}_{ij}\in\mathbb{R}^{2M_j\times 2s_k}$, $\tilde{\mathbf{\Psi}}_{ij}\in\mathbb{R}^{2M_j\times 2\phi_{kj}}$ ($k\neq i$) is defined in (35) and $\hat{\mathbf{T}}_t^1\in\mathbb{R}^{2s_2\times d_{11}^{(IA)}}$, $\hat{\hat{\mathbf{T}}}_t^1\in\mathbb{R}^{2s_2\times d_{12}^{(IA)}}$, $\hat{\mathbf{T}}_t^2\in\mathbb{R}^{2s_1\times d_{21}^{(IA)}}$ are the column-wise selection of matrices $\mathbf{T}_t^1,\mathbf{T}_t^2$ introduced in Definition 1. $\tilde{\mathbf{\Psi}}_{ij}$ is presented in (35). Parameter $m_{ij}$ in (43) comes up because the number of symbol streams using the null-steering basis is not a multiple of the dimension of the null-steering space, with $\phi_{kj}$ defined in Lemma 3.Using similar arguments as in the proofs of Lemma 6 and Lemma 7 matrix $\mathbf{G}_1$ is full-column rank with probability 1. Notice that $\tilde{\mathbf{\Omega}}_{ij}\perp\tilde{\mathbf{\Psi}}_{ij}$ and matrices $\tilde{\mathbf{H}}_{ij}$ are independent, so are matrices $\tilde{\mathbf{H}}_{ij}$, $\tilde{\mathbf{\Omega}}_{ij}$, $\tilde{\mathbf{\Psi}}_{ij}$. The proof for other configurations of $d_{22}^{(IA)},d_{21}^{(IA)},d_{12}^{(NS)},d_{11}^{(NS)}$ follows similar guidelines. □

### *D. Maximum achievable DoF*

In the previous subsections we have derived an achievable scheme for the MIMO *X* channel where precoders are designed first, and afterwards the receivers are obtained. However, as it was discussed in section III.C, this procedure is not always optimal for an arbitrary number of antennas at each node. Consequently, we have to compare the achievable DoF provided by our scheme in the original and reciprocal MIMO *X* channel, as it is indicated by Theorem 1.

### E. Time Varying Channel Extension

Although the proposed precoder design algorithm presented in section VI has dealt with the constant channel case, it can be easily extended to the case where there is a time varying channel, i.e. channel coefficients vary over the $T$ channel extensions. In such a case, our signal model should take into account the following equivalent channel matrices

$$\hat{\mathbf{H}}_{ij} = \begin{bmatrix} \tilde{\mathbf{H}}_{ij}(1) & \mathbf{0} & \mathbf{0} & \mathbf{0} & \mathbf{0} \\ \mathbf{0} & \ddots & \mathbf{0} & \mathbf{0} & \mathbf{0} \\ \mathbf{0} & \mathbf{0} & \tilde{\mathbf{H}}_{ij}(t) & \mathbf{0} & \mathbf{0} \\ \mathbf{0} & \mathbf{0} & \mathbf{0} & \ddots & \mathbf{0} \\ \mathbf{0} & \mathbf{0} & \mathbf{0} & \mathbf{0} & \tilde{\mathbf{H}}_{ij}(T) \end{bmatrix}, \quad \tilde{\mathbf{H}}_{ij}(t) = \begin{bmatrix} \mathrm{Re}\{\mathbf{H}_{ij}(t)\} & -\mathrm{Im}\{\mathbf{H}_{ij}(t)\} \\ \mathrm{Im}\{\mathbf{H}_{ij}(t)\} & \mathrm{Re}\{\mathbf{H}_{ij}(t)\} \end{bmatrix}, \tag{44}$$

where $\tilde{\mathbf{H}}_{ij}(t) \in \mathbb{R}^{2N_i \times 2M_j}$ are the channel matrices obtained after applying the asymmetric complex signaling, $\hat{\mathbf{H}}_{ij} \in \mathbb{R}^{2TN_i \times 2TM_j}$ is the equivalent channel matrix after a $T$-symbol extension and asymmetric complex signaling and $t$ is in channel extension index, $t$=1..$T$.

The lemmas of the generating basis of the overlapping and null-steering spaces have to be reformulated so as to deal with the new channel matrix in equation (44):

*Lemma 8:* The paired-generating basis of the transmit filters associated to symbol streams intended to the $i$th receiver, $\mathbf{s}_{i1}^{(IA)}, \mathbf{s}_{i2}^{(IA)}$, that define a common overlapped space at the $k$th receiver ($i \neq k$) are given by,

$$\tilde{\tilde{\mathbf{\Omega}}}_{i1} = \mathbf{U}_{k1}^{OV}\mathbf{\Gamma}_{k1}^{-1}, \quad \tilde{\tilde{\mathbf{\Omega}}}_{i2} = \mathbf{U}_{k2}^{OV}\mathbf{\Gamma}_{k2}^{-1}, \quad i \neq k, \quad i,k = 1,2 \tag{45}$$

where $\tilde{\tilde{\mathbf{\Omega}}}_{ij} \in \mathbb{C}^{2TM_j \times s_k}$ and $\mathbf{U}_{kj}^{OV} \in \mathbb{C}^{2TM_j \times s_k}, \mathbf{\Gamma}_{kj} \in \mathbb{C}^{s_k \times s_k}$ are obtained by the GSVD over $\hat{\mathbf{H}}_{k,1}, \hat{\mathbf{H}}_{k,2}$. The dimension of this space is,

$$s_k = rank\left(\hat{\mathbf{H}}_{k1}\right) + rank\left(\hat{\mathbf{H}}_{k2}\right) - rank\left(\left[\hat{\mathbf{H}}_{k1}\ \hat{\mathbf{H}}_{k2}\right]\right) \tag{46}$$

*Lemma 9*: The generating basis of the transmit filters associated to symbol streams intended to the $i$th receiver, $\mathbf{s}_{i1}^{(NS)}, \mathbf{s}_{i2}^{(NS)}$, that do not add interference at the $k$th receiver ($i \neq k$) are given by,

$$\tilde{\tilde{\mathbf{\Psi}}}_{i1} = \mathbf{U}_{k1}^{NS}, \quad \tilde{\tilde{\mathbf{\Psi}}}_{i2} = \mathbf{U}_{k2}^{NS}, \quad i \neq k, \quad i,k = 1,2 \tag{47}$$

where $\tilde{\tilde{\mathbf{\Psi}}}_{ij} \in \mathbb{C}^{2TM_j \times \phi_{kj}}, \mathbf{U}_{kj}^{NS} \in \mathbb{C}^{2TM_j \times \phi_{kj}}$ are obtained by the GSVD over $\hat{\mathbf{H}}_{k,1}, \hat{\mathbf{H}}_{k,2}$. The dimensions of the null-steering space is

$$\phi_{kj} = 2TM_j - rank\left(\hat{\mathbf{H}}_{kj}\right) \tag{48}$$

Now, with the generating basis defined by Lemma 8 and Lemma 9 we can apply Definition and the inner bound described in section VI.B. Finally, the achievability proof in this case can be shown using similar steps as in Lemma 5. For example, if $d_{21}^{(IA)} \geq d_{22}^{(IA)}$ the signal space matrix has the following structure,

$$\mathbf{G}_1 = \left[\hat{\mathbf{H}}_{11}\tilde{\tilde{\mathbf{\Omega}}}_{11}\hat{\mathbf{T}}_1^1\ \ \hat{\mathbf{H}}_{12}\tilde{\tilde{\mathbf{\Omega}}}_{12}\hat{\hat{\mathbf{T}}}_1^1\ \ \hat{\mathbf{H}}_{11}\hat{\mathbf{\Psi}}_{11}\ \ \hat{\mathbf{H}}_{12}\hat{\hat{\mathbf{\Psi}}}_{12}\ \ \hat{\mathbf{H}}_{11}\tilde{\tilde{\mathbf{\Omega}}}_{21}\hat{\mathbf{T}}_1^2\right]$$

$$\hat{\mathbf{T}}_1^1 = \left[\mathbf{T}_1^1\right]^{1:d_{11}^{(IA)}}, \quad \hat{\hat{\mathbf{T}}}_1^1 = \left[\mathbf{T}_1^1\right]^{1:d_{12}^{(IA)}}, \quad \hat{\mathbf{\Psi}}_{11} = \left[\tilde{\tilde{\mathbf{\Psi}}}_{11}\right]^{1:d_{11}^{(NS)}}, \quad \hat{\hat{\mathbf{\Psi}}}_{12} = \left[\tilde{\tilde{\mathbf{\Psi}}}_{12}\right]^{1:d_{12}^{(NS)}}, \quad \hat{\mathbf{T}}_1^2 = \left[\mathbf{T}_1^2\right]^{1:d_{21}^{(IA)}}$$

where $\mathbf{G}_1 \in \mathbb{R}^{2TN_1 \times \omega}$ with $\omega = d_{11}^{(IA)} + d_{11}^{(NS)} + d_{12}^{(IA)} + d_{12}^{(NS)} + d_{21}^{(IA)}$.

## VII. Total DoF

The achievable scheme proposed in the previous section is used in Theorem 6 (proof in Appendix B) to show that the outer bound on the total DoF of the MIMO $Z$ channel are attained in a MIMO $X$ channel. The allocation of symbol streams over the transmit filters is obtained by solving problem $P_0$ enunciated in Theorem 10 for different values of symbol extensions ($T$). Although the problem is easily addressed by computational methods, it is hard to be solved analytically. In this regard, here we show for certain antenna configurations and full-rank channel matrices how the outer bound on the total DoF given by Theorem 4 is attained:

- Equal number of antennas at transmitters and at receivers ($M_1$=$M_2$=$M$, $N_1$=$N_2$=$N$) (Lemma 10)
- Equal number of antennas at transmitters ($M_1$=$M_2$=$M$, $N_1$, $N_2$) (Lemma 11)
- Equal number of antennas at receivers ($M_1$, $M_2$, $N_1$=$N_2$=$N$) (Lemma 12)

*Lemma 10* : In a MIMO *X* network equipped with (*M*, *M*, *N*, *N*) antennas the precoders and receivers that attain the total DoF are given by the achievable scheme proposed in section VI, for $M \geq N$. For $M<N$, the linear filters are obtained from the reciprocal MIMO *X* network (*N*, *N*, *M*, *M*).

*Proof.* In such antenna configuration the dimensions of the overlapped and null-steering basis, see Lemma 2 and Lemma 3, are

$$s = s_1 = s_2 = \begin{cases} N & if \quad N \leq M \\ 2M - N & if \quad N/2 \leq M < N \\ 0 & if \quad M < N/2 \end{cases}$$

$$\phi = \phi_{11} = \phi_{12} = \phi_{21} = \phi_{22} = (M-N)^+$$

Furthermore, due to the symmetry of the problem, we can enforce a symmetric solution for the transmitted symbol streams,

$$d^{(IA)} = d_{11}^{(IA)} = d_{12}^{(IA)} = d_{21}^{(IA)} = d_{22}^{(IA)}$$
$$d^{(NS)} = d_{11}^{(NS)} = d_{12}^{(NS)} = d_{21}^{(NS)} = d_{22}^{(NS)}$$

Using the previous variable definition, the problem $P_0$ introduced in Theorem 10 for a given *T*-symbol extension is transformed into,

$$\left(\tilde{P}_0\right) \quad \underset{d^{(IA)},d^{(NS)}}{\text{maximize}} \quad \frac{1}{2T} 4\left(d^{(IA)} + d^{(NS)}\right)$$
$$\textit{subject to}$$
$$\begin{cases} 2d^{(NS)} + 3d^{(IA)} \leq 2TN \\ 2d^{(NS)} + 2d^{(IA)} \leq 2TM \\ d^{(IA)} \leq 2s \\ d^{(NS)} \leq 2T\phi \\ d^{(IA)}, d^{(NS)} \in \mathbb{Z}_+ \end{cases}$$

The number of symbol streams is obtained by solving the dual problem of the integer linear programming problem $\tilde{P}_0$, [23], for different *T*-symbol extensions, which result is presented in Table I. We can observe that when $M<N$ the proposed scheme does not attain the outer bound. However, resorting to Theorem 1 (reciprocal MIMO *X* network) we can claim that inner and outer bounds are tight for all *M* and *N*.□

TABLE I
ACHIEVABLE DOF WHEN $M_1=M_2=M$ TRANSMITTING AND $N_1=N_2=N$ RECEIVING ANTENNAS
OUTER BOUND: $\min\left(2M, 2N, \max(M,N) + \frac{M}{2}, \max(M,N) + \frac{N}{2}, \frac{4}{3}\max(M,N)\right)$, THEOREM 5
THE CASES WHERE THE OUTER BOUND IS MET USING RECIPROCITY ARE MARKED WITH (*)

| | $M \geq 3N/2$ | $N \leq M < 3N/2$ | $N/2 \leq M < N$ | $M < N/2$ |
|---|---|---|---|---|
| $d_{11}^{(IA)}$ | 0 | $6N-4M$ | $4M-2N$ | 0 |
| $d_{11}^{(NS)}$ | $N$ | $6M-6N$ | 0 | 0 |
| $d_{12}^{(IA)}$ | 0 | $6N-4M$ | $4M-2N$ | 0 |
| $d_{12}^{(NS)}$ | $N$ | $6M-6N$ | 0 | 0 |
| $d_{21}^{(IA)}$ | 0 | $6N-4M$ | $4M-2N$ | 0 |
| $d_{21}^{(NS)}$ | $N$ | $6M-6N$ | 0 | 0 |
| $d_{22}^{(IA)}$ | 0 | $6N-4M$ | $4M-2N$ | 0 |
| $d_{22}^{(NS)}$ | $N$ | $6M-6N$ | 0 | 0 |
| $T$ | 1 | 3 | 3 | 1 |
| **DoF** | $\boldsymbol{2N}$ | $\boldsymbol{4/3\,M}$ | $\boldsymbol{4/3(2M-N)}$ (*) | **0** (*) |

*Lemma 11*: In a MIMO *X* network equipped with ($M_1$, $M_2$, $N$, $N$) antennas, the precoders and receivers that attain the total DoF are given by the achievable scheme proposed in section VI , for $M_1+M_2 \geq 2N$ . Otherwise, for $M_1+M_2<2N$ the linear filters must be obtained using the reciprocal MIMO *X* network ($N$, $N$, $M_1$, $M_2$).

*Proof.* For current antenna configuration the dimensions of the overlapped (Lemma 2) and null-steering (Lemma 3) basis are

$$s=s_1=s_2=\begin{cases} N & \text{if} \quad N\leq M_1, N\leq M_2 \\ M_1 & \text{if} \quad M_1<N, N\leq M_2 \\ M_2 & \text{if} \quad N\leq M_1, M_2<N \\ M_1+M_2-N & \text{if} \quad M_1<N, M_2<N, N\leq M_1+M_2 \\ 0 & \text{if} \quad M_1<N, M_2<N, M_1+M_2<N \end{cases}$$

$$\phi_1=\phi_{11}=\phi_{21}=\left(M_1-N\right)^+, \qquad \phi_2=\phi_{12}=\phi_{22}=\left(M_2-N\right)^+$$

Additionally, due to the symmetry of the problem we can define the following variables for the transmitted symbol streams,

$$d^{(IA)}=d_{11}^{(IA)}=d_{12}^{(IA)}=d_{21}^{(IA)}=d_{22}^{(IA)}$$
$$d_1^{(NS)}=d_{11}^{(NS)}=d_{21}^{(NS)}, \quad d_2^{(NS)}=d_{12}^{(NS)}=d_{22}^{(NS)}$$

Using the previous variable definition, problem $P_0$ introduced in Theorem 10 for a given *T*-symbol extension is transformed into,

$$\left(\tilde{\tilde{P}}_0\right) \quad \underset{d^{(IA)},d_1^{(NS)},d_2^{(NS)}}{\text{maximize}} \quad \frac{1}{2T}\left(4d^{(IA)}+2d_1^{(NS)}+2d_2^{(NS)}\right)$$
$$\textit{subject to}$$
$$\begin{cases} d_1^{(NS)}+d_2^{(NS)}+3d^{(IA)}\leq 2TN \\ 2d_1^{(NS)}+2d^{(IA)}\leq 2TM_1 \\ 2d_2^{(NS)}+2d^{(IA)}\leq 2TM_2 \\ d^{(IA)}\leq 2s \\ d_1^{(NS)}\leq 2T\phi_1 \\ d_2^{(NS)}\leq 2T\phi_2 \\ d^{(IA)},d_1^{(NS)},d_2^{(NS)}\in\mathbb{Z}_+ \end{cases}$$

Table II ($M_1 \geq N$, $M_2 \geq N$) and Table III ($M_1<N$, $M_2 \geq N$), ($M_1 \geq N$, $M_2<N$) depict the solution of $\tilde{\tilde{P}}_0$ for different values of $T$. For certain antenna configurations (first and fourth columns in Table III) the upper bound is attained by analyzing the reciprocal MIMO *X* network and using Lemma 1, i.e. $N$ transmitting antennas and $M_1$, $M_2$ receiving antennas, see Table VII (second and third columns). Likewise, the antenna configuration ($M_1<N$, $M_2<N$) attains the outer bound DoF by resorting to MIMO *X* network. □

TABLE II

ACHIEVABLE DOF WHEN $M_1$, $M_2$ TRANSMITTING AND $N_1=N_2=N$ RECEIVING ANTENNAS ($M_1 \geq N$, $M_2 \geq N$)

OUTER BOUND: $\min\left(2N, \frac{M_1+M_2+N}{2}, \frac{2M_1+2M_2}{3}\right)$, SEE THEOREM 5

| | $2N\leq M_1$ | $N\leq M_1<2N$<br>$M_1+M_2<3N$ | $N\leq M_1<2N$<br>$3N\leq M_1+M_2$ |
|---|---|---|---|
| $d_{11}^{(IA)}$ | 0 | $6N-2M_2-2M_1$ | 0 |
| $d_{11}^{(NS)}$ | $2N$ | $6M_1-6N$ | $2M_1-2N$ |
| $d_{12}^{(IA)}$ | 0 | $6N-2M_2-2M_1$ | 0 |
| $d_{12}^{(NS)}$ | 0 | $6M_2-6N$ | $4N-2M_1$ |
| $d_{21}^{(IA)}$ | 0 | $6N-2M_2-2M_1$ | 0 |
| $d_{21}^{(NS)}$ | $2N$ | $6M_1-6N$ | $2M_1-2N$ |
| $d_{22}^{(IA)}$ | 0 | $6N-2M_2-2M_1$ | 0 |
| $d_{22}^{(NS)}$ | 0 | $6M_2-6N$ | $4N-2M_1$ |
| $T$ | 1 | 3 | 1 |
| **DoF** | $\boldsymbol{2N}$ | $\boldsymbol{\frac{2M_1+2M_2}{3}}$ | $\boldsymbol{2N}$ |

TABLE III
ACHIEVABLE DOF WHEN $M_1$, $M_2$ TRANSMITTING AND $N_1$=$N_2$=$N$ RECEIVING ANTENNAS .
($M_1$<$N$, $M_2$≥$N$) OUTER BOUND: $\min\left(M_1+M_2, 2N, \frac{2M_2+M_1}{2}, \frac{2N+M_2}{2}, \frac{2M_2+2N}{3}\right)$

($M_1$≥$N$, $M_2$<$N$) OUTER BOUND: $\min\left(M_1+M_2, 2N, \frac{M_2+2M_1}{2}, \frac{2N+M_1}{2}, \frac{2M_1+2N}{3}\right)$, SEE THEOREM 5

THE CASES WHERE THE OUTER BOUND IS MET USING RECIPROCITY ARE MARKED WITH (*)

| | $M_1<N$<br>$N\le M_1+M_2<2N$ | $M_1<N$<br>$2N\le M_1+M_2$<br>$M_2<2N$ | $M_1<N$<br>$2N\le M_2$ | $M_2<N$<br>$N\le M_1+M_2<2N$ | $M_2<N$<br>$2N\le M_1+M_2$<br>$M_1<2N$ | $M_2<N$<br>$2N\le M_1$ |
|---|---|---|---|---|---|---|
| $d_{11}^{(IA)}$ | $2M_1$ | $4N-2M_2$ | 0 | $2M_2$ | $4N-2M_1$ | 0 |
| $d_{11}^{(NS)}$ | 0 | 0 | 0 | $6M_1-6N$ | $6M_1-6N$ | $2N$ |
| $d_{12}^{(IA)}$ | $2M_1$ | $4N-2M_2$ | 0 | $2M_2$ | $4N-2M_1$ | 0 |
| $d_{12}^{(NS)}$ | $6M_2-6N$ | $6M_2-6N$ | $2N$ | 0 | 0 | 0 |
| $d_{21}^{(IA)}$ | $2M_1$ | $4N-2M_2$ | 0 | $2M_2$ | $4N-2M_1$ | 0 |
| $d_{21}^{(NS)}$ | 0 | 0 | 0 | $6M_1-6N$ | $6M_1-6N$ | $2N$ |
| $d_{22}^{(IA)}$ | $2M_1$ | $4N-2M_2$ | 0 | $2M_2$ | $4N-2M_1$ | 0 |
| $d_{22}^{(NS)}$ | $6M_2-6N$ | $6M_2-6N$ | $2N$ | 0 | 0 | 0 |
| $T$ | 3 | 3 | 1 | 3 | 3 | 1 |
| **DoF** | $\frac{4M_1}{3}+2M_2-2N$ (*) | $\frac{2M_2+2N}{3}$ | $2N$ | $\frac{4M_2}{3}+2M_1-2N$ (*) | $\frac{2M_1+2N}{3}$ | $2N$ |

*Lemma 12*: In a MIMO $X$ network equipped with ($M$, $M$, $N_1$, $N_2$) antennas, the precoders and receivers that attain the total DoF are given by the achievable scheme proposed in section VI when $\max\left(N_1/2+N_2/2, \min\left(N_1,N_2\right)\right)\le M<\min\left(N_1,N_2\right)$. Otherwise, $M>\min\left(N_1,N_2\right)$ or $\max\left(N_1/2+N_2/2, \min\left(N_1,N_2\right)\right)>M$ the linear filters have to be obtained in the reciprocal MIMO $X$ network ($N_1$, $N_2$, $M$, $M$)

*Proof.* The dimensions of the overlapped (Lemma 2) and null-steering (Lemma 3) spaces are given by,

$$s_1=\begin{cases} N_2 & if \quad N_2\le M \\ 2M-N_2 & if \quad N_2/2\le M<N_2 \\ 0 & if \quad M<N_2/2 \end{cases},\quad s_2=\begin{cases} N_1 & if \quad N_1\le M \\ 2M-N_1 & if \quad N_1/2\le M<N_1 \\ 0 & if \quad M<N_1/2 \end{cases}$$

$$\phi_1=\phi_{11}=\phi_{12}=\left(M-N_2\right)^+,\qquad \phi_2=\phi_{21}=\phi_{22}=\left(M-N_1\right)^+$$

Because of the symmetry of the problem we can define the following variables for the transmitted symbol streams,

$$d_1^{(IA)}=d_{11}^{(IA)}=d_{12}^{(IA)},\quad d_2^{(IA)}=d_{21}^{(IA)}=d_{22}^{(IA)}$$
$$d_1^{(NS)}=d_{11}^{(NS)}=d_{12}^{(NS)},\quad d_2^{(NS)}=d_{21}^{(NS)}=d_{22}^{(NS)}$$

The number of transmitted symbols streams are optimized according to problem $P_0$ introduced in Theorem 10 for a given $T$-symbol extension and the previous variable definition,

$$\left(\tilde{\tilde{\mathrm{P}}}_0\right) \quad \underset{d_1^{(IA)},d_2^{(IA)},d_1^{(NS)},d_2^{(NS)}}{\text{maximize}} \quad \frac{1}{2T}\left(2d_1^{(IA)}+2d_2^{(IA)}+2d_1^{(NS)}+2d_2^{(NS)}\right)$$

*subject to*

$$\begin{cases} 2d_1^{(NS)}+2d_1^{(IA)}+d_2^{(IA)} \le 2TN_1 \\ 2d_1^{(NS)}+d_1^{(IA)}+2d_2^{(IA)} \le 2TN_2 \\ d_1^{(NS)}+d_2^{(NS)}+d_1^{(IA)}+d_2^{(IA)} \le 2TM \\ d_1^{(IA)} \le 2s_1 \\ d_2^{(IA)} \le 2s_2 \\ d_1^{(NS)} \le 2T\phi_1 \\ d_2^{(NS)} \le 2T\phi_2 \\ d_1^{(IA)},d_2^{(IA)},d_1^{(NS)},d_2^{(NS)} \in \mathbb{Z}_+ \end{cases}$$

The results for $\tilde{\tilde{\mathrm{P}}}_0$ evaluated for different values of $T$ are presented in Table IV (case $M \ge N_1$, $M \ge N_2$, $N_1 \ge N_2$), Table V (case $M \ge N_1$, $M \ge N_2$, $N_2 \ge N_1$), Table VI (case $M \ge N_1$, $M < N_2$) and Table VII (case $M < N_1$, $M \ge N_2$). As in previous lemmas, wherever the outer bound is not attained, it can be derived by resorting to the reciprocal network. See for example, the first column in Table VI where ($N_1 \le M,\ N_2/2 \le M < N_1/2 + N_2/2$), such condition in the reciprocal network becomes $\underline{M}_1 \le \underline{N},\ \underline{M}_2 \le 2\underline{N},\ 2\underline{N} < \underline{M}_1 + \underline{M}_2$. Observing the second column of Table III we confirm that the upper bound is attained by designing precoders and receivers in the reciprocal MIMO $X$ network and applying Lemma 1. Similar arguments are valid to proof that the case shown in the fourth column of Table VI and first and fourth columns in Table VII, attain the outer bound in the corresponding reciprocal network which is described by third, fifth and sixth columns in Table III, respectively. Finally, let us remark that the case $M < N_1$, $M < N_2$ has not been depicted because it is attained in the reciprocal network when $\underline{M}_1 \ge \underline{N},\ \underline{M}_2 \ge \underline{N}$. □

TABLE IV

ACHIEVABLE DOF WHEN $M_1=M_2=M$ TRANSMITTING $N_1$,$N_2$ RECEIVING ANTENNAS, $M>N_1$, $M>N_2$, $N_1>N_2$

OUTER BOUND: $\min\left(N_1+N_2, \frac{2M+N_2}{2}, \frac{4M}{3}\right)$, SEE THEOREM 5

| | $3N_2/2 \le N_1$<br>$N_1+N_2/2 \le M$ | $3N_2/2 \le N_1$<br>$N_1 \le M < N_1+N_2/2$ | $N_2 \le N_1 < 3N_2/2$<br>$N_1+N_2/3 \le M$ | $N_2 \le N_1 < 3N_2/2$<br>$3N_2/2 \le M < N_1+N_2/3$ | $N_2 \le N_1 < 3N_2/2$<br>$N_1 \le M < 3N_2/2$ |
|---|---|---|---|---|---|
| $d_{11}^{(IA)}$ | 0 | 0 | 0 | 0 | $6N_2-4M$ |
| $d_{11}^{(NS)}$ | $N_1$ | $2M-N_2$ | $N_1$ | $2M-N_2$ | $6M-6N_2$ |
| $d_{12}^{(IA)}$ | 0 | 0 | 0 | 0 | $6N_2-4M$ |
| $d_{12}^{(NS)}$ | $N_1$ | $2M-N_2$ | $N_1$ | $2M-N_2$ | $6M-6N_2$ |
| $d_{21}^{(IA)}$ | 0 | $4N_1+2N_2-4M$ | 0 | $4N_1+2N_2-4M$ | $6N_1-4M$ |
| $d_{21}^{(NS)}$ | $N_2$ | $4M-4N_1$ | $N_2$ | $4M-4N_1$ | $6M-6N_1$ |
| $d_{22}^{(IA)}$ | 0 | $4N_1+2N_2-4M$ | 0 | $4N_1+2N_2-4M$ | $6N_1-4M$ |
| $d_{22}^{(NS)}$ | $N_2$ | $4M-4N_1$ | $N_2$ | $4M-4N_1$ | $6M-6N_1$ |
| $T$ | 1 | 2 | 1 | 2 | 3 |
| **DoF** | $\boldsymbol{N_1+N_2}$ | $\boldsymbol{\frac{2M+N_2}{2}}$ | $\boldsymbol{N_1+N_2}$ | $\boldsymbol{\frac{2M+N_2}{2}}$ | $\boldsymbol{\frac{4M}{3}}$ |

TABLE V

ACHIEVABLE DOF WHEN $M_1=M_2=M$ , TRANSMITTING $N_1,N_2$ RECEIVING ANTENNAS, $M>N_1$, $M>N_2$ $N_1<N_2$

OUTER BOUND: $\min\left(N_1+N_2,\frac{2M+N_1}{2},\frac{4M}{3}\right)$, SEE THEOREM 5

| | $3N_1/2\le N_2$<br>$N_1/2+N_2\le M$ | $3N_1/2\le N_2$<br>$3N_1/2\le M<N_1/2+N_2$ | $N_1\le N_2<3N_1/2$<br>$N_2+N_1/3\le M$ | $N_1\le N_2<3N_1/2$<br>$3N_1/2\le M<N_2+N_1/3$ | $N_1\le N_2<3N_1/2$<br>$N_2\le M<3N_1/2$ |
|---|---|---|---|---|---|
| $d_{11}^{(IA)}$ | 0 | $4N_2+2N_1-4M$ | 0 | $4N_2+2N_1-4M$ | $6N_2-4M$ |
| $d_{11}^{(NS)}$ | $N_1$ | $4M-4N_2$ | $N_1$ | $4M-4N_2$ | $6M-6N_2$ |
| $d_{12}^{(IA)}$ | 0 | $4N_2+2N_1-4M$ | 0 | $4N_2+2N_1-4M$ | $6N_2-4M$ |
| $d_{12}^{(NS)}$ | $N_1$ | $4M-4N_2$ | $N_1$ | $4M-4N_2$ | $6M-6N_2$ |
| $d_{21}^{(IA)}$ | 0 | 0 | 0 | 0 | $6N_1-4M$ |
| $d_{21}^{(NS)}$ | $N_2$ | $2M-N_1$ | $N_2$ | $2M-N_1$ | $6M-6N_1$ |
| $d_{22}^{(IA)}$ | 0 | 0 | 0 | 0 | $6N_1-4M$ |
| $d_{22}^{(NS)}$ | $N_2$ | $2M-N_1$ | $N_2$ | $2M-N_1$ | $6M-6N_1$ |
| $T$ | 1 | 2 | 1 | 2 | 3 |
| **DoF** | $\boldsymbol{N_1+N_2}$ | $\boldsymbol{\frac{2M+N_1}{2}}$ | $\boldsymbol{N_1+N_2}$ | $\boldsymbol{\frac{2M+N_1}{2}}$ | $\boldsymbol{\frac{4M}{3}}$ |

TABLE VI

ACHIEVABLE DOF WHEN $M_1=M_2=M$ TRANSMITTING $N_1,N_2$ RECEIVING ANTENNAS, $M>N_1$, $M<N_2$

OUTER BOUND: $\min\left(2M,N_1+N_2,\frac{2M+N_2}{2},\frac{N_1+2N_2}{2},\frac{2M+2N_2}{3}\right)$, SEE THEOREM 5

THE CASES WHERE THE OUTER BOUND IS MET USING RECIPROCITY MARKED WITH (*)

| | $N_1\le M$<br>$N_2/2\le M<N_1/2+N_2/2$ | $N_1\le M$<br>$N_1/2+N_2/2\le M<3N_1/4+N_2/2$ | $N_1\le M$<br>$3N_1/4+N_2/2\le M<N_2$ | $N_1\le M$<br>$M<N_2/2$ |
|---|---|---|---|---|
| $d_{11}^{(IA)}$ | $4M-2N_2$ | $4M-2N_2$ | $2N_1$ | 0 |
| $d_{11}^{(NS)}$ | 0 | 0 | 0 | 0 |
| $d_{12}^{(IA)}$ | $4M-2N_2$ | $4M-2N_2$ | $2N_1$ | 0 |
| $d_{12}^{(NS)}$ | 0 | 0 | 0 | 0 |
| $d_{21}^{(IA)}$ | $2N_1$ | $6N_1+4N_2-8M$ | 0 | $2N_1$ |
| $d_{21}^{(NS)}$ | $6M-6N_1$ | $6M-6N_1$ | $2N_2-N_1$ | $6M-6N_1$ |
| $d_{22}^{(IA)}$ | $2N_1$ | $6N_1+4N_2-8M$ | 0 | $2N_1$ |
| $d_{22}^{(NS)}$ | $6M-6N_1$ | $6M-6N_1$ | $2N_2-N_1$ | $6M-6N_1$ |
| $T$ | 3 | 3 | 2 | 3 |
| **DoF** | $\boldsymbol{\frac{10M-4N_1-2N_2}{3}}$ $(^*)$ | $\boldsymbol{\frac{2M+2N_2}{3}}$ | $\boldsymbol{\frac{N_1+2N_2}{2}}$ | $\boldsymbol{2M-\frac{4N_1}{3}}$ $(^*)$ |

TABLE VII
ACHIEVABLE DOF WHEN $M_1$=$M_2$=$M$ TRANSMITTING, $N_1$,$N_2$ RECEIVING ANTENNAS, $M<N_1$, $M>N_2$
OUTER BOUND: $\min\left(2M, N_1+N_2, \frac{2M+N_1}{2}, \frac{N_2+2N_1}{2}, \frac{2M+2N_1}{3}\right)$, SEE THEOREM 5
THE CASES WHERE THE OUTER BOUND IS MET USING RECIPROCITY ARE MARKED WITH (*)

| | $N_2 \le M$ <br> $N_1/2 \le M < N_1/2+N_2/2$ | $N_2 \le M$ <br> $N_1/2+N_2/2 \le M < 3N_2/4+N_1/2$ | $N_2 \le M$ <br> $3N_2/4+N_1/2 \le M < N_1$ | $N_2 \le M$ <br> $M < N_1/2$ |
|---|---|---|---|---|
| $d_{11}^{(IA)}$ | $2N_2$ | $6N_2+4N_1-8M$ | 0 | $2N_2$ |
| $d_{11}^{(NS)}$ | $6M-6N_2$ | $6M-6N_2$ | $2N_2-N_1$ | $6M-6N_2$ |
| $d_{12}^{(IA)}$ | $2N_2$ | $6N_2+4N_1-8M$ | 0 | $2N_2$ |
| $d_{12}^{(NS)}$ | $6M-6N_2$ | $6M-6N_2$ | $2N_2-N_1$ | $6M-6N_2$ |
| $d_{21}^{(IA)}$ | $4M-2N_1$ | $4M-2N_1$, | $2N_2$ | 0 |
| $d_{21}^{(NS)}$ | 0 | 0 | 0 | 0 |
| $d_{22}^{(IA)}$ | $4M-2N_1$ | $4M-2N_1$, | $2N_2$ | 0 |
| $d_{22}^{(NS)}$ | 0 | 0 | 0 | 0 |
| $T$ | 3 | 3 | 2 | 3 |
| **DoF** | $\frac{10M-4N_1-2N_2}{3}$ (*) | $\frac{2M+2N_1}{3}$ | $\frac{2N_1+N_2}{2}$ | $2M-\frac{4N_2}{3}$ (*) |

## VIII. OTHER MULTIUSER MIMO CHANNELS

The MIMO $X$ channel is a general multiuser channel that subsumes conventional multiuser channels like the MIMO BC, MAC and IC, see Fig. 4. Likewise, the MIMO $Z$ channel, see Fig. 3, is an additional multiuser channel that can be described by the MIMO $X$ channel when one of the channels matrices is set to zero: Theorem 6 elucidates that the DoF of the MIMO $Z(ij)$ channel are attained in the MIMO $X$ channel by setting message $W_{ij}=\varnothing$ (channel referenced by MIMO X($ij$)), without requiring that one of the channels should be zero, thanks to the efficient use of the interference alignment concept and null-steering transmission. For example, in the MIMO $X$(21) channel depicted in Fig. 4, the receiver $R_2$ observes the transmitted signal associated to message $W_{12}$ (intended to receiver $R_1$) as an interfering signal. That receiver does not observe any additional interference in case transmitter $T_1$ designs its transmit filter in order to align the transmitted signal (message $W_{11}$) with the received signal associated to message $W_{12}$ at receiver $R_2$.

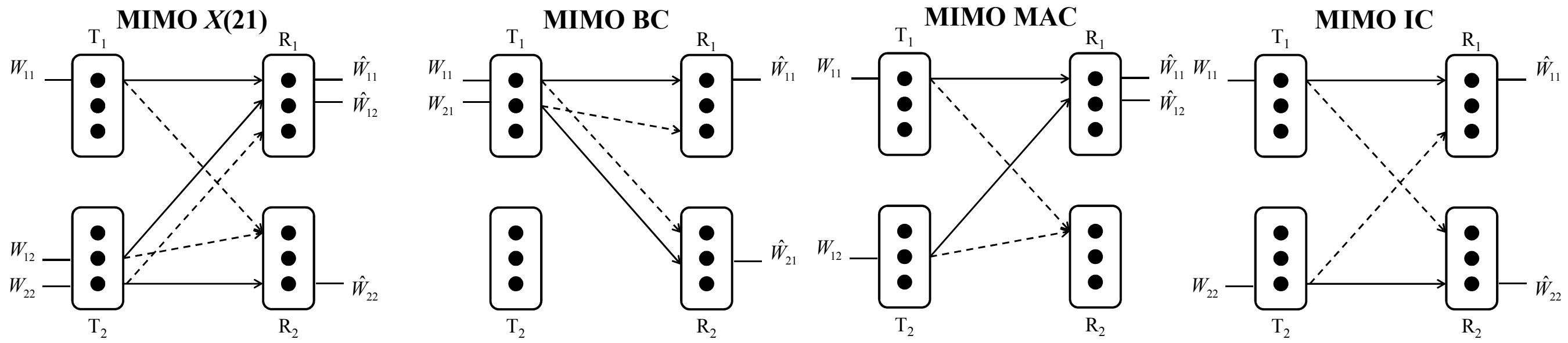

Fig. 4. Channels subsumed in the MIMO $X$ channel. The MIMO $X$(21) channel stands for the MIMO $X$ channel transmitting only three messages, similar to the MIMO $Z$ channel, but in that case one of the links is set to zero.

Therefore, the GSVD-based precoding scheme introduced in section VI provides a suitable framework for designing the precoders and receivers that get the optimal DoF on the two-user MIMO channels. Table VIII depicts the total DoF and the symbol extension ($T$) for the MIMO $X$(21) channel when transmitters and receivers are equipped with $M$ and $N$ antennas. The obtained total DoF coincide with the total DoF of the MIMO $Z$ channel, Theorem 4. On the other hand, the DoF of the BC, MAC and IC for the configuration of 2 transmitters and 2 receivers can be attained using linear filters and zero-forcing strategy with a symbol extension $T$=1, without requiring the interference alignment concept. The DoF of the MIMO BC are attained using the proposed precoding scheme for the MIMO $X$ channel when one of the sources does not transmit any message, i.e $d_{12}^{(IA)}=d_{12}^{(NS)}=0$, $d_{22}^{(IA)}=d_{22}^{(NS)}=0$ in the example depicted in Fig. 4. In such a case, the null-steering precoders are designed in order to not generate interference, i.e. $\mathbf{Z}_{11} \perp \mathbf{H}_{21}$, $\mathbf{Z}_{21} \perp \mathbf{H}_{11}$. Nevertheless, the transmit precoders using the overlapped space are obtained using the GSVD over the set of matrices $(\mathbf{H}_{21}, \mathbf{H}_{22})$, $(\mathbf{H}_{11}, \mathbf{H}_{12})$. Notice that the overlapped space is not exploited as in the general MIMO *X* case because transmitter $T_2$ does not send any message, but we keep the same nomenclature.

Likewise, the MIMO MAC, where all messages are intended to a single receiver, can be seen as a reciprocal network of the MIMO BC. In the example depicted in Fig. 4 $d_{21}^{(IA)} = d_{21}^{(NS)} = 0$, $d_{22}^{(IA)} = d_{22}^{(NS)} = 0$. Finally, the MIMO IC is obtained by imposing that each transmitter only has one message to a single receiver, $d_{12}^{(IA)} = d_{12}^{(NS)} = 0$, $d_{21}^{(IA)} = d_{21}^{(NS)} = 0$.

TABLE VIII

ACHIEVABLE DOF IN THE MIMO $X(21)$ WHEN $M_1=M_2=M$ TRANSMITTING, $N_1=N_2=N$ RECEIVING ANTENNAS, $M>N$

OUTER BOUND MIMO $Z(21)$: $\min(2N, M)$, SEE THEOREM 4.

| | $N \le M < 4N/3$ | $4N/3 \le M < 5N/3$ | $5N/3 \le M < 2N$ | $2N \le M$ |
|---|---|---|---|---|
| $d_{11}^{(IA)}$ | $2N$ | $2N$ | $2N - M$ | 0 |
| $d_{11}^{(ZI)}$ | $6M - 6N$ | $6M - 8N$ | $2M - 2N$ | $2N$ |
| $d_{12}^{(IA)}$ | $2N$ | $2N$ | $2N - M$ | 0 |
| $d_{12}^{(ZI)}$ | 0 | 0 | 0 | 0 |
| $d_{22}^{(IA)}$ | $8N - 6M$ | $10N - 6M$ | 0 | 0 |
| $d_{22}^{(ZI)}$ | $6M - 6N$ | $6M - 6N$ | $2M - 2N$ | $2N$ |
| $T$ | 3 | 3 | 1 | 1 |
| **DoF** | $\boldsymbol{M}$ | $\boldsymbol{M}$ | $\boldsymbol{M}$ | $\boldsymbol{2N}$ |

## IX. RANK DEFICIENCY

Achievable DoF have been derived for the MIMO *X* channel when the channel matrix coefficients are obtained from a continuous probability distribution. It remains to see what results can be obtained in the rank-deficient case. While no outer bounds are known in this case, the GSVD-based scheme naturally addresses this case because it derives the overlapped and null-steering dimensions as a function of rank of the different channel matrices involved in the communication as it is shown in Definition 1. Notice that the IA and null-steering precoders are subspaces of the overlapped and null-steering spaces defined by the GSVD. The problem $P_0$ introduced in Theorem 10, must be updated to also satisfy the following condition: $d_{ij}^{(NS)} + d_{ij}^{(IA)} \le 2T \cdot rank(\mathbf{H}_{ij})$, assuming a *T* symbol extension along with the asymmetric complex signaling. This new constraint comes up as a result of the maximum number of symbol streams devoted to the intended receiver as a function of the available spatial dimensions in that channel. Consequently, we enunciate a new Theorem accordingly,

*Theorem 11* (*Symbol stream optimization in rank-deficient channels*): Given a $2T$ channel extension using asymmetric complex signaling plus *T*-symbol extension with rank-deficient channels, the number of transmitted symbol streams per precoder, denoted by $d_{ij}^{(IA)}, d_{ij}^{(NS)} \in \mathbb{Z}_+$, are obtained as the solution of the following integer linear programming problem which maximizes the weighted sum achievable DoF of the MIMO *X* channel

$$
\begin{aligned}
(P_1)\quad & \underset{\{d_{ij}^{(IA)}\},\{d_{ij}^{(NS)}\}}{\text{maximize}} \quad \frac{1}{2T}\sum_{i=1}^{2}\sum_{j=1}^{2}\mu_{ij}\left(d_{ij}^{(IA)} + d_{ij}^{(NS)}\right) \\
& \textit{subject to} \\
& \begin{cases}
d_{i1}^{(NS)} + d_{i2}^{(NS)} + d_{i1}^{(IA)} + d_{i2}^{(IA)} + d_{k1}^{(IA)} \le 2TN_i \\
d_{i1}^{(NS)} + d_{i2}^{(NS)} + d_{i1}^{(IA)} + d_{i2}^{(IA)} + d_{k2}^{(IA)} \le 2TN_i \\
d_{1j}^{(NS)} + d_{2j}^{(NS)} + d_{1j}^{(IA)} + d_{2j}^{(IA)} \le 2TM_j \\
d_{ij}^{(IA)} \le 2s_k \\
d_{ij}^{(NS)} \le 2T\phi_{kj} \\
d_{ij}^{(NS)} + d_{ij}^{(IA)} \le 2T \cdot rank(\mathbf{H}_{ij}) \\
d_{ij}^{(IA)}, d_{ij}^{(NS)} \in \mathbb{Z}_+
\end{cases} \quad , \quad i \ne k, \quad i,k,j = 1,2
\end{aligned}
\tag{49}
$$

where $\mu_{ij}$ are the weight factors for the different messages, $s_k$ is the dimension of the overlapped space according to Lemma 2 and $\phi_{kj}$ is the dimension of the null-steering space used by the $j$th transmitter (Lemma 3).

*Proof*: It is the extension of Theorem 10 to deal with rank-deficient channels. Although, the GSVD already takes into account the type of channels when deriving the overlapped ($s_k$) and null-steering ($\phi_{kj}$) spaces, we should add a constraint of the maximum number of symbols streams that can be received through a given channel, $d_{ij}^{(NS)} + d_{ij}^{(IA)} \leq 2T \cdot rank\left(\mathbf{H}_{ij}\right)$ □

Let us remark that the new constraint is meaningless when channels matrices are full rank, in such case problem $P_1$ becomes $P_0$.

*Theorem 12*: (*Inner bound DoF region in rank deficient channels*) The achievable DoF region for the 2-user MIMO *X* channel with $M_1$, $M_2$ transmitting and $N_1$, $N_2$ receiving antennas when precoding is based on null-steering and interference alignment over a constant channel with *T* symbol extensions is

$$\tilde{D}_{in}^{X}(T) = \tilde{Q}_{in}^{X}(T) \cup \tilde{\underleftarrow{Q}}_{in}^{X}(T) \tag{50}$$

where *T* denotes the symbol extension and $Q_{in}^{X}(T)$, $\underleftarrow{Q}_{in}^{X}(T)$ stand for the following DoF region obtained when the precoding is designed in the original or reciprocal MIMO *X* network

$$\begin{aligned}
\tilde{Q}_{in}^{X}(T) = \Big\{ & \left(d_{11}^{(IA)}, d_{12}^{(IA)}, d_{21}^{(IA)}, d_{22}^{(IA)}\right) \in \mathbb{Z}_+^4, \quad \left(d_{11}^{(NS)}, d_{12}^{(NS)}, d_{21}^{(NS)}, d_{22}^{(NS)}\right) \in \mathbb{Z}_+^4, \\
& d_{ij}^{(IA)} \leq 2\left(rank\left(\mathbf{H}_{k1}\right) + rank\left(\mathbf{H}_{k2}\right) - rank\left(\left[\mathbf{H}_{k1}\ \mathbf{H}_{k2}\right]\right)\right), \\
& d_{ij}^{(NS)} \leq 2T\left(M_j - rank\left(\mathbf{H}_{kj}\right)\right), \\
& d_{ij}^{(IA)} + d_{ij}^{(NS)} \leq 2T\ rank\left(\mathbf{H}_{ij}\right) \\
& \tilde{d}_{ij} = \frac{d_{ij}^{(IA)} + d_{ij}^{(NS)}}{2T}, \qquad i,j,k = 1,2, \quad i \neq k \\
& \tilde{d}_{11} + \tilde{d}_{12} + \tilde{d}_{21} \leq N_1 + M_1 - rank\left(\mathbf{H}_{11}\right), \\
& \tilde{d}_{11} + \tilde{d}_{12} + \tilde{d}_{22} \leq N_1 + M_2 - rank\left(\mathbf{H}_{12}\right), \\
& \tilde{d}_{21} + \tilde{d}_{22} + \tilde{d}_{11} \leq N_2 + M_1 - rank\left(\mathbf{H}_{21}\right), \\
& \tilde{d}_{21} + \tilde{d}_{22} + \tilde{d}_{12} \leq N_2 + M_2 - rank\left(\mathbf{H}_{22}\right), \\
& \tilde{d}_{11} + \tilde{d}_{21} \leq M_1, \\
& \tilde{d}_{12} + \tilde{d}_{22} \leq M_2 \\
\Big\} &
\end{aligned} \tag{51}$$

$$\begin{aligned}
\tilde{\underleftarrow{Q}}_{in}^{X}(T) = \Big\{ & \left(d_{11}^{(IA)}, d_{12}^{(IA)}, d_{21}^{(IA)}, d_{22}^{(IA)}\right) \in \mathbb{Z}_+^4, \quad \left(d_{11}^{(NS)}, d_{12}^{(NS)}, d_{21}^{(NS)}, d_{22}^{(NS)}\right) \in \mathbb{Z}_+^4, \\
& d_{ij}^{(IA)} \leq 2\left(rank\left(\mathbf{H}_{1k}\right) + rank\left(\mathbf{H}_{2k}\right) - rank\left(\begin{bmatrix}\mathbf{H}_{1k} \\ \mathbf{H}_{2k}\end{bmatrix}\right)\right), \\
& d_{ij}^{(NS)} \leq 2T\left(N_i - rank\left(\mathbf{H}_{ik}\right)\right), \\
& d_{ij}^{(IA)} + d_{ij}^{(NS)} \leq 2T\ rank\left(\mathbf{H}_{ij}\right) \\
& \tilde{d}_{ij} = \frac{d_{ij}^{(IA)} + d_{ij}^{(NS)}}{2T}, \qquad i,j,k = 1,2, \quad j \neq k \\
& \tilde{d}_{11} + \tilde{d}_{12} + \tilde{d}_{21} \leq N_1 + M_1 - rank\left(\mathbf{H}_{11}\right), \\
& \tilde{d}_{11} + \tilde{d}_{12} + \tilde{d}_{22} \leq N_1 + M_2 - rank\left(\mathbf{H}_{12}\right), \\
& \tilde{d}_{21} + \tilde{d}_{22} + \tilde{d}_{11} \leq N_2 + M_1 - rank\left(\mathbf{H}_{21}\right), \\
& \tilde{d}_{21} + \tilde{d}_{22} + \tilde{d}_{12} \leq N_1 + M_2 - rank\left(\mathbf{H}_{22}\right), \\
& \tilde{d}_{11} + \tilde{d}_{12} \leq N_1, \\
& \tilde{d}_{21} + \tilde{d}_{22} \leq N_2 \\
\Big\} &
\end{aligned} \tag{52}$$

*Proof.* Similar steps as for Theorem 8 have to be followed, but taking into account the dimensions of the overlapped and null-steering spaces (see Lemma 2 and Lemma 3), along with the constraint of the rank of the channel. □

Therefore, the transmit-receive scheme introduced in section VI that optimizes the number of symbol according to Theorem 11, can be used for an achievable scheme in the rank-deficient MIMO *X* channel. Problem $P_1$ can be solved with efficient computational methods for different values of symbol extensions (*T*) to get the maximum weighted sum DoF. Nevertheless, in the following lemmas we provide a closed-form solution of the achievable total DoF for the MIMO *X* channel, BC and IC under rank-deficiency when all terminals are equipped with the same number of antennas:

*Lemma 13* (*Rank-deficient MIMO X*) : In a rank-deficient MIMO *X* channel with *M* antennas at all nodes where channel coefficients satisfy $rank(\mathbf{H}_{ii}) = r_d$ and $rank(\mathbf{H}_{ij}) = r_c$ with *i≠j*, the achievable total DoF using the proposed GSVD-based scheme and Theorem 11 are

$$\bar{\eta}_X = \begin{cases} \frac{4}{3}\left(2M - \frac{r_c + r_d}{2}\right) & if \quad r_c + r_d \geq M \\ 2(r_c + r_d) & if \quad r_c + r_d < M \end{cases} \tag{53}$$

*Proof.* We solve problem $P_1$ with $\mu_{ij}$=1 as an integer linear programming problem for different values of *T* symbol extension in order to find an integer solution that maximizes the achievable sum DoF.

$$d_{11}^{(IA)} = d_{22}^{(IA)} = d_{12}^{(IA)} = d_{12}^{(IA)} = \begin{cases} 2(r_c + r_d - M) & if \quad r_c + r_d \geq M \\ 0 & otherwise \end{cases}$$

$$d_{11}^{(NS)} = d_{22}^{(NS)} = \begin{cases} 6(M - r_c) & if \quad r_c + r_d \geq M \\ 2r_d & otherwise \end{cases}$$

$$d_{12}^{(NS)} = d_{21}^{(NS)} = \begin{cases} 6(M - r_d) & if \quad r_c + r_d \geq M \\ 2r_c & otherwise \end{cases}$$

$$T = \begin{cases} 3 & r_c + r_d \geq M \\ 1 & otherwise \end{cases}$$

□

*Lemma 14* (*Rank-deficient MIMO IC*) : In a rank-deficient MIMO IC channel with *M* antennas at all nodes where channel coefficients satisfy $rank(\mathbf{H}_{ii}) = r_d$ and $rank(\mathbf{H}_{ij}) = r_c$ with *i≠j*, the achievable total DoF using the proposed GSVD-based scheme and Theorem 11 turns out to be,

$$\bar{\eta}_{IC} = \begin{cases} 2M - r_c & if \quad r_c/2 + r_d \geq M, \quad r_c + r_d \geq M \\ 2r_d & if \quad r_c/2 + r_d < M, \quad r_c + r_d \geq M \\ 2r_d & if \quad r_c + r_d < M \end{cases} \tag{54}$$

*Proof.* We solve problem $P_1$ with $\mu_{ij}$=1 as an integer linear programming problem for *T*=1 symbol extension. For the two-user MIMO IC there is not interference alignment, but we keep the same variable definition used in Theorem 11, so that, $d_{12}^{(IA)} = d_{12}^{(NS)} = d_{21}^{(IA)} = d_{21}^{(NS)} = 0$ and

$$d_{11}^{(IA)} = d_{22}^{(IA)} = \begin{cases} r_c & f \quad r_c/2 + r_d \geq M, \quad r_c + r_d \geq M \\ 2(r_c + r_d - M) & if \quad r_c/2 + r_d < M, \quad r_c + r_d \geq M \\ 0 & otherwise \end{cases}$$

$$d_{11}^{(NS)} = d_{22}^{(NS)} = \begin{cases} 2(M - r_c) & if \quad r_c + r_d \geq M \\ 2r_d & otherwise \end{cases}$$

□

Note that this inner bound coincides with the optimal sum DoF given in [14].

*Lemma 15* (*Rank-deficient MIMO BC*) : In a rank-deficient MIMO BC channel with $M$ antennas at all nodes where channel coefficients satisfy $rank(\mathbf{H}_{11}) = r_d$ and $rank(\mathbf{H}_{12}) = r_c$, the achievable total DoF using the proposed GSVD-based scheme and Theorem 11 are,

$$\overline{\eta}_{BC} = \begin{cases} M & if \quad r_c + r_d \geq M \\ r_c + r_d & if \quad r_c + r_d < M \end{cases} \tag{55}$$

*Proof.* We solve problem $P_1$ with $\mu_{ij}$=1 as an integer linear programming problem for $T$=1 symbol extension. We keep the same variable definition used in Theorem 11, so that, $d_{12}^{(IA)} = d_{12}^{(NS)} = d_{22}^{(IA)} = d_{22}^{(NS)} = 0$ and

$$d_{21}^{(IA)} = 0$$

$$d_{11}^{(IA)} = \begin{cases} 2(r_c + r_d - M) & if \quad r_c + r_d \geq M \\ 0 & otherwise \end{cases}$$

$$d_{11}^{(NS)} = \begin{cases} 2(M - r_c) & if \quad r_c + r_d \geq M \\ 2r_d & otherwise \end{cases}$$

$$d_{21}^{(NS)} = \begin{cases} 2(M - r_d) & if \quad r_c + r_d \geq M \\ 2r_c & otherwise \end{cases}$$

. □

With the objective of comparing the total DoF obtained by MIMO BC, IC X channel we have considered a terminal deployment akin to Fig. 1, where all terminals have $M$ antenna elements and channels matrices are rank-deficient ($rank(\mathbf{H}_{11})$=$rank(\mathbf{H}_{22})$=$r_d$, $rank(\mathbf{H}_{12})$=$rank(\mathbf{H}_{22})$=$r_c$). Likewise, we have considered the case when both sources cooperate to transmit to both destinations (Coop BC), for which the total DoF are given by Lemma 13 and assuming one transmitter with 2$M$ antennas and channel matrices to each receiver are equal to [$\mathbf{H}_{11}$ $\mathbf{H}_{12}$], [$\mathbf{H}_{12}$ $\mathbf{H}_{22}$].

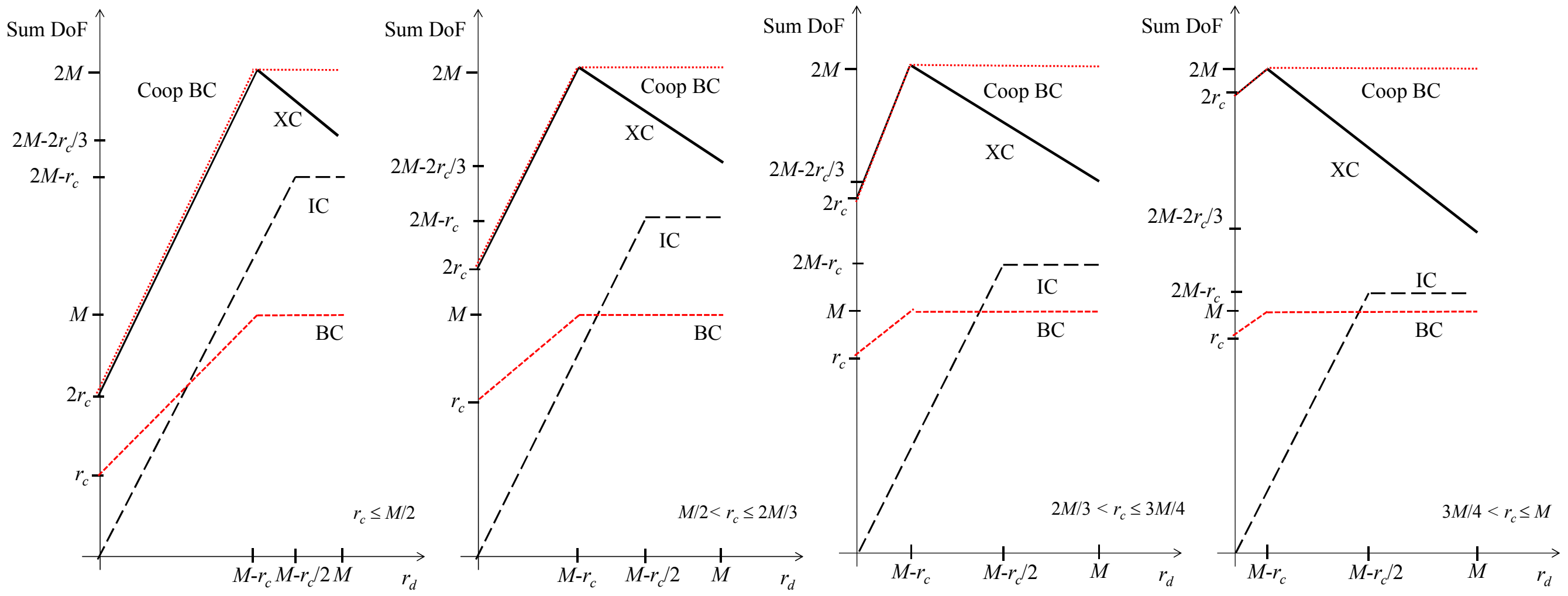

Fig. 5. Total achievable DoF in the BC, IC, Coop BC and $X$ channels for the terminal configuration shown in Fig. 1 with rank-deficient channels and $M$ antennas at all terminals, in four different configurations for the value of $r_c$ $rank(\mathbf{H}_{11}) = rank(\mathbf{H}_{22}) = r_d$, $rank(\mathbf{H}_{12}) = rank(\mathbf{H}_{21}) = r_c$

Fig. 5 presents the achievable DoF of the different transmission strategies as a function of variables $r_c$ and $r_d$ (rank of cross and direct channel matrices). It has to be remarked that the outer bound on the total DoF with rank-deficient channels is only known for the MIMO IC, case that coincides with the achievable total DoF, Lemma 14. The following remarks can be concluded:

- The maximum total DoF of rank-deficient BC and Coop BC are attained whenever $r_c + r_d \geq M$ .
- The total DoF in the rank-deficient IC is superior to full-rank channel case whenever $r_d \geq M/2$
- The total DoF in the rank-deficient MIMO $X$ channel is superior to the full-rank channel case whenever $r_c + r_d \geq 2/3M$
- The MIMO $X$ transmission gets the same DoF than cooperative BC when $r_c + r_d \leq M$ . Otherwise, the total DoF

decreases as $r_c$+$r_d$ increases. This degradation is due to the nature of the MIMO $X$ and the dimension of the overlapping and null-steering spaces. When $r_c + r_d \le M$ the dimension of null-steering space is enough to get the maximum achievable DoF of the GSVD-based algorithm, because each receiver does not observe any interference signal. However, when $r_c + r_d > M$, the dimension of the null-steering space reduces with $r_c$, $r_d$ and hence, the interference alignment concept have to be exploited. It turns out that the aligned interference consumes space dimensions at each receiver and consequently, reduces the total achievable DoF.

## APPENDIX

### A. Lemma 16

In this section we enunciate the ensuing lemma needed in section VI.

*Lemma 16*: The matrix $\mathbf{D} \in \mathbb{R}^{2NT \times 2d}$ with $d \le 2T \cdot \min(N,s)$ is full-column rank with probability one.

$$\mathbf{D} = \begin{bmatrix} \mathbf{\Phi}_A \widehat{\mathbf{T}}_1 & \mathbf{\Phi}_B \widehat{\mathbf{T}}_1 \\ \vdots & \vdots \\ \mathbf{\Phi}_A \widehat{\mathbf{T}}_t & \mathbf{\Phi}_B \widehat{\mathbf{T}}_t \\ \vdots & \vdots \\ \mathbf{\Phi}_A \widehat{\mathbf{T}}_T & \mathbf{\Phi}_B \widehat{\mathbf{T}}_T \end{bmatrix} \tag{56}$$

where matrices $\widehat{\mathbf{T}}_t \in \mathbb{R}^{2s \times d}$ have i.i.d elements and matrices $\mathbf{\Phi}_A \in \mathbb{R}^{2N \times 2s}$, $\mathbf{\Phi}_B \in \mathbb{R}^{2N \times 2s}$ present the following structure,

$$\mathbf{U}(z) = \begin{bmatrix} \cos(z) & -\sin(z) \\ \sin(z) & \cos(z) \end{bmatrix}$$

$$\mathbf{\Phi}_q = \begin{bmatrix} |h_{1,1}^q|\mathbf{U}(\phi_{1,1}^q) & \cdots & |h_{1,s}^q|\mathbf{U}(\phi_{1,s}^q) \\ \vdots & \ddots & \vdots \\ |h_{N,1}^q|\mathbf{U}(\phi_{N,1}^q) & \cdots & |h_{N,s}^q|\mathbf{U}(\phi_{N,s}^q) \end{bmatrix}, \qquad q = A, B \tag{57}$$

where $h_{u,v}^q$ can denote the complex channel coefficient between the $u$th transmitting and $v$th receiving antennas with phase $\phi_{u,v}^q$.

*Proof.* All columns of matrix $\mathbf{D} \in \mathbb{R}^{2NT \times 2d}$ are linearly independent if,

$$\mathbf{D}\begin{bmatrix} \lambda_1 \\ \vdots \\ \lambda_{2d} \end{bmatrix} = \sum_{n=1}^{d} \left(\lambda_n (\mathbf{I}_T \otimes \mathbf{\Phi}_A) + \lambda_{d+n} (\mathbf{I}_T \otimes \mathbf{\Phi}_B)\right) \begin{bmatrix} \widehat{\mathbf{T}}_1 \\ \vdots \\ \widehat{\mathbf{T}}_T \end{bmatrix}^n = \mathbf{0},$$
$$iif \quad \lambda_n = \lambda_{d+n} = 0, \quad n = 1 \ldots d \tag{58}$$

where $\lambda_n \in \mathbb{R}$ and operator $[\mathbf{A}]^i$ selects the $i$th columns of $\mathbf{A}$. Since matrices $\widehat{\mathbf{T}}_t$ are independently generated and $d \le 2T \cdot \min(N,s)$, all vectors in the sum (58) associated to different values of $n$ are linearly independent with probability 1. It remains to check what happens for those vectors associated to the same value of $n$, that is, if:

$$\left(\lambda_n (\mathbf{I}_T \otimes \mathbf{\Phi}_A) + \lambda_{d+n} (\mathbf{I}_T \otimes \mathbf{\Phi}_B)\right) \begin{bmatrix} \widehat{\mathbf{T}}_1 \\ \vdots \\ \widehat{\mathbf{T}}_T \end{bmatrix}^n = 0, \quad \forall \quad n = 1 \ldots d \tag{59}$$

Without loss of generality, let us assume the following structure for matrix $\widehat{\mathbf{T}}_t$

$$\widehat{\mathbf{T}}_t = \begin{bmatrix} \mathrm{Re}\{\overline{\mathbf{T}}_{t,1}\} \\ \mathrm{Im}\{\overline{\mathbf{T}}_{t,1}\} \\ \vdots \\ \mathrm{Re}\{\overline{\mathbf{T}}_{t,s}\} \\ \mathrm{Im}\{\overline{\mathbf{T}}_{t,s}\} \end{bmatrix} \tag{60}$$

where $\overline{\mathbf{T}}_{t,s} \in \mathbb{C}^{1 \times d}$. It turns out that with simple matrix manipulations condition in (59) can be rewritten as,

$$\sum_{m=1}^{s}\left(\lambda_n\left|h_{i,m}^{A}\right|\left(\mathbf{I}_T\otimes\mathbf{U}\left(\phi_{i,m}^{A}\right)\right)+\lambda_{d+n}\left|h_{i,m}^{B}\right|\left(\mathbf{I}_T\otimes\mathbf{U}\left(\phi_{i,m}^{B}\right)\right)\right)\begin{bmatrix}\mathrm{Re}\left\{\overline{\mathbf{T}}_{1,m}\right\}\\ \mathrm{Im}\left\{\overline{\mathbf{T}}_{1,m}\right\}\\ \vdots\\ \mathrm{Re}\left\{\overline{\mathbf{T}}_{T,m}\right\}\\ \mathrm{Im}\left\{\overline{\mathbf{T}}_{T,m}\right\}\end{bmatrix}^{n}=0,\quad \forall i=1..N \tag{61}$$

Furthermore, exploiting the properties of matrix **U**, (57), and the complex variable considered in (60), the previous real equation can be transformed into a complex one,

$$\left(\begin{bmatrix}\left(\lambda_n\left|h_{1,1}^{A}\right|+\lambda_{d+n}\left|h_{1,1}^{B}\right|\exp\left(j\left(\phi_{1,1}^{B}-\phi_{1,1}^{A}\right)\right)\right)\exp\left(j\phi_{1,1}^{A,}\right) & \cdots & \left(\lambda_n\left|h_{1,s}^{A}\right|+\lambda_{d+n}\left|h_{1,s}^{B}\right|\exp\left(j\left(\phi_{1,s}^{B}-\phi_{1,s}^{A}\right)\right)\right)\exp\left(j\phi_{1,s}^{A,}\right)\\ \vdots & \ddots & \vdots\\ \left(\lambda_n\left|h_{N,1}^{A}\right|+\lambda_{d+n}\left|h_{N,1}^{B}\right|\exp\left(j\left(\phi_{N,1}^{B}-\phi_{N,1}^{A}\right)\right)\right)\exp\left(j\phi_{N,1}^{A,}\right) & \cdots & \left(\lambda_n\left|h_{1,s}^{A}\right|+\lambda_{d+n}\left|h_{N,s}^{B}\right|\exp\left(j\left(\phi_{N,s}^{B}-\phi_{N,s}^{A}\right)\right)\right)\exp\left(j\phi_{N,s}^{A,}\right)\end{bmatrix}\otimes\mathbf{I}_T\right)\begin{bmatrix}\overline{\mathbf{T}}_{1,1}\\ \vdots\\ \overline{\mathbf{T}}_{T,1}\\ \vdots\\ \overline{\mathbf{T}}_{1,s}\\ \vdots\\ \overline{\mathbf{T}}_{T,s}\end{bmatrix}^{n}=0$$

which can be written in compact form as,

$$\left(\lambda_n\sum_{j=m}^{s}\left|h_{i,m}^{A}\right|\exp\left(j\left(\phi_{1,m}^{A}-\phi_{i,1}^{A}\right)\right)\left[\overline{\mathbf{T}}_{t,j}\right]^{(n)}+\lambda_{d+n}\sum_{j=m}^{s}\left|h_{i,m}^{B}\right|\exp\left(j\left(\phi_{1,m}^{B}-\phi_{i,1}^{A}\right)\right)\left[\overline{\mathbf{T}}_{t,m}\right]^{(n)}\right)\exp\left(j\phi_{i,1}^{A}\right)=0,\quad \forall n, i=1..N, t=1..T \tag{62}$$

Let us assume that each complex value $\left[\overline{\mathbf{T}}_{t,m}\right]^{n}$ is written as,

$$\left[\overline{\mathbf{T}}_{t,m}\right]^{n}=\left|\kappa_{t,m}^{(n)}\right|^{2}\exp\left(j\varphi_{t,m}^{(n)}\right) \tag{63}$$

The condition shown in (62) becomes,

$$\left|\kappa_{t,m}^{(n)}\right|^{2}\left(\lambda_n\sum_{m=1}^{s}\left|h_{i,m}^{A}\right|\exp\left(j\left(\phi_{i,m}^{A}-\phi_{i,1}^{A}+\varphi_{t,m}^{(n)}\right)\right)+\lambda_{d+n}\sum_{m=1}^{s}\left|h_{i,m}^{B}\right|\exp\left(j\left(\phi_{i,m}^{B}-\phi_{i,1}^{A}+\varphi_{t,m}^{(n)}\right)\right)\right)\exp\left(j\phi_{i,1}^{A}\right)=0,\quad \forall n, i=1..N, t=1..T \tag{64}$$

Since $\lambda_d$, $\lambda_{d+n}$ are real coefficients, the conditions to satisfy (64) can be recast as,

$$\begin{cases}\lambda_n\sum_{m=1}^{s}\left|h_{i,m}^{A}\right|\sin\left(\phi_{i,m}^{A}-\phi_{i,1}^{A}+\varphi_{t,m}^{(n)}\right)+\lambda_{d+n}\sum_{m=1}^{s}\left|h_{i,m}^{B}\right|\sin\left(\phi_{i,m}^{B}-\phi_{i,1}^{A}+\varphi_{t,m}^{(n)}\right)=0\\ \lambda_n\sum_{m=1}^{s}\left|h_{i,m}^{A}\right|\cos\left(\phi_{i,m}^{A}-\phi_{i,1}^{A}+\varphi_{t,m}^{(n)}\right)+\lambda_{d+n}\sum_{m=1}^{s}\left|h_{i,m}^{B}\right|\cos\left(\phi_{i,m}^{B}-\phi_{i,1}^{A}+\varphi_{t,m}^{(n)}\right)=0\end{cases},\quad \forall n, i=1..N, t=1..T \tag{65}$$

When $s>1$ due to the random nature of all modules and phases, the only solution to satisfy conditions in (65) with probability one is $\lambda_n=\lambda_{n+d}=0$. Nevertheless, when $s=1$, the previous conditions are reduced to

$$\begin{cases}\lambda_{d+n}\left|h_{i,1}^{B}\right|\sin\left(\phi_{i,1}^{B}-\phi_{i,1}^{A}\right)=0\\ \lambda_n\left|h_{i,1}^{A}\right|+\lambda_{d+n}\left|h_{i,1}^{B}\right|\cos\left(\phi_{i,1}^{B}-\phi_{i,1}^{A}\right)=0\end{cases},\quad \forall n,\ i=1..N \tag{66}$$

Notice that in case $\phi_{i,1}^{B}-\phi_{i,1}^{A}=0 \bmod (\pi)$ then the previous equation can be satisfied with $\lambda_n\neq 0, \lambda_{n+d}\neq 0$, therefore matrix **D** is not full column rank. However the channel that satisfy such property has zero measure. Hence, we can conclude that with probability 1 matrix **D** is full column rank. □

## B. Proof of Theorem 6

In order to prove Theorem 6 we explicitly solve the optimization problem of Theorem 10 in (40) assuming that one of the messages is set to zero and we compare the obtained DoF with the outer bound derived for the MIMO *Z* channel in Theorem 3. Without loss of generality, we consider the outer bound of MIMO *Z*(21) channel depicted in Fig. 3, where the channel coefficients of the link $\mathbf{H}_{21}$ are null, and we optimize the MIMO *X* precoding when $W_{21}=\varnothing$ ( $d_{21}^{(IA)} = d_{21}^{(NS)} = 0$ ), i.e. MIMO *X*(21). Since we are comparing the total sum DoF we can enforce the following variable definition,

$$d_1^{(IA)} = d_{11}^{(IA)}$$
$$d_2^{(IA)} = d_{21}^{(IA)} = d_{22}^{(IA)}$$

It turns out that the maximization of the number of DoF in the MIMO *X*(21) channel is equivalent to,

$$\left(\mathrm{P}_{X(21)}\right) \quad \underset{d_1^{(IA)},d_2^{(IA)},d_{11}^{(NS)},d_{12}^{(NS)},d_{22}^{(NS)}}{\text{maximize}} \quad \frac{1}{2T}\left(2d_1^{(IA)} + d_2^{(IA)} + d_{11}^{(NS)} + d_{12}^{(NS)} + d_{22}^{(NS)}\right)$$

$$\textit{subject to}$$

$$\begin{cases} d_{11}^{(NS)} + d_{12}^{(NS)} + 2d_1^{(IA)} + d_2^{(IA)} \le 2TN_1 \\ d_{22}^{(NS)} + 2d_2^{(IA)} + d_1^{(IA)} \le 2TN_2 \\ d_{11}^{(NS)} + d_1^{(IA)} \le 2TM_1 \\ d_{12}^{(NS)} + d_{22}^{(NS)} + d_1^{(IA)} + d_2^{(IA)} \le 2TM_2 \\ d_1^{(IA)} \le 2s_2 \\ d_2^{(IA)} \le 2s_1 \\ d_{ij}^{(NS)} \le 2T\phi_{ij} \qquad i,j = 1,2 \\ d_1^{(IA)}, d_2^{(IA)}, d_{11}^{(NS)}, d_{12}^{(NS)}, d_{22}^{(NS)} \in \mathbb{Z}_+ \end{cases} \tag{67}$$

where the dimensions of the overlapped and null-steering subspaces according to Lemma 2 and Lemma 3 are given by,

$$\begin{cases} s_k = \min\left(N_k, M_1\right) + \min\left(N_k, M_2\right) - \min\left(N_k, M_1 + M_2\right) \\ \phi_{ij} = \left(M_j - N_k\right), \quad k \ne i, \quad i,j = 1,2 \end{cases}$$

Table IX- Table XVI (at the end of this section) present the solution of the problem depicted in (67) evaluated for different values of *T* in order to find an integer solution for the optimizing variables. The following antenna configuration cases have been considered:

- $M_1 \ge N_1$, $M_1 \ge N_2$, $M_2 \ge N_1$, $M_2 \ge N_2$ (Table IX and Table X)
- $M_1 \ge N_1$, $M_1 < N_2$, $M_2 \ge N_1$, $M_2 \ge N_2$ (Table XI)
- $M_1 \ge N_1$, $M_1 \ge N_2$, $M_2 \ge N_1$, $M_2 < N_2$ (Table XII)
- $M_1 < N_1$, $M_1 \ge N_2$, $M_2 \ge N_1$, $M_2 \ge N_2$ (Table XIII)
- $M_1 < N_1$, $M_1 < N_2$, $M_2 \ge N_1$, $M_2 \ge N_2$ (Table XIV)
- $M_1 \ge N_1$, $M_1 \ge N_2$, $M_2 < N_1$, $M_2 \ge N_2$ (Table XV)
- $M_1 \ge N_1$, $M_1 \ge N_2$, $M_2 < N_1$, $M_2 < N_2$ (Table XVI)

In all investigated cases the outer bound defined by Theorem 3 is tight, except when $N_1+N_2 \ge M_1+M_2$ for the cases $\{M_1<N_1, M_1<N_2, M_2 \ge N_1, M_2 \ge N_2\}$ and $\{M_1 \ge N_1, M_1 \ge N_2, M_2<N_1, M_2<N_2\}$, see fourth and second rows of Table XIV and Table XVI, respectively. In such a case we analyze the reciprocal MIMO *Z*(21) channel which is shown in Fig. 6 and it is obtained by exchanging transmitters and receivers with respect Fig. 3. Notice that due to the nomenclature employed for defining the messages and channel matrices ( $\mathbf{H}_{ij} = \underleftarrow{\mathbf{H}}_{ji}^T$ ), the reciprocal MIMO *Z*(21) becomes a MIMO *Z*(12), with $\underleftarrow{M}_1 = N_1, \underleftarrow{M}_2 = N_2$ and $\underleftarrow{N}_1 = M_1, \underleftarrow{N}_2 = M_2$ transmitting and receiving antennas, respectively. The outer bound on DoF are the same in reciprocal and original MIMO *Z* channels, see Theorem 3.

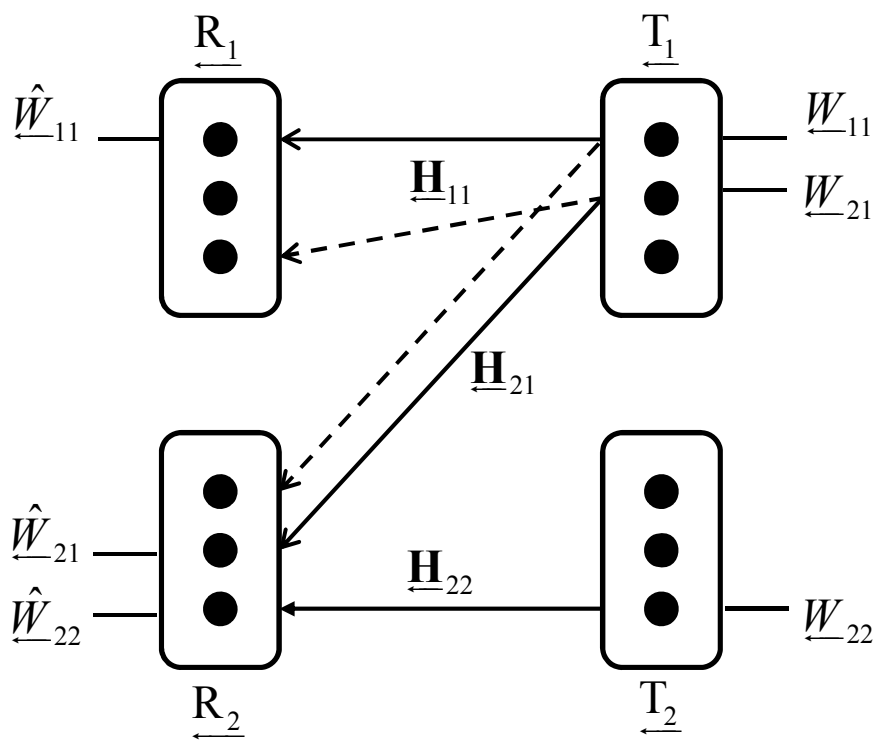

Fig. 6. Reciprocal MIMO $Z$ (21) channel with 2 sources-destinations, the original MIMO $Z$ (21) channel is presented in Fig. 3. Intended signals are referenced by solid lines, while unintended signals are denoted by dotted lines. The antenna configuration is connected to the original network by $\underleftarrow{M}_1 = N_1, \underleftarrow{M}_2 = N_2$ and $\underleftarrow{N}_1 = M_1, \underleftarrow{N}_2 = M_2$ .

In the following we show that outer bound on the total DoF of the MIMO $Z(12)$ for antenna configuration $N_1+N_2 \geq M_1+M_2$ , whenever $\{M_1<N_1,\ M_1<N_2,\ M_2 \geq N_1,\ M_2 \geq N_2\}$ or $\{M_1 \geq N_1,\ M_1 \geq N_2,\ M_2<N_1,\ M_2<N_2\}$ is attained with equality by solving the reciprocal MIMO $X(12)$ network ( $\underleftarrow{W}_{12} = \varnothing$ ) and assuming that

$$\begin{aligned} \underleftarrow{d}_1^{(IA)} &= \underleftarrow{d}_{11}^{(IA)} = \underleftarrow{d}_{21}^{(IA)} \\ \underleftarrow{d}_2^{(IA)} &= \underleftarrow{d}_{22}^{(IA)} \end{aligned}$$

The symbol streams to be transmitted by the different precoders are obtained as a solution of the following linear programming problem, derived from Theorem 10, when $\underleftarrow{d}_{12}^{(IA)} = \underleftarrow{d}_{12}^{(NS)} = 0$ ,

$$\left(\mathrm{P}_{X(12)}\right) \quad \underset{\underleftarrow{d}_1^{(IA)}, \underleftarrow{d}_2^{(IA)}, \underleftarrow{d}_{21}^{(NS)}, \underleftarrow{d}_{22}^{(NS)}}{\text{maximize}} \quad \frac{1}{2T}\left(\underleftarrow{d}_1^{(IA)} + 2\underleftarrow{d}_2^{(IA)} + \underleftarrow{d}_{21}^{(NS)} + \underleftarrow{d}_{22}^{(NS)}\right)$$

*subject to*

$$\begin{cases} \underleftarrow{d}_1^{(IA)} + \underleftarrow{d}_2^{(IA)} \leq 2T\underleftarrow{N}_1 \\ \underleftarrow{d}_{21}^{(NS)} + \underleftarrow{d}_{22}^{(ZI)} + 2\underleftarrow{d}_2^{(IA)} + \underleftarrow{d}_1^{(IA)} \leq 2T\underleftarrow{N}_2 \\ \underleftarrow{d}_{21}^{(NS)} + \underleftarrow{d}_1^{(IA)} + \underleftarrow{d}_2^{(IA)} \leq 2T\underleftarrow{M}_1 \\ \underleftarrow{d}_{22}^{(NS)} + \underleftarrow{d}_2^{(IA)} \leq 2T\underleftarrow{M}_2 \\ \underleftarrow{d}_1^{(IA)} \leq 2\left(\underleftarrow{M}_1 + \underleftarrow{M}_2 - \underleftarrow{N}_2\right) \\ \underleftarrow{d}_2^{(IA)} \leq 2\underleftarrow{N}_1 \\ \underleftarrow{d}_{21}^{(NS)} \leq 2T\left(\underleftarrow{M}_1 - \underleftarrow{N}_1\right)^+ \\ \underleftarrow{d}_{22}^{(NS)} \leq 2T\left(\underleftarrow{M}_2 - \underleftarrow{N}_1\right)^+ \\ \underleftarrow{d}_1^{(IA)}, \underleftarrow{d}_2^{(IA)}, \underleftarrow{d}_{21}^{(NS)}, \underleftarrow{d}_{22}^{(NS)} \in \mathbb{Z}_+ \end{cases} \qquad (68)$$

In case $N_1+N_2 \geq M_1+M_2$ and $\{M_1<N_1,\ M_1<N_2,\ M_2 \geq N_1,\ M_2 \geq N_2\}$ the conditions on the reciprocal network becomes $\underleftarrow{M}_1 + \underleftarrow{M}_2 \geq \underleftarrow{N}_1 + \underleftarrow{N}_2$ and $\left\{\underleftarrow{N}_1 < \underleftarrow{M}_1, \underleftarrow{N}_1 < \underleftarrow{M}_2, \underleftarrow{N}_2 \geq \underleftarrow{M}_1, \underleftarrow{N}_2 \geq \underleftarrow{M}_2\right\}$ and the solution to $\mathrm{P}_{X(12)}$ is equal to

$$\begin{cases} \mathrm{DoF} = \underleftarrow{M}_1 \\ \underleftarrow{d}_1^{(IA)} = \underleftarrow{d}_2^{(IA)} = 2\underleftarrow{N}_2 \\ \underleftarrow{d}_{11}^{(NS)} = 6\underleftarrow{M}_1 - 6\underleftarrow{N}_2 \\ \underleftarrow{d}_{12}^{(NS)} = \underleftarrow{d}_{22}^{(ZI)} = 0 \\ T = 3 \end{cases}$$

The number of transmitted symbol streams of the different precoders in the original network are obtained by applying $d_{ij} = \underleftarrow{d}_{ji}$, see section II.C. Furthermore, the results obtained when $N_1+N_2 \geq M_1+M_2$ and $\{M_1 \geq N_1,\ M_1 \geq N_2,\ M_2<N_1,\ M_2<N_2\}$ are presented in Table XVII. For both configurations the outer bound on the total DoF of the MIMO $Z(12)$ (or reciprocal MIMO $Z(21)$) is attained tightly. We would like to remark that the antenna cases not considered in this proof, naturally comes up when we solve reciprocal channel, for example $\{M_1<N_1,\ M_1<N_2,\ M_2<N_1,\ M_2<N_2\}$ which is the case $\{\underline{M}_1 > \underline{N}_1, \underline{M}_2 > \underline{N}_1, \underline{M}_1 > \underline{N}_2, \underline{M}_2 > \underline{N}_2\}$ for a MIMO $X(12)$. The same approach can be followed to show that the outer DOF of the MIMO Z(11), MIMO Z(21) and MIMO Z(22) are attained by the achievable schemes of the MIMO X(11), MIMO X(21), MIMO X(22) and their respective reciprocal channels. Hence we can conclude that the outer bound on the total DoF of the MIMO *Z* are attained by the achievable scheme derived for the MIMO *X*. □

TABLE IX
ACHIEVABLE DOF AFTER SOLVING PROBLEM P$X$(21) IN (67) WHEN $M_1 \geq N_1$, $M_1 \geq N_2$, $M_2 \geq N_1$, $M_2 \geq N_2$, $N_2 \geq N_1$.
OUTER BOUND: MIN($N_1+N_2$, $M_2$), SEE THEOREM 4

| | $d_1^{(IA)}$ | $d_2^{(IA)}$ | $d_{11}^{(NS)}$ | $d_{12}^{(NS)}$ | $d_{22}^{(NS)}$ | $T$ | **DoF** |
|---|---|---|---|---|---|---|---|
| $N_1 \geq 3M_2 + 3M_1 - 6N_2$ | $2N_1$ | $2N_1 - 6M_2 - 6M_1 + 12N_2$ | $6M_1 - 6N_2$ | $6M_2 - 6N_2$ | $6M_2 - 6N_1$ | 3 | $\boldsymbol{M_2}$ |
| $N_1 \geq M_2 + M_1 - 2N_2$<br>$N_1 < 3M_2 + 3M_1 - 6N_2$ | $3N_1 - 3M_2 - 3M_1 + 6N_2$ | 0 | $6M_1 - 6N_2$ | $6M_2 - 6N_2$ | $6M_2 - 6N_1$ | 3 | $\boldsymbol{M_2}$ |
| $N_1 \geq M_2 - N_2$<br>$N_1 < M_2 + M_1 - 2N_2$ | 0 | 0 | $2N_1 - 3M_2 + 6N_2$ | $2M_2 - 2N_2$ | $2M_2 - 2N_1$ | 1 | $\boldsymbol{M_2}$ |
| $N_1 < M_2 - N_2$ | 0 | 0 | 0 | $2N_2$ | $2N_1$ | 1 | $\boldsymbol{N_1 + N_2}$ |

TABLE X
ACHIEVABLE DOF AFTER SOLVING PROBLEM P$X$(21) IN (67) WHEN $M_1 \geq N_1$, $M_1 \geq N_2$, $M_2 \geq N_1$, $M_2 \geq N_2$, $N_2 < N_1$.
OUTER BOUND: MIN($N_1+N_2$, $M_2$), SEE THEOREM 4

| | $d_1^{(IA)}$ | $d_2^{(IA)}$ | $d_{11}^{(NS)}$ | $d_{12}^{(NS)}$ | $d_{22}^{(NS)}$ | $T$ | **DoF** |
|---|---|---|---|---|---|---|---|
| $N_1 \geq 3/2 M_2 - N_2$ | $2N_1$ | $2N_2$ | 0 | $2N_1 - 2N_2$ | $6M_2 - 6N_1$ | 3 | $\boldsymbol{M_2}$ |
| $N_1 < 3/2 M_2 - N_2$ | $2N_1$ | $4N_1 - 6M_2 + 6N_2$ | 0 | $6M_2 - 2N_1 - 6N_2$ | $6M_2 - 6N_1$ | 3 | $\boldsymbol{M_2}$ |

TABLE XI
ACHIEVABLE DOF AFTER SOLVING PROBLEM P$X$(21) IN (67) WHEN $M_1 \geq N_1$, $M_1 < N_2$, $M_2 \geq N_1$, $M_2 \geq N_2$.
OUTER BOUND: MIN($N_1+N_2$, $M_2$), SEE THEOREM 4

| | $d_1^{(IA)}$ | $d_2^{(IA)}$ | $d_{11}^{(NS)}$ | $d_{12}^{(NS)}$ | $d_{22}^{(NS)}$ | $T$ | **DoF** |
|---|---|---|---|---|---|---|---|
| $N_1 \geq 3M_2 - 6N_2$ | $2N_1$ | $2N_1$ | 0 | 0 | $6M_2 - 6N_1$ | 3 | $\boldsymbol{M_2}$ |
| $N_1 \geq 3/2 M_2 - 3/2 N_2$<br>$N_1 < 3M_2 - 6N_2$ | $2N_1$ | $2N_1$ | 0 | $6M_2 - 2N_1 - 6N_2$ | $6M_2 - 6N_1$ | 3 | $\boldsymbol{M_2}$ |
| $N_1 \geq M_2 - N_2$<br>$N_1 \geq 3/2 M_2 - 3/2 N_2$ | $2N_1 - 2M_2 + 2N_2$ | 0 | 0 | $4M_2 - 2N_1 - 4N_2$ | $2M_2 - 2N_1$ | 1 | $\boldsymbol{M_2}$ |
| $N_1 < M_2 - N_2$ | 0 | 0 | 0 | $2N_2$ | $2N_1$ | 1 | $\boldsymbol{N_1 + N_2}$ |

TABLE XII
ACHIEVABLE DOF AFTER SOLVING PROBLEM P$X$(21) IN (67) WHEN $M_1 \geq N_1$, $M_1 \geq N_2$, $M_2 \geq N_1$, $M_2 < N_2$.
OUTER BOUND: MIN($N_1+N_2$, $M_2$), SEE THEOREM 4

| | $d_1^{(IA)}$ | $d_2^{(IA)}$ | $d_{11}^{(NS)}$ | $d_{12}^{(NS)}$ | $d_{22}^{(NS)}$ | $T$ | **DoF** |
|---|---|---|---|---|---|---|---|
| $N_1 \geq 3M_2 - 3N_2$ | $2N_1$ | $2N_1$ | 0 | 0 | $6M_2 - 6N_1$ | 3 | $\boldsymbol{M_2}$ |
| $N_1 \geq 3/2 M_2 - 3/2 N_2$<br>$N_1 < 2M_2 - 3/2 N_2$ | $2N_1$ | $4N_1 - 6M_2 + 6N_2$ | $6M_2 - 2N_1 - 6N_2$ | 0 | $6M_2 - 6N_1$ | 3 | $\boldsymbol{M_2}$ |
| $N_1 \geq 2M_2 - 3/2 N_2$<br>$N_1 < 1/3 M_2$ | 0 | $2N_1$ | 0 | 0 | $2M_2 - 2N_1$ | 1 | $\boldsymbol{M_2}$ |
| $N_1 \geq 1/3 M_2$<br>$N_1 < 3/2 M_2 - 3/2 N_2$ | $3N_1 - M_2$ | $2M_2$ | 0 | 0 | $6M_2 - 6N_1$ | 3 | $\boldsymbol{M_2}$ |
| $N_1 < 3/2 M_2 - 3/2 N_2$ | $2N_1$ | $2N_1$ | 0 | 0 | $6M_2 - 6N_1$ | 3 | $\boldsymbol{M_2}$ |

TABLE XIII

ACHIEVABLE DOF AFTER SOLVING PROBLEM P$x$(21) IN (67) WHEN $M_1<N_1$, $M_1\geq N_2$, $M_2\geq N_1$, $M_2\geq N_2$.
OUTER BOUND: MIN($N_1+N_2$, $M_2$), SEE THEOREM 4

| | $d_1^{(IA)}$ | $d_2^{(IA)}$ | $d_{11}^{(NS)}$ | $d_{12}^{(NS)}$ | $d_{22}^{(NS)}$ | $T$ | **DoF** |
|---|---|---|---|---|---|---|---|
| $N_1 < M_2 - N_2$ | 0 | 0 | 0 | $2N_1$ | $2N_1$ | 1 | $\boldsymbol{N_1 + N_2}$ |
| $N_1 \geq M_2 - N_2$<br>$N_1 < 1/3M_1 + M_2 - N_2$ | $2N_1 - 2M_2 + 2N_2$ | 0 | 0 | $4M_2 - 2N_1 + 4N_2$ | $2M_2 - 2N_1$ | 1 | $\boldsymbol{M_2}$ |
| $N_1 \geq 1/3M_1 + M_2 - N_2$<br>$N_1 < 1/3M_1 + M_2 - 2/3N_2$<br>$N_2 < M_2 - 1/3M_1$ | $2M_1$ | $6N_1 - 6M_2 + 6N_2 - 2M_1$ | 0 | $6M_2 - 2M_1 - 6N_2$ | $6M_2 - 6N_1$ | 3 | $\boldsymbol{M_2}$ |
| $N_1 \geq 1/3M_1 + M_2 - N_2$<br>$N_1 < 1/3M_1 + M_2 - 2/3N_2$<br>$N_2 \geq M_2 - 1/3M_1$ | $2M_1$ | 0 | 0 | $6N_1 - 4M_1$ | $6M_2 - 6N_1$ | 3 | $\boldsymbol{M_2}$ |
| $N_1 \geq 1/3M_1 + M_2 - 2/3N_2$<br>$N_1 < 2/3M_1 + M_2 - 2/3N_2$ | $2M_1$ | $2N_2$ | 0 | $6N_1 - 4M_1 - 2N_2$ | $6M_2 - 6N_1$ | 3 | $\boldsymbol{M_2}$ |
| $N_1 \geq 2/3M_1 + M_2 - 2/3N_2$<br>$N_1 < M_1 + M_2 - 2N_2$ | $2M_1$ | $2N_2$ | $6N_1 - 6M_2 - 4M_1 + 2N_2$ | $6M_2 - 6N_2$ | $6M_2 - 6N_1$ | 3 | $\boldsymbol{M_2}$ |
| $N_1 \geq M_1 + M_2 - 2N_2$<br>$N_1 < M_2 - 2/3N_2$ | $6N_1 - 6M_2 + 4N_2$ | $2N_2$ | 0 | $6M_2 - 6N_2$ | $6M_2 - 6N_1$ | 3 | $\boldsymbol{M_2}$ |
| $N_1 \geq M_2 - 2/3N_2$ | 0 | $2N_1 - 2M_2 + 2N_2$ | 0 | $2M_2 - 2N_2$ | $2M_2 - 2N_1$ | 1 | $\boldsymbol{M_2}$ |

TABLE XIV

ACHIEVABLE DOF AFTER SOLVING PROBLEM P$x$(21) IN (67) WHEN $M_1<N_1$, $M_1<N_2$, $M_2\geq N_1$, $M_2\geq N_2$.
OUTER BOUND: MIN($N_1+N_2$, $M_2$), SEE THEOREM 4
THE CASES WHERE THE OUTER BOUND IS MET USING RECIPROCITY ARE MARKED WITH (*)

| | $d_1^{(IA)}$ | $d_2^{(IA)}$ | $d_{11}^{(NS)}$ | $d_{12}^{(NS)}$ | $d_{22}^{(NS)}$ | $T$ | **DoF** |
|---|---|---|---|---|---|---|---|
| $N_1 < 1/3M_1 + M_2 - N_2$ | $2N_1 - 2M_2 + 2N_2$ | 0 | 0 | $4M_2 - 2N_1 - 4N_2$ | $2M_2 - 2N_1$ | 1 | $\boldsymbol{M_2}$ |
| $N_1 \geq 1/3M_1 + M_2 - N_2$<br>$N_1 < 2/3M_1 + M_2 - N_2$ | $2M_1$ | $6N_1 - 6M_2 - 2M_1 + 6N_2$ | 0 | $6M_2 - 2M_1 - 6N_2$ | $6M_2 - 6N_1$ | 3 | $\boldsymbol{M_2}$ |
| $N_1 \geq 2/3M_1 + M_2 - N_2$<br>$N_1 < M_1 + M_2 - N_2$ | $2M_1$ | $2M_1$ | 0 | $6M_2 - 6N_2$ | $6M_2 - 6N_1$ | 3 | $\boldsymbol{M_2}$ |
| $N_1 \geq M_1 + M_2 - N_2$ | $2M_1$ | $2M_1$ | 0 | $6N_1 - 6M_2 - 6M_1 + 6N_2$ | $6M_2 - 6N_1$ | 3 | $\boldsymbol{M_1 + 2M_2 - N_1 - N_2}$ $(^*)$ |

TABLE XV

ACHIEVABLE DOF AFTER SOLVING PROBLEM P$x$(21) IN (67) WHEN $M_1\geq N_1$, $M_1\geq N_2$, $M_2<N_1$, $M_2\geq N_2$.
OUTER BOUND: MIN($M_1+M_2$, $N_1$), SEE THEOREM 4

| | $d_1^{(IA)}$ | $d_2^{(IA)}$ | $d_{11}^{(NS)}$ | $d_{12}^{(NS)}$ | $d_{22}^{(NS)}$ | $T$ | **DoF** |
|---|---|---|---|---|---|---|---|
| $N_1 < 1/3M_2 + N_2$ | $2M_2$ | $6N_1 - 4M_2$ | 0 | 0 | 0 | 3 | $\boldsymbol{N_1}$ |
| $N_1 \geq 1/3M_2 + N_2$<br>$N_2 < 1/3M_2$ | $2N_2$ | 0 | $2N_1 - 4N_2$ | 0 | 0 | 1 | $\boldsymbol{N_1}$ |
| $N_1 \geq 1/3M_2 + N_2$<br>$1/3M_2 \leq N_2 < 1/2M_2$<br>$N_1 < 1/3M_2 + M_1$ | $2M_2$ | $6N_2 - 2M_2$ | $6N_1 - 2M_2 - 6N_2$ | 0 | 0 | 3 | $\boldsymbol{N_1}$ |
| $N_1 \geq 1/3M_2 + N_2$<br>$1/3M_2 \leq N_2 < 1/2M_2$<br>$N_1 \geq 1/3M_2 + M_1$ | $2M_2$ | $6N_2 - 2M_2$ | $6N_1 - 6M_2 + 6N_2$ | $6N_1 - 2M_2 - 6M_1$ | 0 | 3 | $\boldsymbol{N_1}$ |
| $N_1 \geq 1/3M_2 + N_2$<br>$N_2 \geq 1/2M_2$ | $2M_2$ | $2N_2$ | $6N_1 - 4M_2 - 2N_2$ | 0 | 0 | 3 | $\boldsymbol{N_1}$ |

TABLE XVI

ACHIEVABLE DOF AFTER SOLVING PROBLEM P$X$(21) IN (67) WHEN $M_1 \geq N_1$, $M_1 \geq N_2$, $M_2 < N_1$, $M_2 < N_2$.
OUTER BOUND: MIN($M_1+M_2$, $N_1$), SEE THEOREM 4
THE CASES WHERE THE OUTER BOUND IS MET USING RECIPROCITY ARE MARKED WITH (*)

| | $d_1^{(IA)}$ | $d_2^{(IA)}$ | $d_{11}^{(NS)}$ | $d_{12}^{(NS)}$ | $d_{22}^{(NS)}$ | $T$ | **DoF** |
|---|---|---|---|---|---|---|---|
| $N_1 < M_2 + M_1 - N_2$ | $2M_2$ | $2M_2$ | $6N_1 - 6M_2$ | 0 | 0 | 3 | $\boldsymbol{N_1}$ |
| $N_1 \geq M_2 + M_1 - N_2$ | $2M_2$ | $2M_2$ | $6M_1 - 6N_2$ | 0 | 0 | 3 | $\boldsymbol{M_1 + M_2 - N_2}$ $(^*)$ |

TABLE XVII

ACHIEVABLE DOF AFTER SOLVING PROBLEM P$X$(12) IN (68) WHEN $\underline{M}_1 + \underline{M}_2 > \underline{N}_1 + \underline{N}_2$ AND $\{\underline{M}_1 > \underline{N}_1, \underline{M}_1 < \underline{N}_2, \underline{M}_2 > \underline{N}_1, \underline{M}_2 < \underline{N}_2\}$.
OUTER BOUND: MIN($N_1+N_2$, $M_2$)=MIN( $\underline{M}_1 + \underline{M}_2$ , $\underline{N}_2$ ), SEE THEOREM 4

| | $\underline{d}_1^{(IA)}$ | $\underline{d}_2^{(IA)}$ | $\underline{d}_{11}^{(NS)}$ | $\underline{d}_{12}^{(NS)}$ | $\underline{d}_{22}^{(NS)}$ | $T$ | **DoF** |
|---|---|---|---|---|---|---|---|
| $\underline{M}_1 < 4/3\,\underline{N}_1 + \underline{N}_2 - \underline{M}_2$ | $8\underline{N}_1 + 6\underline{N}_2 - 6\underline{M}_1 - 6\underline{M}_2$ | $2\underline{N}_1$ | 0 | $6\underline{M}_1 - 6\underline{N}_1$ | $6\underline{M}_2 - 6\underline{N}_1$ | 3 | $\boldsymbol{\underline{N}_2}$ |
| $\underline{M}_1 \geq 4/3\,\underline{N}_1 + \underline{N}_2 - \underline{M}_2$ <br> $\underline{M}_1 < 2\underline{N}_1 + \underline{N}_2 - \underline{M}_2$ | 0 | $2\underline{N}_1 - \underline{M}_1 - \underline{M}_2 + \underline{N}_2$ | 0 | $2\underline{M}_1 - 2\underline{N}_1$ | $2\underline{M}_2 - 2\underline{N}_1$ | 1 | $\boldsymbol{\underline{N}_2}$ |
| $\underline{M}_1 \geq 2\underline{N}_1 + \underline{N}_2 - \underline{M}_2$ | 0 | 0 | 0 | $2\underline{N}_1 - 2\underline{M}_2 + 2\underline{N}_2$ | $2\underline{M}_2 - 2\underline{N}_1$ | 1 | $\boldsymbol{\underline{N}_2}$ |

### *C. Proof of Theorem 8*

The outer bound region of DoF for the full-rank MIMO $X$ channel is defined by,

$$
\begin{aligned}
D_{out}^{X} = \Big\{ & \left(\hat{d}_{11}, \hat{d}_{12}, \hat{d}_{21}, \hat{d}_{22}\right) \in \mathbb{R}_+^4 \\
& \text{a) } \hat{d}_{11} + \hat{d}_{12} + \hat{d}_{21} \leq \max\left(N_1, M_1\right), \\
& \text{b) } \hat{d}_{11} + \hat{d}_{12} + \hat{d}_{22} \leq \max\left(N_1, M_2\right), \\
& \text{c) } \hat{d}_{21} + \hat{d}_{22} + \hat{d}_{11} \leq \max\left(N_2, M_1\right), \\
& \text{d) } \hat{d}_{21} + \hat{d}_{22} + \hat{d}_{12} \leq \max\left(N_2, M_2\right), \\
& \text{e) } \hat{d}_{11} + \hat{d}_{12} \leq N_1 \\
& \text{f) } \hat{d}_{11} + \hat{d}_{21} \leq M_1 \\
& \text{g) } \hat{d}_{21} + \hat{d}_{22} \leq N_2 \\
& \text{h) } \hat{d}_{12} + \hat{d}_{22} \leq M_2 \\
\Big\} &
\end{aligned}
\tag{69}
$$

The proposed scheme presented in section VI optimizes the symbol streams per message taking into account that are contained in the following symbol stream region for a given $T$ symbol extension, see Theorem 10,

$$
\begin{aligned}
\Upsilon_{in}^{X}(T) = \Big\{ & \left(d_{11}^{(IA)}, d_{11}^{(NS)}, d_{12}^{(IA)}, d_{12}^{(NS)}, d_{21}^{(IA)}, d_{21}^{(NS)}, d_{22}^{(IA)}, d_{22}^{(NS)}\right) \in \mathbb{Z}_+^8 \\
& \text{a) } d_{11}^{(NS)} + d_{12}^{(NS)} + d_{11}^{(IA)} + d_{12}^{(IA)} + d_{21}^{(IA)} \leq 2TN_1, \\
& \text{b) } d_{11}^{(NS)} + d_{12}^{(NS)} + d_{11}^{(IA)} + d_{12}^{(IA)} + d_{22}^{(IA)} \leq 2TN_1, \\
& \text{c) } d_{21}^{(NS)} + d_{22}^{(NS)} + d_{21}^{(IA)} + d_{22}^{(IA)} + d_{11}^{(IA)} \leq 2TN_2, \\
& \text{d) } d_{21}^{(NS)} + d_{22}^{(NS)} + d_{21}^{(IA)} + d_{22}^{(IA)} + d_{12}^{(IA)} \leq 2TN_2, \\
& \text{e) } d_{11}^{(IA)} \leq 2s_2,\ d_{12}^{(IA)} \leq 2s_2, d_{11}^{(NS)} \leq 2T\phi_{21}, d_{12}^{(NS)} \leq 2T\phi_{22} \\
& \text{f) } d_{11}^{(NS)} + d_{21}^{(NS)} + d_{11}^{(IA)} + d_{21}^{(IA)} \leq 2TM_1, \\
& \text{g) } d_{21}^{(IA)} \leq 2s_1,\ d_{22}^{(IA)} \leq 2s_1, d_{21}^{(NS)} \leq 2T\phi_{11}, d_{22}^{(NS)} \leq 2T\phi_{12} \\
& \text{h) } d_{12}^{(NS)} + d_{22}^{(NS)} + d_{12}^{(IA)} + d_{22}^{(IA)} \leq 2TM_2 \\
\Big\} &
\end{aligned}
\tag{70}
$$

where $d_{ij}^{(IA)}, d_{ij}^{(NS)}$ denote the symbol streams used by the IA-based and null-steering based precoders of message $W_{ij}$, $M_1$, $M_2$ stand for the transmitting antennas while $N_1$, $N_2$ are the receiving antennas. Furthermore, according to Lemma 2 and Lemma 3 the dimension of the overlapped and null-steering spaces are given by,

$$\begin{aligned} s_k &= \min\left(N_k, M_1\right) + \min\left(N_k, M_2\right) - \min\left(N_k, M_1 + M_2\right) \\ \phi_{kj} &= M_j - \min\left(N_k, M_j\right) = \left(M_j - N_k\right)^+ \end{aligned} \tag{71}$$

According to (15) the symbol streams and the achievable DoF are connected by,

$$\tilde{d}_{ij} = \tilde{d}_{ij}^{(IA)} + \tilde{d}_{ij}^{(NS)} = \frac{d_{ij}^{(IA)}}{2T} + \frac{d_{ij}^{(NS)}}{2T} \tag{72}$$

Therefore, the inner bound of DoF defined by the achievable scheme is defined by,

$$\begin{aligned} Q_{in}^X(T) = \Big\{ & \left(\tilde{d}_{11}^{(IA)}(T), \tilde{d}_{11}^{(NS)}(T), \tilde{d}_{12}^{(IA)}(T), \tilde{d}_{12}^{(NS)}(T), \tilde{d}_{21}^{(IA)}(T), \tilde{d}_{21}^{(NS)}(T), \tilde{d}_{22}^{(IA)}(T), \tilde{d}_{22}^{(NS)}(T)\right) \in \mathbb{R}_+^8 \\ & \text{a) } \tilde{d}_{11} + \tilde{d}_{12} + \tilde{d}_{21}^{(IA)} \le N_1, \\ & \text{b) } \tilde{d}_{11} + \tilde{d}_{12} + \tilde{d}_{22}^{(IA)} \le N_1, \\ & \text{c) } \tilde{d}_{21} + \tilde{d}_{22} + \tilde{d}_{11}^{(IA)} \le N_2, \\ & \text{d) } \tilde{d}_{21} + \tilde{d}_{22} + \tilde{d}_{12}^{(IA)} \le N_2, \\ & \text{e) } \tilde{d}_{11}^{(IA)} \le \frac{s_2}{T},\ \tilde{d}_{12}^{(IA)} \le \frac{s_2}{T}, \tilde{d}_{11}^{(NS)} \le \left(M_1 - N_2\right)^+, \tilde{d}_{12}^{(NS)} \le \left(M_2 - N_2\right)^+ \\ & \text{f) } \tilde{d}_{11} + \tilde{d}_{21} \le M_1, \\ & \text{g) } \tilde{d}_{21}^{(IA)} \le \frac{s_1}{T},\ \tilde{d}_{22}^{(IA)} \le \frac{s_1}{T}, \tilde{d}_{21}^{(NS)} \le \left(M_1 - N_1\right)^+, \tilde{d}_{22}^{(NS)} \le \left(M_2 - N_1\right)^+ \\ & \text{h) } \tilde{d}_{12} + \tilde{d}_{22} \le M_2 \\ \Big\} & \end{aligned} \tag{73}$$

Notice that we get the same constraints f), g) in the inner bound region that in the outer bound, (69). Moreover, since conditions a), b), c), d) of the inner bound are satisfied, that means that conditions e) and g) of the outer bound , (69), also are fulfilled. Furthermore, let us combine constraints a), b), c), d) with the constraints associated to $\tilde{d}_{21}^{(NS)}, \tilde{d}_{22}^{(NS)}$, $\tilde{d}_{11}^{(NS)}$ , $\tilde{d}_{12}^{(NS)}$, respectively, and use the following equality $\max\left(A, B\right) = B + \left(B - A\right)^+$ . The inner bound region is

$$\begin{aligned} Q_{in}^X(T) = \Big\{ & \left(\tilde{d}_{11}^{(IA)}(T), \tilde{d}_{11}^{(NS)}(T), \tilde{d}_{12}^{(IA)}(T), \tilde{d}_{12}^{(NS)}(T), \tilde{d}_{21}^{(IA)}(T), \tilde{d}_{21}^{(NS)}(T), \tilde{d}_{22}^{(IA)}(T), \tilde{d}_{22}^{(NS)}(T)\right) \in \mathbb{R}_+^8 \\ & \text{a) } \tilde{d}_{11} + \tilde{d}_{12} + \tilde{d}_{21}^{(IA)} \le \max\left(N_1, M_1\right), \\ & \text{b) } \tilde{d}_{11} + \tilde{d}_{12} + \tilde{d}_{22}^{(IA)} \le \max\left(N_1, M_2\right), \\ & \text{c) } \tilde{d}_{21} + \tilde{d}_{22} + \tilde{d}_{11}^{(IA)} \le \max\left(N_2, M_1\right), \\ & \text{d) } \tilde{d}_{21} + \tilde{d}_{22} + \tilde{d}_{12}^{(IA)} \le \max\left(N_2, M_2\right), \\ & \text{e) } \tilde{d}_{11}^{(IA)} \le \frac{s_2}{T},\ \tilde{d}_{12}^{(IA)} \le \frac{s_2}{T}, \tilde{d}_{11}^{(NS)} \le \left(M_1 - N_2\right)^+, \tilde{d}_{12}^{(NS)} \le \left(M_2 - N_2\right)^+ \\ & \text{f) } \tilde{d}_{11} + \tilde{d}_{21} \le M_1, \\ & \text{g) } \tilde{d}_{21}^{(IA)} \le \frac{s_1}{T},\ \tilde{d}_{22}^{(IA)} \le \frac{s_1}{T}, \tilde{d}_{21}^{(NS)} \le \left(M_1 - N_1\right)^+, \tilde{d}_{22}^{(NS)} \le \left(M_2 - N_1\right)^+ \\ & \text{h) } \tilde{d}_{12} + \tilde{d}_{22} \le M_2 \\ \Big\} & \end{aligned} \tag{74}$$

It turns out, that constraints a), b), c), d), f) and h) in the inner and outer bound are the same, and constraints e), g) of the outer bound are satisfied by the inner bound. Hence we can define the inner bound

$$Q_{in}^{X}(T) = D_{out}^{X} \cap \Delta(T) \tag{75}$$

where $\Delta(T)$ is defined in (24) and obtained from the constraints e) and g) in (74).

According to Theorem 1 we have to evaluate the proposed achievable scheme in the original and the reciprocal network, because the imposed design might lead us to different number of DoF. In such a case, exchanging transmitters and receivers and following a similar procedure, taking into account that $d_{ij} = \underleftarrow{d}_{ji}$ we get the following inner bound,

$$\underleftarrow{Q}_{in}^{X}(T) = D_{out}^{X} \cap \underleftarrow{\Delta}(T) \tag{76}$$

where $\underleftarrow{\Delta}(T)$ is defined in (25).

Therefore, the inner bound is defined by

$$D_{in}^{X} = Q_{in}^{X}(T) \cup \underleftarrow{Q}_{in}^{X}(T) = \left(D_{out}^{X} \cap \Delta(T)\right) \cup \left(D_{out}^{X} \cap \underleftarrow{\Delta}(T)\right) \tag{77}$$

## REFERENCES

[1] S. A. Jafar, M. Fakhereddin, "Degrees of freedom on the MIMO interference channel", *IEEE Trans. on Information Theory*, vol. 53, no. 7, July 2007.

[2] Y. Birk. T. Kol, "Informed-source coding-on-demand (ISCOD) over broadcast channels", *Proc. IEEE 17th Annual Joint Conf. on Computer Communications* (INFOCOM), March 1998.

[3] M.A. Maddah-Ali, A.S. Motahari, A.K. Khandani, "Communication over MIMO X channels: Interference Alignment, Decomposition and performance analysis", *IEEE Trans. on Information Theory*, vol.54, no. 8, Aug. 2008

[4] V.R.Cadambe, S.A.Jafar, "Interference Alignment and Degrees of Freedom of the K-user Interference Channel", *IEEE Trans. on Information Theory*, vol.54, no.8, Aug. 2008.

[5] C. Wang, T. Gou, S. Jafar, "Subspace Alignment Chains and the Degrees of Freedom of the Three-User MIMO Interference Channel", arXiv:1109.4350, Information Theory.

[6] S. A. Jafar, *Interference Alignment: A New Look at Signal Dimensions in a Communication Network*. Foundations and Trends in Communications and Information Theory, vol. 7, no. 1, 2010.

[7] S.A. Jafar, S.Shamai, "Degrees of Freedom Region of the MIMO *X* channel", *IEEE Trans. on Information Theory*, vol. 54, no.1, Jan. 2008.

[8] V.R. Cadambe, S.A. Jafar, C. Wang, "Interference Alignment with Asymmetric Complex signalling –settling the Høst-Madsen-Nosratinia conjecture", *IEEE Trans. on Information Theory*, vol.56, no.9, Sep.2010

[9] V. R. Cadambe, S. A. Jafar, "Interference Alignment and the Degrees of Freedom of Wireless *X* Networks", *IEEE Trans. on Information Theory*, vol. 55, no. 9, Sep. 2009

[10] S. H. Mahboubi, A. S. Motahari, A. K. Khandani, "Layered Interference Alignment: Achieving the total DOF of MIMO X channels", in *Proc. IEEE Intl. Symp. Information Theory* (ISIT), Austin, USA, June 2010.

[11] Y. Kwon, T. Hwang, "Sum-Rate Improved Interference Alignment for the *M*×2 MIMO *X* Channel" accepted for publication *in IEEE Communication Letters*. Available: http://ieeexplore.ieee.org/stamp/stamp.jsp?tp=&arnumber=6200379&isnumber=5534602.

[12] A. Agustin, J. Vidal, "Improved Interference Alignment Precoding for the MIMO *X* channel", in *Proc. IEEE Intl. Conf. on Communications* (ICC), Kyoto, Japan, June 2011.

[13] S. H. Chae, S.Y. Chung, "On the degrees of freedom of rank deficient interference channels", in *Proc. IEEE Intl. Symp. Information Theory* (ISIT), St. Petersburg, Russia, August 2011.

[14] S. R. Krishnamurthy, S. A. Jafar, "Degrees of Freedom of 2-user and 3-user Rank-Deficient MIMO Interference Channels", Recent work. Center for Pervasive Communications and Computing, UC Irvine, 2012. Available: http://escholarship.org/uc/item/3gc1z8kn

[15] G.H. Golub, C.F.Van Loan, *Matrix Computations*, 3rd ed. Baltimore: Johns Hopkins Univ. Press, 1996.

[16] C.C. Paige, M.A. Saunders, "Towards a generalized singular value decomposition", *SIAM Journal on Numerical Analysis*, vol.18, no.3, pp. 398-405, Jun. 1981.

[17] J. Park, J. Chun, H. Park, "Efficient GSVD Based multi-user MIMO Linear Precoding and antenna selection scheme", in *Proc. IEEE Intl. Conf. on Communications* (ICC), Dresden, Germany, June 2009.

[18] D. Seneratne, C. Tellambura, "Generalized singular value decomposition for Coordinated Beamforming in MIMO systems", in *Proc. IEEE Global Communications Conf.* (GLOBECOM), Miami, USA, Dec. 2010.

[19] A. Agustin, J. Vidal, "Weighted Sum Rate Maximization for the MIMO X channel through MMSE Precoding", in *Proc. IEEE Intl. Symp. Information Theory* (ISIT), St. Petersburg, Russia, August 2011

[20] L. Ke, Z. Wang, "Degrees of Freedom Regions of Two-User MIMO Z and Full Interference Channels: The Benefit of Reconfigurable Antennas", *IEEE Trans. on Information Theory*, vol. 58, no. 6, June 2012.

[21] D. Tse, P. Vishwanath, Fundamentals of Wireless Communication. Cambridge, U.K., Cambridge Univ. Press, 2005.

[22] K. Gomadam, V. R. Cadambe, S. A. Jafar, "A Distributed Numerical Approach to Interference Alignment and Applications to Wireless Interference Networks", *IEEE Trans. on Information Theory*, vol. 57, no. 6, June 2011.

[23] M. S. Bazaraa, J. J. Jarvis, *Linear Programming and Network Flows*, John Wiley & Sons, Inc., 1977.